\documentclass[floatfix,twocolumn,showpacs,preprintnumbers,amsmath,amssymb,pra,superscriptaddress,longbibliography]{revtex4-1}
\usepackage{color}
\usepackage[usenames,dvipsnames,svgnames,table]{xcolor}
\usepackage[colorlinks=true,linkcolor=blue,urlcolor=blue,citecolor=blue]{hyperref}
\usepackage{mathtools}
\usepackage{graphicx}
\usepackage{dcolumn}
\usepackage{array}
\usepackage{lipsum}
\usepackage{bm}
\usepackage{subfigure}
\usepackage{amssymb}
\usepackage{multirow}
\usepackage{tabularx}
\usepackage{amsmath}
\usepackage{braket}
\usepackage{csquotes}
\graphicspath{{plots/}}
 \usepackage{lipsum}
\usepackage{mathrsfs}
\usepackage{MnSymbol}

\newcommand{\beq}{\begin{equation}}
\newcommand{\eeq}{\end{equation}}
\newcommand{\bea}{\begin{eqnarray}}
\newcommand{\eea}{\end{eqnarray}}

\begin{document}

\title{Dynamic exchange--correlation effects in the strongly coupled electron liquid}

\author{Tobias Dornheim}
\email{t.dornheim@hzdr.de}
\affiliation{Center for Advanced Systems Understanding (CASUS), D-02826 G\"orlitz, Germany}
\affiliation{Helmholtz-Zentrum Dresden-Rossendorf (HZDR), D-01328 Dresden, Germany}

\author{Panagiotis Tolias}
\affiliation{Space and Plasma Physics, Royal Institute of Technology (KTH), Stockholm, SE-100 44, Sweden}

\author{Fotios Kalkavouras}
\affiliation{Space and Plasma Physics, Royal Institute of Technology (KTH), Stockholm, SE-100 44, Sweden}

\author{Zhandos A.~Moldabekov}
\affiliation{Center for Advanced Systems Understanding (CASUS), D-02826 G\"orlitz, Germany}
\affiliation{Helmholtz-Zentrum Dresden-Rossendorf (HZDR), D-01328 Dresden, Germany}

\author{Jan Vorberger}
\affiliation{Helmholtz-Zentrum Dresden-Rossendorf (HZDR), D-01328 Dresden, Germany}

\begin{abstract}
We present the first quasi-exact \textit{ab initio} path integral Monte Carlo (PIMC) results for the  dynamic local field correction $\widetilde{G}(\mathbf{q},z_l;r_s,\Theta)$ in the imaginary Matsubara frequency domain, focusing on the strongly coupled finite temperature uniform electron gas. These allow us to investigate the impact of dynamic exchange--correlation effects onto the static structure factor. Our results provide a straightforward explanation for previously reported spurious effects in the so-called \textit{static approximation} [Dornheim \textit{et al.}, \textit{Phys.~Rev.~Lett.}~\textbf{125}, 235001 (2020)], where the frequency-dependence of the local field correction is neglected. Our findings hint at the intriguing possibility of constructing an analytical four-parameter
representation of $\widetilde{G}(\mathbf{q},z_l;r_s,\Theta)$ valid for a substantial part of the phase diagram, which would constitute key input for thermal density functional theory simulations.
\end{abstract}
\maketitle

\section{Introduction\label{sec:introduction}}

The uniform electron gas (UEG)~\cite{quantum_theory,loos,review} constitutes the archetypal model system of interacting electrons in physics, quantum chemistry, and related fields. Having originally been introduced as a qualitative description of the conduction electrons in simple metals~\cite{mahan1990many}, the UEG exhibits a surprising wealth of intrinsically interesting physical effects such as the \textit{roton}-type negative dispersion relation of its dynamic structure factor~\cite{Takada_PRB_2016,dornheim_dynamic,dynamic_folgepaper,Dornheim_Nature_2022,Dornheim_JCP_Force_2022,koskelo2023shortrange} and Wigner crystallization at very low densities~\cite{Wigner_PhysRev_1934,Drummond_PRB_2004,Azadi_Wigner_2022}. In addition, fast and reliable parametrizations~\cite{vwn,Perdew_Wang_PRB_1992,Perdew_Zunger_PRB_1981,Gori-Giorgi_PRB_2000,cdop} of highly accurate quantum Monte Carlo simulations in the electronic ground state~\cite{Ceperley_Alder_PRL_1980,Spink_PRB_2013,moroni,moroni2} have been of paramount importance for the remarkable success of density functional theory (DFT)~\cite{Jones_RMP_2015}, and a host of other applications.

Very recently, the high interest in inertial confinement nuclear fusion experiments~\cite{Betti2016,Hurricane_RevModPhys_2023,NIF_PRL_2024} and dense astrophysical bodies (giant planet interiors~\cite{Benuzzi_Mounaix_2014}, brown dwarfs~\cite{becker}, white dwarf atmospheres~\cite{Kritcher_Nature_2020}) has triggered similar developments~\cite{Malone_PRL_2016,dornheim_prl,groth_prl,ksdt,Karasiev_status_2019,dornheim_pre,groth_jcp} in the warm dense matter (WDM) regime~\cite{review,new_POP,wdm_book}, which is characterized by the complex interplay of Coulomb coupling, thermal excitations, and partially degenerate electrons. 
The rigorous description of these extreme states thus requires a holistic description, which is challenging~\cite{new_POP}. In practice, the most accurate method is given by the \emph{ab initio} path integral Monte Carlo (PIMC) approach~\cite{cep,Berne_JCP_1982}, which is capable of delivering quasi-exact results for $\Theta\gtrsim1$, with $\Theta=k_\textnormal{B}T/E_\textnormal{F}$ the reduced temperature~\cite{Ott2018} and $E_\textnormal{F}$ the Fermi energy.

A fundamental field of investigation concerns the linear density response of the UEG~\cite{Dornheim_review}, which is fully characterized by the dynamic density response function~\cite{kugler1}
\begin{eqnarray}\label{eq:define_G}
    \chi(\mathbf{q},\omega) = \frac{\chi_0(\mathbf{q},\omega)}{1-4\pi/q^2\left[1-G(\mathbf{q},\omega)\right]\chi_0(\mathbf{q},\omega)}\ ,
\end{eqnarray}
where $\chi_0(\mathbf{q},\omega)$ is the Lindhard function describing the density response of an ideal (i.e., noninteracting) Fermi gas at the same density and temperature. Note that Hartree atomic units are employed throughout this work. Setting $G(\mathbf{q},\omega)\equiv0$ in Eq.~(\ref{eq:define_G}) corresponds to the \emph{random phase approximation} (RPA), where the density response is described on the mean-field level. Therefore, the full wave-vector and frequency resolved information about electronic exchange--correlation (XC) effects is contained in the dynamic local field correction (LFC) $G(\mathbf{q},\omega)$. Naturally, the LFC constitutes important input for a gamut of applications, including the estimation of ionization potential depression~\cite{Zan_PRE_2021}, calculation of the stopping power~\cite{Moldabekov_PRE_2020}, and interpretation of X-ray Thomson scattering (XRTS) measurements~\cite{Fortmann_PRE_2010}. Moreover, it is directly related to the dynamic XC-kernel $K_\textnormal{xc}(\mathbf{q},\omega)$, which is the key quantity in linear-response time-dependent density functional theory (LR-TDDFT) simulations~\cite{book_Ullrich, Moldabekov_PRR_2023, Moldabekov_JCTC_2023}.

A particularly useful result of linear response theory is the well-known fluctuation--dissipation theorem~\cite{quantum_theory}
\begin{eqnarray}\label{eq:FDT}
S(\mathbf{q},\omega) = - \frac{\textnormal{Im}\chi(\mathbf{q},\omega)}{\pi n (1-e^{-\beta\omega})}\ ,
\end{eqnarray}
where $n$ is the electronic number density and $\beta=1/k_\textnormal{B}T$ is the inverse temperature in energy units. It relates $\chi(\mathbf{q},\omega)$ --- an effective single-electron property --- with the dynamic structure factor $S(\mathbf{q},\omega)$, that is defined as the Fourier transform of the intermediate scattering function $F(\mathbf{q},t)=\braket{\hat{n}(\mathbf{q},t)\hat{n}(-\mathbf{q},0)}$~\cite{siegfried_review} and, thus, constitutes an electron--electron correlation function. We note that $S(\mathbf{q},\omega)$ is routinely (though indirectly due to the inevitable convolution with the source-and-instrument function) probed in scattering experiments~\cite{siegfried_review}; for example, XRTS measurements are a key diagnostic technique for experiments with high energy density matter~\cite{Tilo_Nature_2023,Dornheim_T_2022,Dornheim_T2_2022,Frydrych2020,kraus_xrts,boehme2023evidence,Gregori_PRE_2003}. In addition, an integration over the frequency yields the static structure factor (SSF)
\begin{eqnarray}\label{eq:SSF}
    S(\mathbf{q}) = \int_{-\infty}^\infty \textnormal{d}\omega\ S(\mathbf{q},\omega)\,,
\end{eqnarray}
which is connected to the electronic pair correlation function $g(\mathbf{r})$. Finally, the combination of Eqs.~(\ref{eq:define_G}), (\ref{eq:FDT}) and (\ref{eq:SSF}) with the adiabatic connection formula gives one access to the XC-free energy of a given system, which can be used as a promising route for the construction of advanced, nonlocal and explicitly thermal XC-functionals for DFT simulations~\cite{pribram}.

In practice, obtaining accurate results for the dynamic LFC of the UEG is difficult. Consequently, previous efforts were mostly based on various approximations and interpolations, e.g.~\cite{dynamic1,Constantin_PRB_2007,PhysRevB.65.235121,Dabrowski_PRB_1986,Adrienn_PRB_2020,dynamic_ii}. Fairly recently, Dornheim \emph{et al.}~\cite{dornheim_dynamic,dynamic_folgepaper,Dornheim_PRE_2020,Hamann_PRB_2020} have presented the first highly accurate results for $S(\mathbf{q},\omega)$ based on the \emph{analytic continuation}~\cite{JARRELL1996133} of the imaginary-time density--density correlation function (ITCF) $F(\mathbf{q},\tau)=\braket{\hat{n}(\mathbf{q},\tau)\hat{n}(-\mathbf{q},0)}$, cf.~Eq.~(\ref{eq:Laplace}) below. This was achieved through the stochastic sampling of $G(\mathbf{q},\omega)$, which has rendered the analytic continuation practical in the case of the UEG. For completeness, we note that very recently LeBlanc \emph{et al.}~\cite{LeBlanc_PRL_2022} have presented complementary results for lower temperatures. Remarkably, it has been reported~\cite{dornheim_dynamic} that the \emph{static approximation}, i.e., setting $G(\mathbf{q},\omega)\equiv G(\mathbf{q},0)$ in Eq.~(\ref{eq:define_G}) gives very accurate results for $S(\mathbf{q},\omega)$ and related observables~\cite{Hamann_PRB_2020,Hamann_CPP_2020} over a broad range of parameters.
Unfortunately, the situation turned out to be somewhat more complex: while observables such as $S(\mathbf{q},\omega)$ are indeed very accurate for a given wave vector, integrating over $\mathbf{q}$ as it is required to estimate e.g.~the interaction energy leads to a substantial accumulation of individual small errors~\cite{Dornheim_PRL_2020_ESA}; these originate from a systematic overestimation of $S(\mathbf{q})$ for large wave numbers that is connected to the diverging on-top pair correlation function $g(0)$. 

In lieu of a full dynamic LFC, it was subsequently suggested to replace the exact static limit of $G(\mathbf{q},\omega)$ by an effectively frequency-averaged and, therefore, inherently static LFC $G(\mathbf{q})$. The resulting \emph{effective static approximation} (ESA) combines available PIMC results for $G(\mathbf{q},0)$~\cite{dornheim_ML} with a parametrization of restricted PIMC results for $g(0)$~\cite{Brown_PRL_2013} and removes the deficiencies of the static approximation by design.
While being computationally cheap and easy to use for various applications, the ESA scheme is somewhat empirical. Furthermore, it is incapable of providing detailed insights concerning the actual importance of dynamic XC-effects, since the entire frequency dependence has been averaged out for $g(0)$.

To overcome this limitation, we first represent $S(\mathbf{q})$ as a Matsubara series~\cite{IIT}
\begin{eqnarray}\label{eq:Matsubara_Series}
    S(\mathbf{q}) = -\frac{1}{n\beta}
\sum_{l=-\infty}^\infty \widetilde{\chi}(\mathbf{q},z_l)\ ,
\end{eqnarray}
where $z_l=i2\pi l/\beta$ are the imaginary bosonic Matsubara frequencies and where the tilde symbol signifies dynamic quantities whose definition has been extended in the complex frequency domain by means of analytic continuation. We note that Eq.(\ref{eq:Matsubara_Series}) is a key expression in the finite temperature dielectric formalism~\cite{stls,stls_original,stls2,IIT,dornheim_electron_liquid,tolias2024_VS,Tolias_JCP_2021,Tolias_JCP_2023,tanaka_hnc,Tanaka_CPP_2017,arora}. To unambiguously resolve the impact of dynamic XC-effects, we utilize the very recent Fourier-Matsubara series representation of the ITCF $F(\mathbf{q},\tau)$ derived by Tolias~\emph{et al.}~\cite{tolias2024fouriermatsubara} to obtain quasi-exact PIMC results for the dynamic LFC $\widetilde{G}(\mathbf{q},z_l)$ of the UEG in the discrete Matsubara frequency domain.
Our results provide new insights into the validity of the ESA and the deficiencies of the \emph{static approximation}, and further elucidate the complex interplay of quantum delocalization with electronic XC-effects. Moreover, the presented approach opens up new opportunities for the systematic development of fully dynamic dielectric theories and the construction of improved XC-functionals for DFT simulations of WDM and beyond.

The paper is organized as follows. In Sec.~\ref{sec:theory}, we introduce the theoretical background, including the imaginary-time correlation functions~(\ref{sec:ITCF}), the exact Fourier-Matsubara expansion (\ref{sec:matsubara}), the exact high frequency behavior of the dynamic LFC (\ref{sec:asymptoticfrequency}), and the effective static approximation (\ref{sec:ESA}).
In Sec.~\ref{sec:results}, we present our new results for the dynamic density response (\ref{sec:density_response}), for the impact of dynamic XC-effects onto the static structure factor (\ref{sec:SSF}), and an analysis of the high-frequency limit of the LFC (\ref{sec:asymptotic_results}). The paper is concluded by a summary and outlook in Sec.~\ref{sec:summary}.

\section{Theory\label{sec:theory}}

The \emph{ab initio} PIMC method is based on the Feynman imaginary-time path integral formalism of statistical mechanics~\cite{kleinert2009path}, which exploits the isomorphism between the canonical density $\hat{\varrho}=e^{-\beta\hat{H}}$ and time evolution operators within an interval $t=-i\hbar\beta$. Detailed derivation of the method~\cite{cep}, information on efficient path sampling schemes~\cite{boninsegni1,boninsegni2,Dornheim_PRB_nk_2021}, and discussion of the estimation of various imaginary-time correlation functions~\cite{Dornheim_JCP_ITCF_2021} has been presented in the literature and need not be repeated here.

\subsection{Imaginary-time correlation function\label{sec:ITCF}}

The PIMC method is capable of providing asymptotically exact estimates for any well-defined thermodynamic observable. This includes integrated quantities such as the pressure and the energies, various many-body correlation functions, and even the off-diagonal density matrix and the momentum distribution~\cite{MILITZER201913,Dornheim_PRB_nk_2021,Dornheim_PRE_2021}. Moreover, PIMC gives one access to the full \emph{equilibrium dynamics} of the simulated system, but in the imaginary-time domain.
There exist a variety of interesting imaginary-time correlation functions that are connected with dynamic properties; for example, the Matsubara Green function is related to the well-known single-particle spectral function $A(\mathbf{q},\omega)$. In the present work, we focus on the ITCF $F(\mathbf{q},\tau)$ introduced above.

The ITCF $F(\mathbf{q},\tau)$ is connected to the dynamic structure factor $S(\mathbf{q},\omega)$ via a two-sided Laplace transform~\cite{dornheim_dynamic}
\begin{eqnarray}\label{eq:Laplace}
   F(\mathbf{q},\tau) = \mathcal{L}\left[S(\mathbf{q},\omega)\right] =\int_{-\infty}^\infty \textnormal{d}\omega\ S(\mathbf{q},\omega)\ e^{-\tau\omega} \ .
\end{eqnarray}
In principle, the Laplace transform constitutes a unique mapping and $F(\mathbf{q},\tau)$ contains, by definition, the same information as $S(\mathbf{q},\omega)$, only in a different representation. Nevertheless, the numerical inversion of Eq.~(\ref{eq:Laplace}) to solve for $S(\mathbf{q},\omega)$ is an ill-posed problem~\cite{JARRELL1996133}, and the associated difficulties are further exacerbated by the error bars in the ITCF. The inversion problem was successfully overcome for the UEG, for parts of the WDM regime, based on the stochastic sampling of $G(\mathbf{q},\omega)$ taking into account a number of rigorous constraints~\cite{dornheim_dynamic,dynamic_folgepaper,Hamann_PRB_2020}. Similar relations that connect higher-order dynamic structure factors with higher-order imaginary-time correlation functions have been reported in the literature~\cite{Dornheim_JCP_ITCF_2021}.

Another important application of the ITCF is given by the imaginary-time version of the fluctuation--dissipation theorem~\cite{Dornheim_insight_2022}, which relates $F(\mathbf{q},\tau)$ to the static limit of the density response function, see Eq.~(\ref{eq:define_G}),
\begin{eqnarray}\label{eq:static_chi}
    \chi(\mathbf{q},0) = -2n \int_0^{\beta/2} \textnormal{d}\tau\ F(\mathbf{q},\tau)\ .
\end{eqnarray}
In practice, Eq.~(\ref{eq:static_chi}) implies that it is possible to get the full wave number dependence of the static linear density response from a single simulation of the unperturbed system. This relation provided the basis for a number of investigations of the uniform electron gas covering a broad range of parameters~\cite{dornheim_ML,dynamic_folgepaper,dornheim_electron_liquid,dornheim_HEDP,Dornheim_HEDP_2022,Tolias_JCP_2021}. These efforts have recently been extended to the case of warm dense hydrogen~\cite{dornheim2024ab}, providing the first results for the species-resolved static local field factors of a real WDM system. Moreover, there exist nonlinear generalizations of Eq.~(\ref{eq:static_chi}) that relate higher-order ITCFs with different nonlinear density response functions, see Refs.~\cite{Dornheim_JCP_ITCF_2021,Dornheim_CPP_2022,Dornheim_review}.

Furthermore, we point out that the ITCF has emerged as an important quantity in the interpretation of XRTS experiments~\cite{Dornheim_T_2022,Dornheim_T2_2022,dornheim2023xray,Schoerner_PRE_2023,Dornheim_Science_2024,Vorberger_PLA_2024,Dornheim_PRB_2023}, as the deconvolution with respect to the source-and-instrument function is substantially more stable in the Laplace domain compared to the usual frequency domain. For example, the symmetry relation $F(\mathbf{q},\tau)=F(\mathbf{q},\beta-\tau)$ gives one model-free access to the temperature for arbitrarily complicated materials in thermal equilibrium.

For completeness, we note that the fixed node approximation~\cite{Ceperley1991} that is often imposed to circumvent the numerical fermion sign problem~\cite{dornheim_sign_problem} breaks the usual imaginary-time translation invariance of PIMC, and thus access to imaginary-time properties is lost. Hence, we use the direct PIMC method throughout this work, making our simulations computationally costly, but exact within the statistical error bars.

\subsection{Fourier--Matsubara expansion\label{sec:matsubara}}

Tolias and coworkers recently derived an exact Fourier--Matsubara series expansion for the ITCF that reads~\cite{tolias2024fouriermatsubara}
\begin{eqnarray}\label{eq:FMITCF}
    F(\boldsymbol{q},\tau)=-\frac{1}{n\beta}\sum_{l=-\infty}^{+\infty}\widetilde{\chi}(q,z_l)e^{-z_l\tau}\,,
\end{eqnarray}
which constitutes the generalization of the Matsubara series for the SSF, see Eq.(\ref{eq:Matsubara_Series}), since $F(\boldsymbol{q},\tau=0)=S(\boldsymbol{q})$. The coefficients of the Fourier--Matsubara series are given by~\cite{tolias2024fouriermatsubara}
\begin{eqnarray}\label{eq:MDR}
    \widetilde{\chi}(\mathbf{q},z_l) = -2n\int_0^{\beta/2}\textnormal{d}\tau\ F(\mathbf{q},\tau)\ \textnormal{cos}\left(i z_l \tau\right)\,,
\end{eqnarray}
which constitutes the generalization of the imaginary-time version of the fluctuation--dissipation theorem, see Eq.(\ref{eq:static_chi}), since $\widetilde{\chi}(\boldsymbol{q},z_l\to0)=\chi(\mathbf{q},0)$. Finally, solving  Eq.~(\ref{eq:define_G}) for the dynamic Matsubara LFC, we obtain
\begin{eqnarray}\label{eq:LFC}
\widetilde{G}(\mathbf{q},z_l) = 1 - \frac{q^2}{4\pi}\left[
\frac{1}{\widetilde{\chi}_0(\mathbf{q},z_l)} - \frac{1}{\widetilde{\chi}(\mathbf{q},z_l)}
    \right]\,.
\end{eqnarray}

Utilizing these expressions, in order to characterize the impact of dynamic XC-effects, we switch to the Matsubara imaginary frequency domain. We extract the dynamic Matsubara density response function $\widetilde{\chi}(\mathbf{q},z_l)$ from our highly accurate PIMC results for the ITCF via Eq.~(\ref{eq:MDR}), we compute the SSF via the Matsubara series of Eq.~(\ref{eq:Matsubara_Series}), and we calculate the dynamic Matsubara LFC via Eq.~(\ref{eq:LFC}). Such a dynamic Matsubara LFC extraction becomes problematic at high Matsubara orders and at high wave numbers. Thus, exact asymptotic results need to be invoked.

\subsection{Exact high frequency behavior of the Matsubara local field correction}\label{sec:asymptoticfrequency}

The spectral representation of the dynamic LFC formally extends its domain of definition from real frequencies $\omega$ to complex frequencies $z$. It reads~\cite{kugler1}
\begin{eqnarray}\label{eq:Gspectral}
   \widetilde{G}(\mathbf{q},z)=G(\mathbf{q},\infty)-\frac{1}{\pi}\int_{-\infty}^{+\infty}\frac{\Im\{G(\mathbf{q},\omega)\}}{z-\omega}d\omega\,,
\end{eqnarray}
where $G(\mathbf{q},\infty)=\Re\{G(\mathbf{q},\infty)\}$ due to $\Im\{G(\mathbf{q},\infty)\}=0$~\cite{quantum_theory}. This directly leads to $\widetilde{G}(\mathbf{q},\imath\infty)=G(\mathbf{q},\infty)$ for $z\to\imath\infty$.

The high frequency behavior of the dynamic LFC is obtained by combining the high frequency expansion of the real part of Eq.(\ref{eq:define_G}) for the dynamic density response with the f-sum rule and third frequency moment sum rule for the imaginary part of the dynamic density response. It reads~\cite{quantum_theory,kugler1,holas_limit}
\begin{eqnarray}\label{eq:Ginfty}
   G(\mathbf{q},\infty) = I(\mathbf{q}) - \frac{2q^2}{m\omega_\textnormal{p}^2}T_\textnormal{xc}\,,
\end{eqnarray}
where $\omega_\textnormal{p}$ is the electron plasma frequency, $T_\textnormal{xc}=T-T_0$ is the XC contribution to the kinetic energy, and $I(\mathbf{q})$ is a functional of the static structure factor defined as
\begin{align}\label{eq:Iq}
    I(q)&=\frac{1}{8\pi^2n}\int_0^{\infty}dkk^2[S(k)-1]\left[\frac{5}{3}-\frac{k^2}{q^2}+\frac{(k^2-q^2)^2}{2kq^3}\times\right.\nonumber\\&\quad\left.\ln{\left|\frac{k+q}{k-q}\right|}\right]\,,
\end{align}
that is sometimes referred to as the Pathak-Vashishta functional~\cite{PathakVashishtaScheme,NiklassonLimit,SingwiTosi_Review}. Combining the above, in normalized units, the imaginary-frequency LFC at the limit of infinite Matsubara order becomes
\begin{eqnarray}\label{eq:Ginftynorm}
   \widetilde{G}(\mathbf{x},l\to\infty) = I(\mathbf{x})-\frac{3}{2}\pi\lambda{r}_{\mathrm{s}}\tau_{\mathrm{xc}}x^2\,,
\end{eqnarray}
where $\lambda=1/(dq_{\mathrm{F}})=[4/(9\pi)]^{1/3}$ for the numerical constant, $\tau_{\mathrm{xc}}$ is the XC kinetic energy in Hartree units and the Pathak-Vashishta functional is given by
\begin{align}\label{eq:Iqnorm}
    I(x)&=\frac{3}{8}\int_0^{\infty}dyy^2[S(y)-1]\left[\frac{5}{3}-\frac{y^2}{x^2}+\frac{(y^2-x^2)^2}{2yx^3}\times\right.\nonumber\\&\quad\left.\ln{\left|\frac{y+x}{y-x}\right|}\right]\,.
\end{align}
After straightforward Taylor expansions with respect to the small $x/y$ and $y/x$ arguments, respectively, the long wavelength and short wavelength limits of the Pathak-Vashishta functional are found to be~\cite{kugler1}
\begin{align}
    I(x\to0)&=-\frac{1}{5}\pi\lambda{r}_{\mathrm{s}}v_{\mathrm{int}}x^2\,,\label{eq:Iqlong}\\
    I(x\to\infty)&=\frac{2}{3}\left[1-g(0)\right]\,,\label{eq:Iqshort}
\end{align}
where the interaction energy in Hartree units is given by $v_{\mathrm{int}}=(\pi\lambda{r}_{\mathrm{s}})^{-1}\int_0^{\infty}dy\left[S(y)-1\right]$ and the on-top pair correlation function by $g(0)=1+(3/2)\int_0^{\infty}dyy^2\left[S(y)-1\right]$. Combining the above, the long and short wavelength limits of the imaginary-frequency LFC at the limit of infinite Matsubara order become~\cite{kugler1}
\begin{align}
    \widetilde{G}(\mathbf{x}\to0,l\to\infty)&=-\frac{1}{10}\pi\lambda{r}_{\mathrm{s}}\left(2v_{\mathrm{int}}+15\tau_{\mathrm{xc}}\right)x^2\,,\label{eq:Ginflong}\\
    \widetilde{G}(\mathbf{x}\to\infty,l\to\infty)&=-\frac{3}{2}\pi\lambda{r}_{\mathrm{s}}\tau_{\mathrm{xc}}x^2\,.\label{eq:Ginftyshort}
\end{align}

\subsection{Effective static approximation\label{sec:ESA}}

A neural-network representation of the static limit of the fully dynamic LFC, i.e., $G(\mathbf{q},0)=\lim_{\omega\to0}G(\mathbf{q},\omega)=\lim_{l\to0}\widetilde{G}(\mathbf{q},z_l)$, of the warm dense UEG is available~\cite{dornheim_ML}. The corresponding \emph{static approximation} is obtained by setting $G(\mathbf{q},\omega)\equiv{G}(\mathbf{q},0)\equiv{G}(\mathbf{q})$ in Eq.~(\ref{eq:define_G}) and leads to an explicitly static dielectric scheme, in which the large wave number limit of the LFC is connected to the on-top pair correlation function via~\cite{stls,Dornheim_PRL_2020_ESA}
\begin{eqnarray}\label{eq:ontop}
    \lim_{q\to\infty} G(\mathbf{q}) = 1-g(0)\,.
\end{eqnarray}

In practice, the static limit of the fully dynamic LFC quadratically diverges for large $q$~\cite{holas_limit,cdop,Hou_PRB_2022}, which implies an unphysical divergence of $g(0)$.
To correct this error, it has been subsequently suggested to combine the exact static limit for small to intermediate wave numbers $q\lesssim3q_\textnormal{F}$ ($q_\textnormal{F}$ the Fermi wave number~\cite{quantum_theory}) with restricted PIMC results for $g(0)$~\cite{Brown_PRL_2013}.
The resulting \emph{effective static approximation} (ESA) can be expressed as~\cite{Dornheim_PRL_2020_ESA}
\begin{eqnarray}\label{eq:ESA}
    G_\textnormal{ESA}(\mathbf{q}) = A(\mathbf{q})\left[1-g(0)\right] + G(\mathbf{q},0)\left[1-A(\mathbf{q})\right]\ ,
\end{eqnarray}
where $A(\mathbf{q})$ is a suitable activation function for which $A(0)=0$ and $A(q\to\infty)=1$.
A corresponding analytical parametrization of Eq.(\ref{eq:ESA}) has also been proposed~\cite{Dornheim_PRB_ESA_2021}.



\section{Results\label{sec:results}}

\begin{figure*}\centering
\includegraphics[width=0.45\textwidth]{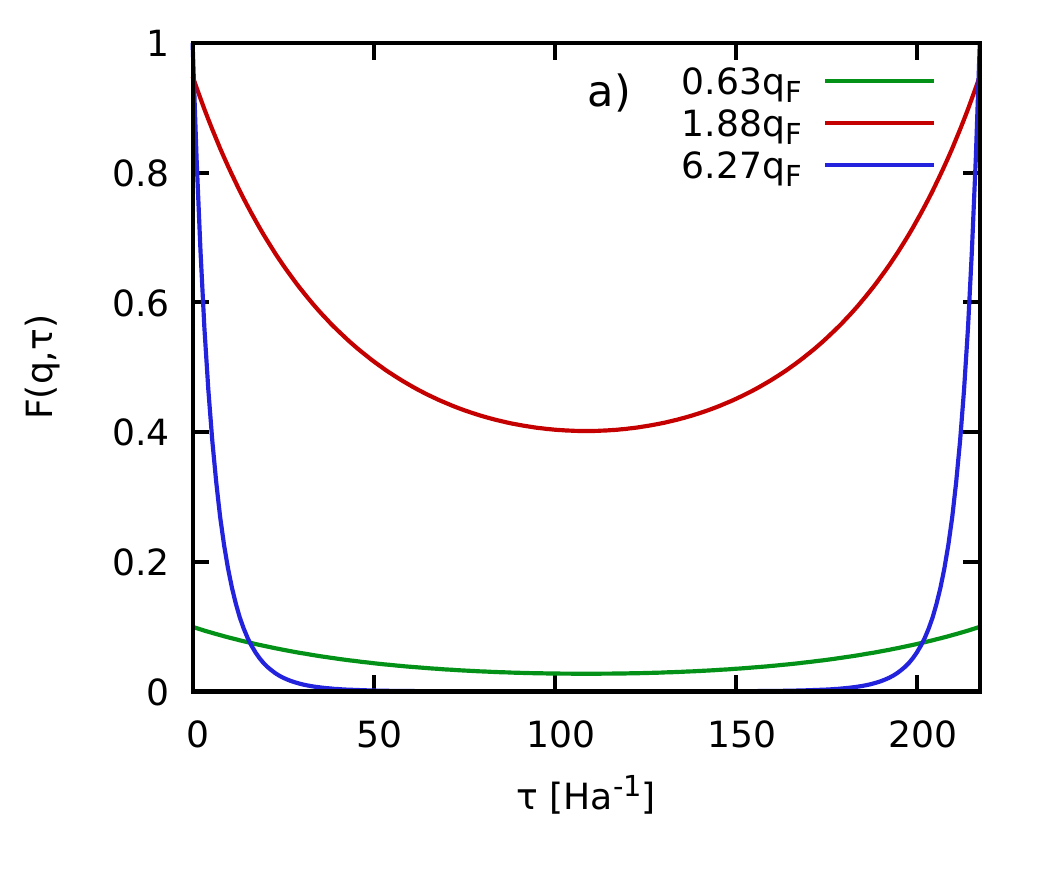}
\includegraphics[width=0.45\textwidth]{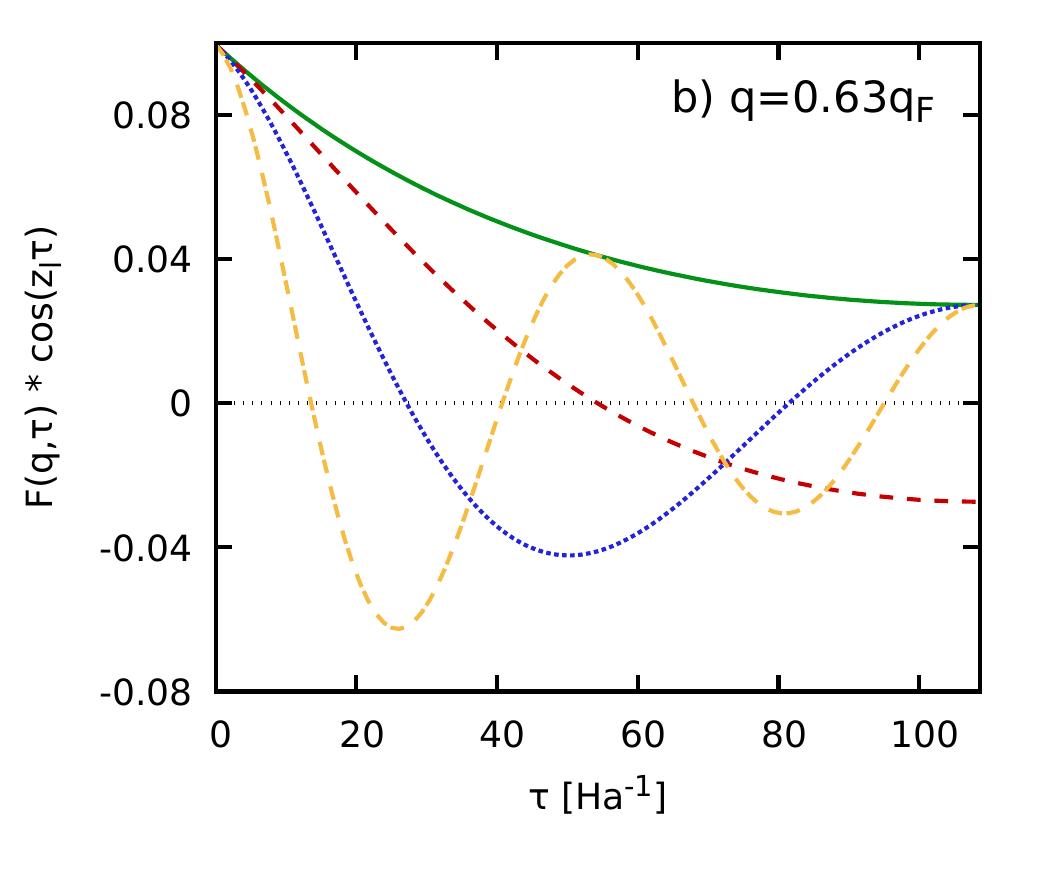}\\
\includegraphics[width=0.45\textwidth]{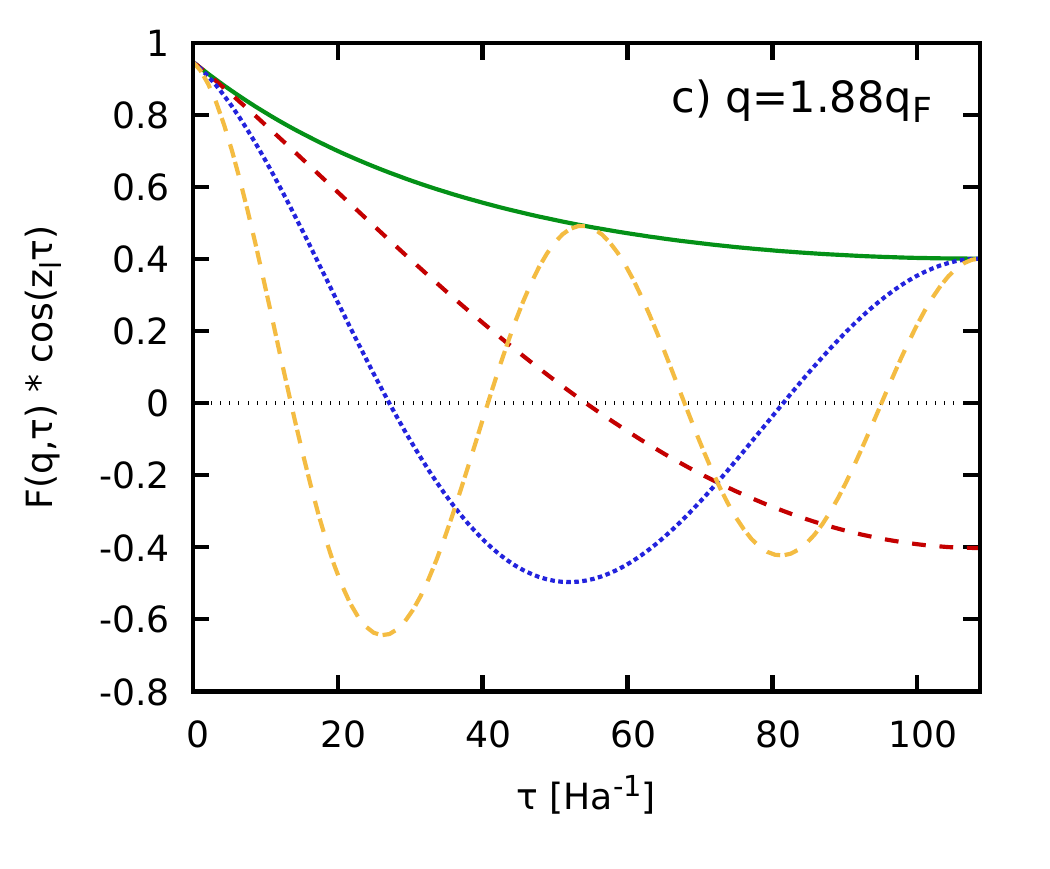}
\includegraphics[width=0.45\textwidth]{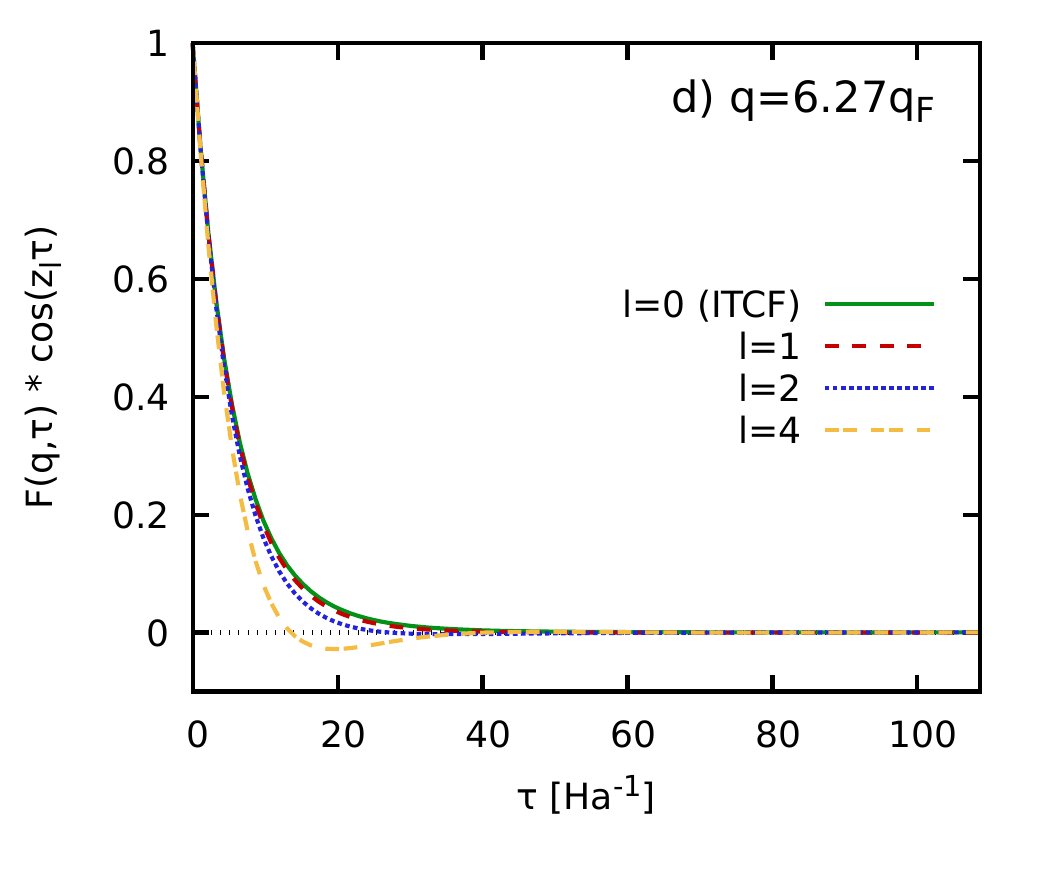}
\caption{\label{fig:contribution} a) PIMC results for the ITCF of the UEG at $r_s=20$ and $\Theta=1$ for three wave numbers; b-d) $\tau$-resolved contribution to $\widetilde{\chi}(\mathbf{q},z_l)$ [cf.~Eq.~(\ref{eq:MDR})] for the three wave numbers and for $l=0$ (solid green), $l=1$ (dashed red), $l=2$ (dotted blue), and $l=4$ (double-dashed yellow).}
\end{figure*} 

We use the extended ensemble PIMC sampling algorithm~\cite{Dornheim_PRB_nk_2021} as implemented in the \texttt{ISHTAR} code~\cite{ISHTAR}. All PIMC results are freely available online~\cite{repo}.

\subsection{Dynamic density response and local field correction\label{sec:density_response}}

We have carried out PIMC simulations of the strongly coupled electron liquid at $r_s=20$ [where $r_s=d/a_\textnormal{B}$, $d$ being the Wigner-Seitz radius and $a_\textnormal{B}$ the Bohr radius] and $\Theta=1$. These conditions give rise to an interesting \emph{roton-type} feature in the dynamic structure factor~\cite{Dornheim_Nature_2022,Dornheim_JCP_Force_2022,Takada_PRB_2016}, which is related to an effective electron--electron attraction in the medium~\cite{Dornheim_JCP_Force_2022}. In addition, they constitute an interesting test bed for dielectric theories~\cite{dornheim_electron_liquid,Tolias_JCP_2021,Tolias_JCP_2023} and have been used to illustrate the shortcomings of the \emph{static approximation}~\cite{Dornheim_PRB_ESA_2021}.

In Fig.~\ref{fig:contribution}a), we show PIMC results for the ITCF for three representative wave numbers. Since the physical meaning of its $\tau$-dependence has been discussed in the existing literature~\cite{Dornheim_insight_2022,Dornheim_PTR_2022,dornheim2024ab}, we here restrict ourselves to a very brief summary of the main trends. First, it is easy to see that it holds that $F(\mathbf{q},0)=S(\mathbf{q})$. For the UEG, it also holds~\cite{kugler_bounds}
\begin{eqnarray}
    \lim_{q\to0}S(\mathbf{q}) &=& \frac{q^2}{2\omega_p}\textnormal{coth}\left(\frac{\beta\omega_p}{2}\right)\ \textnormal{and}\\
    \lim_{q\gg q_\textnormal{F}}S(\mathbf{q}) &=& 1\ ,
\end{eqnarray}
where $\omega_p=\sqrt{3/r_s^3}$ is the plasma frequency, which explains the observed trends in Fig.~\ref{fig:contribution}a) for $\tau=0$.
Second, the slope of $F(\mathbf{q},\tau)$ around $\tau=0$ is governed by the well-known f-sum rule~\cite{Dornheim_insight_2022}
\begin{eqnarray}
    \left.\frac{\partial}{\partial\tau}F(\mathbf{q},\tau)\right|_{\tau=0} = -\frac{q^2}{2}\ ,
\end{eqnarray}
which explains the increasingly steep $\tau$-decay for large wave numbers. This trend originates from quantum delocalization effects, as correlations eventually vanish along the imaginary-time propagation when the wave length $\lambda_q=2\pi/q$ becomes comparable to the thermal wavelength $\lambda_\beta=\sqrt{2\pi\beta}$~\cite{Dornheim_insight_2022,Dornheim_PTR_2022}. Finally, we again note the symmetry of $F(\mathbf{q},\tau)$ around $\tau=\beta/2$, which is equivalent to the detailed balance relation $S(\mathbf{q},-\omega)=e^{-\beta\omega}S(\mathbf{q},\omega)$ of the dynamic structure factor~\cite{quantum_theory,Dornheim_T_2022}.

In Figs.~\ref{fig:contribution}b)-d), we show the corresponding contributions to the dynamic Matsubara density response function $\widetilde{\chi}(\mathbf{q},z_l)$ [cf.~Eq.~(\ref{eq:MDR})] for these three $q$-values. Obviously, the case of $l=0$ corresponds to the ITCF itself, and Eq.~(\ref{eq:MDR}) reverts to the usual imaginary-time version of the fluctuation--dissipation theorem, Eq.~(\ref{eq:static_chi}).
With increasing Matsubara frequency order $l$, the contributions to $\widetilde{\chi}(\mathbf{q},z_l)$ oscillate around zero, leading to cancellations. Evidently, these cancellations are reduced for larger wave numbers $q$, where the ITCF exhibits a steeper decay.

\begin{figure}\centering
\includegraphics[width=0.49\textwidth]{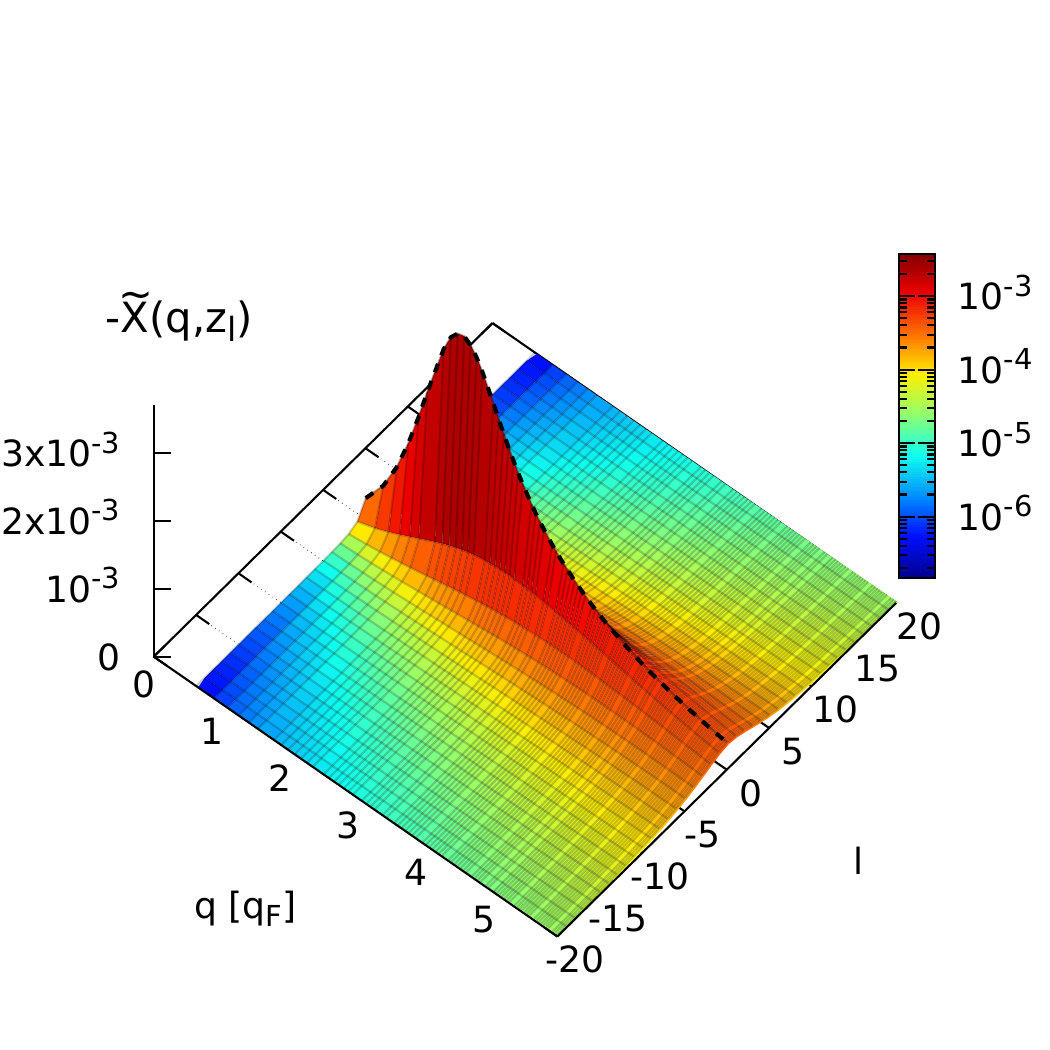}
\caption{\label{fig:UEG_N34_rs20_theta1} The dynamic Matsubara density response $\widetilde{\chi}(\mathbf{q},z_l)$ [cf.~Eq.~(\ref{eq:MDR})] of the UEG at $r_s=20$ and $\Theta=1$, with $l$ being the integer index (order) of the Matsubara frequency.}
\end{figure} 

In Fig.~\ref{fig:UEG_N34_rs20_theta1}, we show the full dynamic Matsubara density response function in the $q$-$l$-plane. The dashed black line corresponds to the static limit $\chi(\mathbf{q},0)$, which has been discussed extensively in the literature, e.g.~\cite{dornheim_ML,dynamic_folgepaper,dornheim_electron_liquid,Tolias_JCP_2021,Tolias_JCP_2023}. For any classical system, this cutout would already completely determine the static structure factor owing to the exact connection~\cite{kugler_classical,tolias2021modes}
\begin{eqnarray}
    S^\textnormal{cl}(\mathbf{q}) = - \frac{\chi^\textnormal{cl}(\mathbf{q},0)}{n\beta} \ .
\end{eqnarray}
In other words, all contributions to Eq.~(\ref{eq:Matsubara_Series}) for $|l|>0$ are quantum mechanical in origin and would, indeed, vanish for classical point particles where the ITCF is constant with respect to $\tau$. From Fig.~\ref{fig:UEG_N34_rs20_theta1}, it is evident that these quantum contributions become increasingly important with increasing $q$, i.e., on smaller length scales, as it is expected. Indeed, any system exhibits quantum mechanical behaviour on sufficiently small length scales; the large wave number limit thus always requires a full quantum mechanical description.
In practice, this leads to increasingly large truncation parameters $l_\textnormal{max}$ that are required to converge the Matsubara series Eq.~(\ref{eq:Matsubara_Series}), which is a well known issue in dielectric theories~\cite{stls,stlspra,Tolias_JCP_2021,Tolias_JCP_2023}.

\begin{figure}\centering
\includegraphics[width=0.49\textwidth]{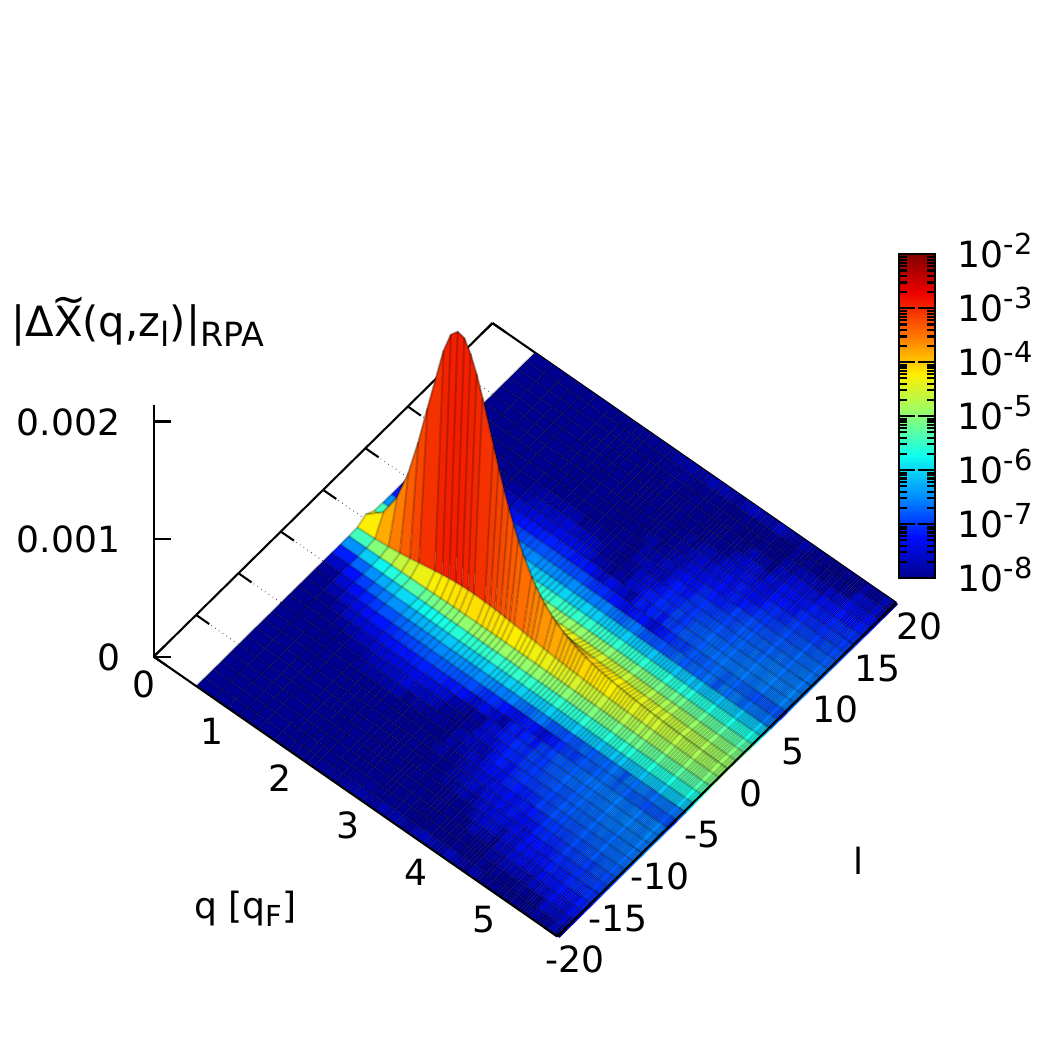}
\caption{\label{fig:Diff_UEG_N34_rs20_theta1} Deviations between the exact PIMC and RPA results for the dynamic Matsubara frequency response $\widetilde{\chi}(\mathbf{q},z_l)$ of the UEG at $r_s=20$ and $\Theta=1$.}
\end{figure} 

Let us next consider the difference between our quasi-exact PIMC results for $\widetilde{\chi}(\mathbf{q},z_l)$ and the RPA, which is shown in Fig.~\ref{fig:Diff_UEG_N34_rs20_theta1}.
Remarkably, by far the most pronounced difference occurs in the static limit of $l=0$, which explains the previously observed good performance of the \emph{static approximation}; for the latter, the $l=0$ limit is, by definition, exactly reproduced, even though this comes at the expense of a small systematic error for $|l|>0$.
In addition, we find that dynamic XC-effects are mostly limited to the small-$l$ regime, and the difference between PIMC and RPA decays substantially faster with the Matsubara frequency than $\widetilde{\chi}(\mathbf{q},z_l)$ itself.

\begin{figure}\centering
\includegraphics[width=0.49\textwidth]{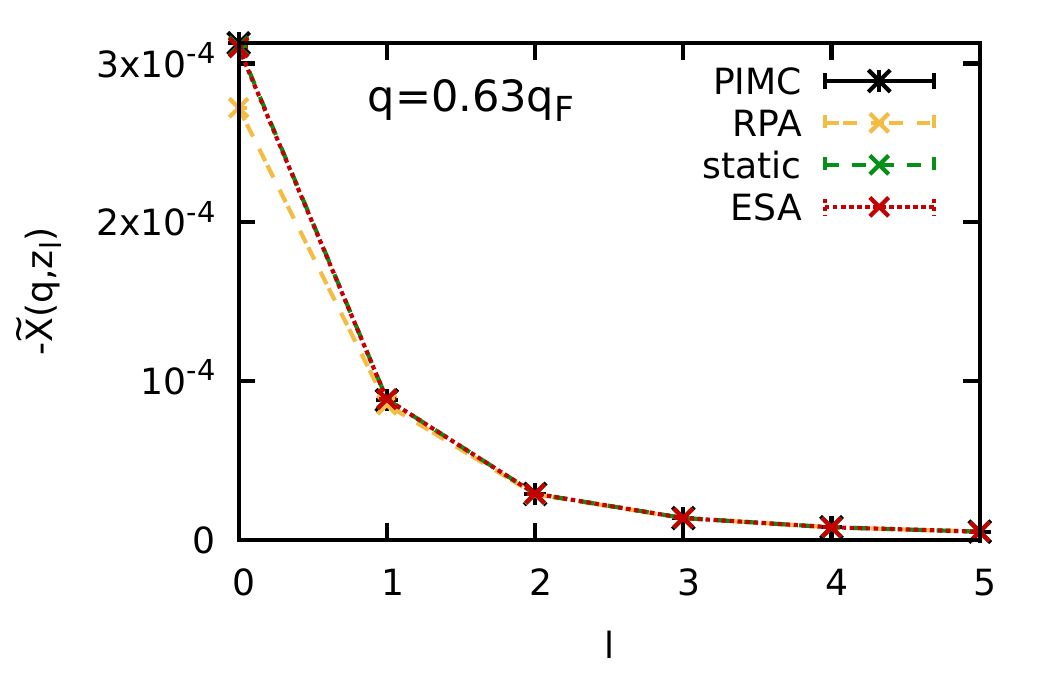}\\\vspace*{-1cm}
\includegraphics[width=0.49\textwidth]{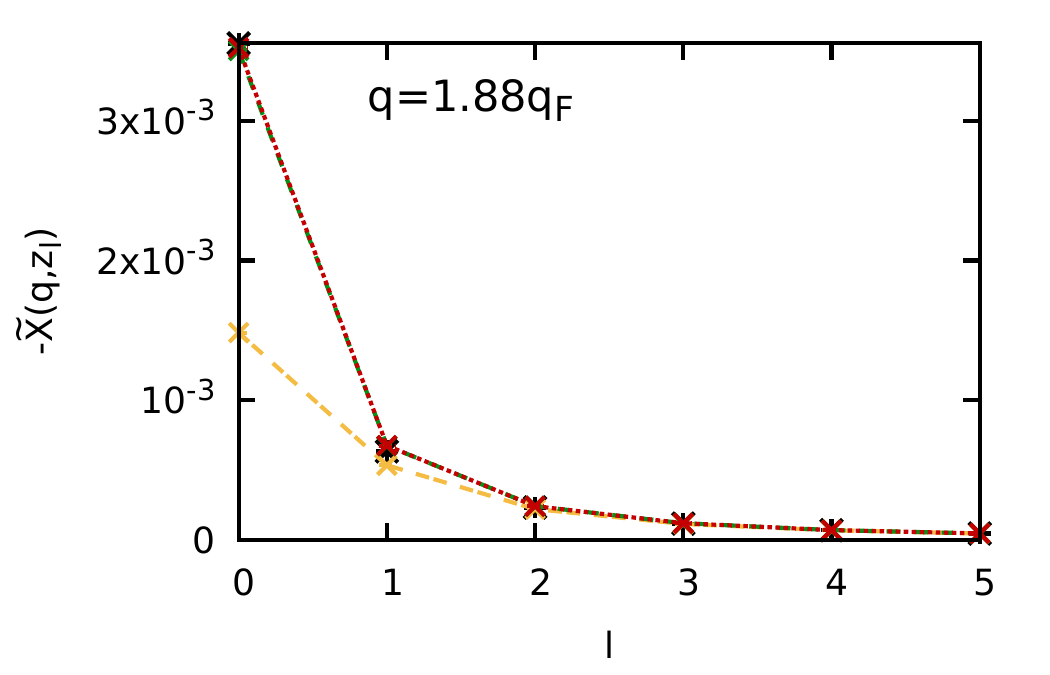}\\\vspace*{-1cm}
\includegraphics[width=0.49\textwidth]{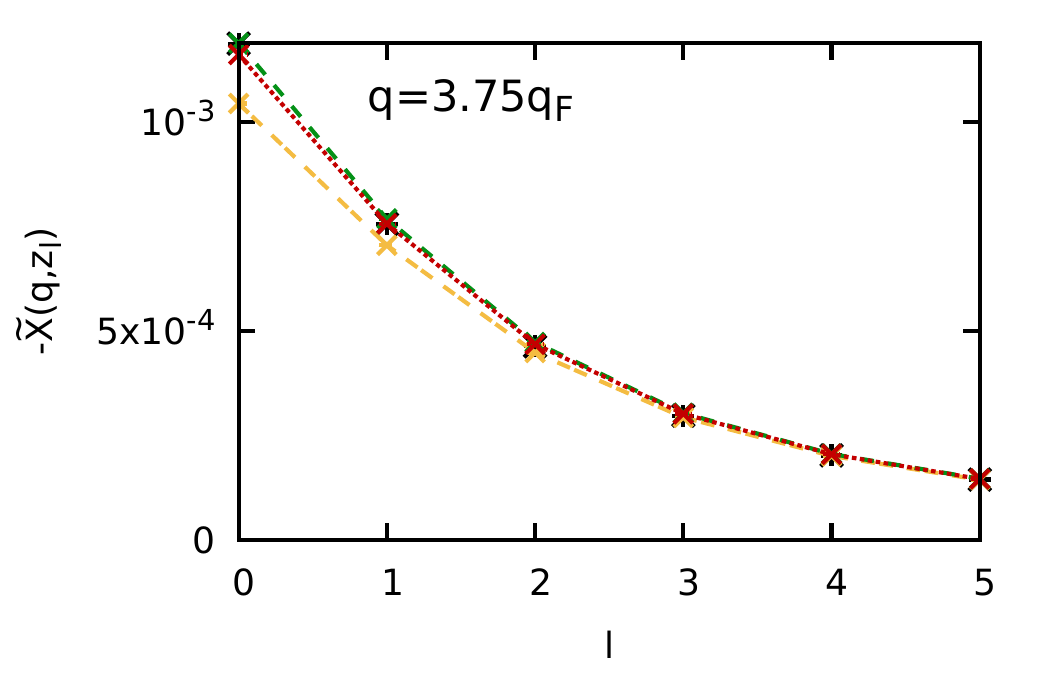}
\caption{\label{fig:chi_l_UEG_N34_rs20_theta1} The dynamic Matsubara density response $\widetilde{\chi}(\mathbf{q},z_l)$ [cf.~Eq.~(\ref{eq:MDR})] of the UEG at $r_s=20$ and $\Theta=1$ as a function of the Matsubara order $l$ for three wave numbers $q$. Black stars: PIMC results for $N=34$; double-dashed yellow: RPA; dashed green: \emph{static approximation}; dotted red: ESA.
}
\end{figure} 

This can be seen particularly well in Fig.~\ref{fig:chi_l_UEG_N34_rs20_theta1}, where we show $\widetilde{\chi}(\mathbf{q},z_l)$ as a function of $l$ for three relevant wave numbers.
For $q=0.63q_\textnormal{F}$, differences between our new PIMC results (black stars) and the mean-field based RPA are mainly limited to $l=0$. This changes for $q=1.88q_\textnormal{F}$ and $q=3.75q_\textnormal{F}$, where we find small yet significant contributions up to $l=2$ and $l=3$, respectively.
In addition, we include the \emph{static approximation $G(\mathbf{q},0)$} and ESA as the dashed green and dotted red lines. For the two smallest depicted wave numbers, both data sets are in excellent agreement with the PIMC reference data for all $l$.
For $q=3.75q_\textnormal{F}$, the \emph{static approximation $G(\mathbf{q},0)$} reproduces the PIMC results for $l=0$ by design, whereas the ESA somewhat deviates. Interestingly, we find opposite trends for $l\geq1$, which has consequences for the computation of the static structure factor as we explain below.

\begin{figure}\centering\vspace*{-1cm}
\includegraphics[width=0.49\textwidth]{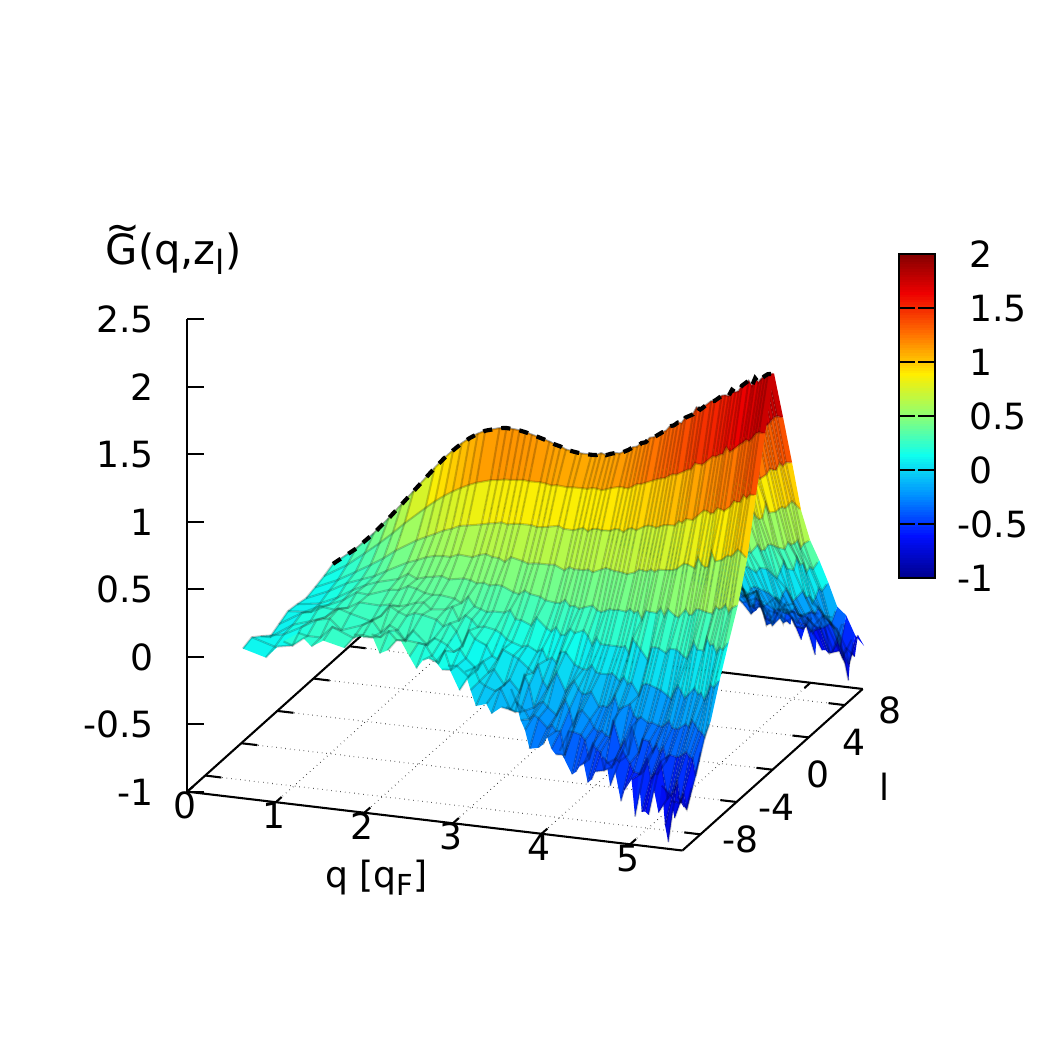}\vspace*{-1cm}
\caption{\label{fig:LFC_UEG_N34_rs20_theta1} PIMC results for the dynamic Matsubara local field correction $\widetilde{G}(\mathbf{q},z_l)$ [cf.~Eq.~(\ref{eq:LFC})] of the UEG at $r_s=20$ and $\Theta=1$, with $l$ the Matsubara order.}
\end{figure} 

Let us next consider the dynamic Matsubara local field correction, which is shown in Fig.~\ref{fig:LFC_UEG_N34_rs20_theta1} in the relevant $q$-$l$-plane. We stress the remarkable stability of the inversion via Eq.~(\ref{eq:LFC}) in this case due to the high quality (i.e., small error bars) of the PIMC results for $F(\mathbf{q},\tau)$ and $\widetilde{\chi}(\mathbf{q},z_l)$. The dashed black lines show the usual static limit, $G(\mathbf{q},0)$, which exhibits the well known form given by the compressibility sum-rule at small $q$~\cite{Dornheim_PRB_ESA_2021}, a maximum around twice the Fermi wave number, and again a quadratic increase in the limit of large $q$~\cite{holas_limit}. The full dynamic LFC exhibits a fairly complex behaviour that is more difficult to interpret than the density response shown in Fig.~\ref{fig:UEG_N34_rs20_theta1} above. Indeed, $\widetilde{G}(\mathbf{q},z_l)$ attains negative values for large $l$, and exhibits an increasingly steep decay with $l$ for large wave numbers. Obviously, this decay is not captured by the \emph{static approximation $G(\mathbf{q},0)$}, which is the root cause of the spurious behaviour in the static structure factor reported in Refs.~\cite{Dornheim_PRL_2020_ESA,Dornheim_PRB_ESA_2021}. Overall, the dynamic Matsubara local field correction exhibits a smooth behaviour without any sharp peaks or edges, which renders its accurate parametrization with respect to $q$, $z_l$, but also $r_s$ and $\Theta$ a promising possibility that will be explored in dedicated future works.

\begin{figure}\centering
\includegraphics[width=0.49\textwidth]{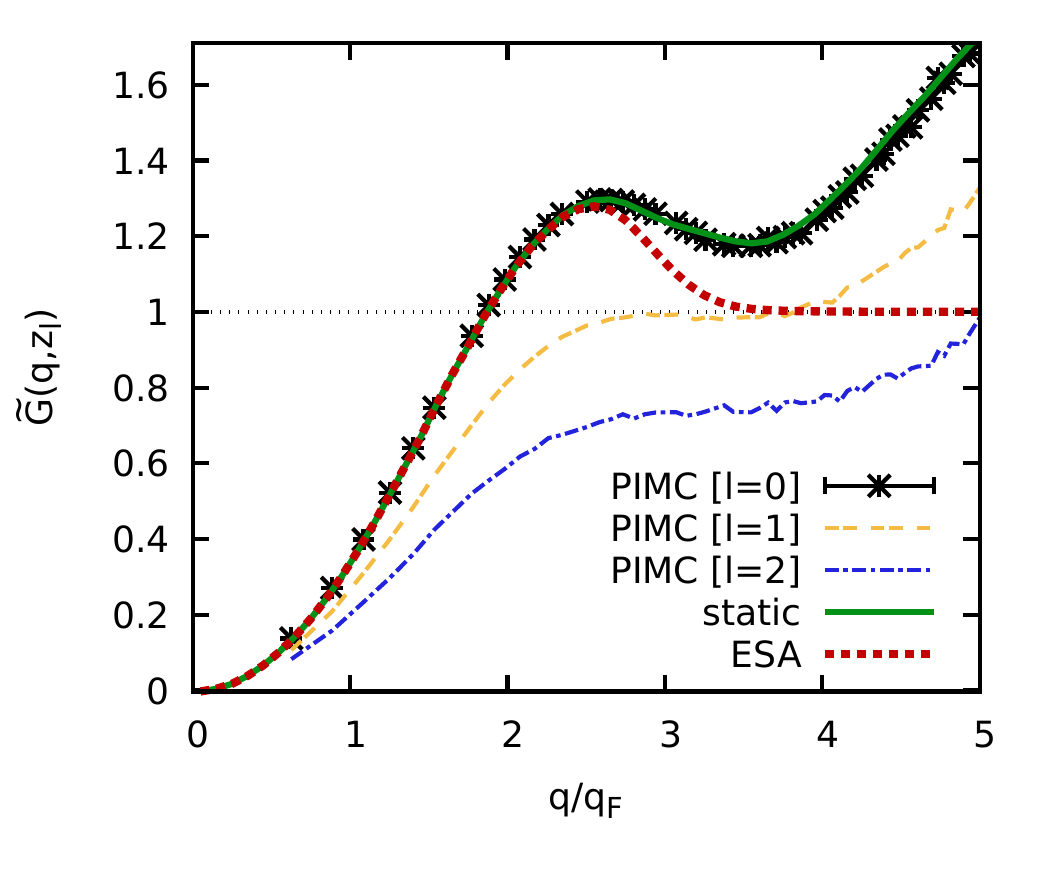}
\caption{\label{fig:q_LFC_UEG_N34_rs20_theta1} Dynamic Matsubara local field correction $\widetilde{G}(\mathbf{q},z_l)$ of the UEG as a function of the wave number $q$ for different Matsubara orders at $r_s=20$ and $\Theta=1$. The solid green line has been obtained using the neural network representation from Ref.~\cite{dornheim_ML} and the dotted red line has been obtained from the analytical ESA parametrization presented in Ref.~\cite{Dornheim_PRB_ESA_2021}.
}
\end{figure} 

In Fig.~\ref{fig:q_LFC_UEG_N34_rs20_theta1}, we show the local field correction as a function of the wave number; the black stars correspond to $G(\mathbf{q},0)$, which is in excellent agreement with the neural network representation from Ref.~\cite{dornheim_ML}, as it is expected. The ESA, on the other hand, is in perfect agreement with both data sets, but starts to deviate around $q=2.5q_\textnormal{F}$. It converges towards unity for large $q$ as the on-top pair correlation function vanishes in the electron liquid regime, $g(0)=0$. The double-dashed yellow and dash-dotted blue curves show our new PIMC results for $l=1$ and $l=2$, respectively. They are systematically lower than the static limit over the entire $q$-range. The ESA curve thus does indeed constitute an effectively frequency-averaged LFC, which explains its good performance.

\subsection{Static structure factor\label{sec:SSF}}

\begin{figure}\centering
\includegraphics[width=0.49\textwidth]{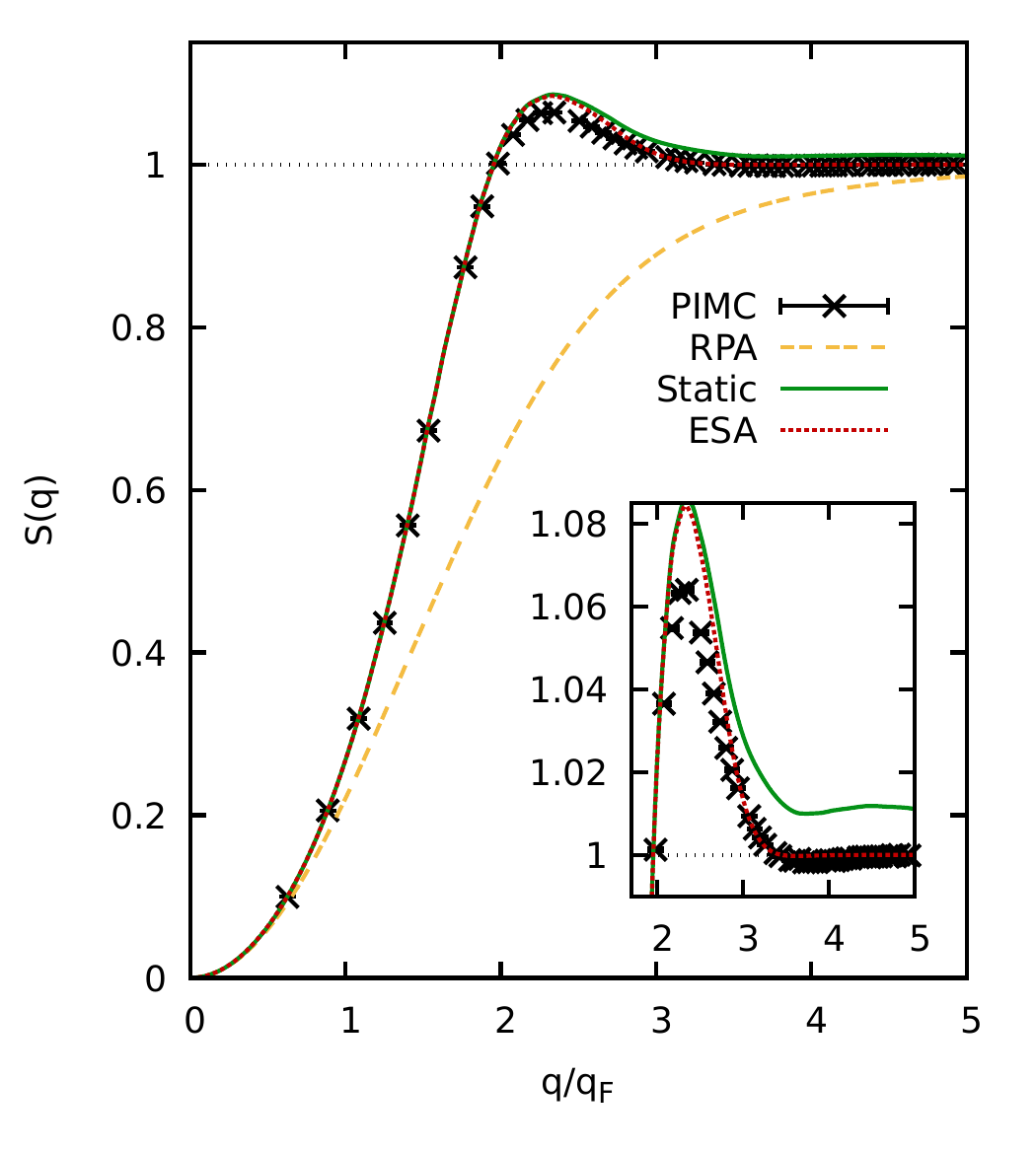}
\caption{\label{fig:SSF_UEG_N34_rs20_theta1} Static structure factor $S(\mathbf{q})$ of the UEG at $r_s=20$ and $\Theta=1$. Black crosses: PIMC results for $N=34$; double-dashed yellow: RPA; solid green: \emph{static approximation}; dotted red: ESA. The inset shows a magnified segment around the peak.
}
\end{figure} 

In Fig.~\ref{fig:SSF_UEG_N34_rs20_theta1}, we show the static structure factor $S(\mathbf{q})$ at the same conditions as in the previous sections. We note that a similar investigation has been provided in Ref.~\cite{Dornheim_PRB_ESA_2021}. Overall, both the \emph{static approximation $G(\mathbf{q},0)$} (solid green) and ESA (dotted red) qualitatively reproduce the PIMC results (black crosses) over the entire $q$-range. The peak position is reproduced with high accuracy, whereas the peak height is equally overestimated by both approaches. Yet, ESA basically becomes exact for $q\gtrsim3q_\textnormal{F}$, whereas the \emph{static approximation $G(\mathbf{q},0)$} does not converge towards the expected limit of unity in the depicted range of wave numbers.

\begin{figure}\centering
\includegraphics[width=0.49\textwidth]{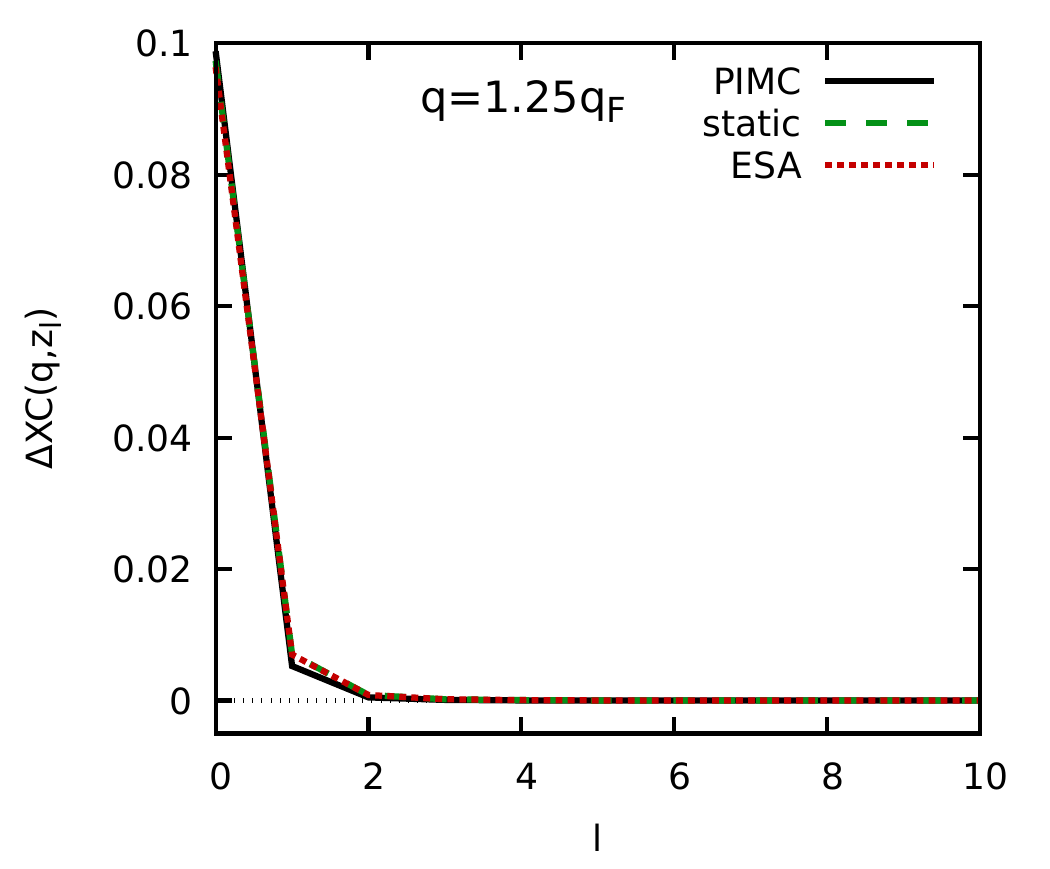}\\\vspace*{-1cm}\includegraphics[width=0.49\textwidth]{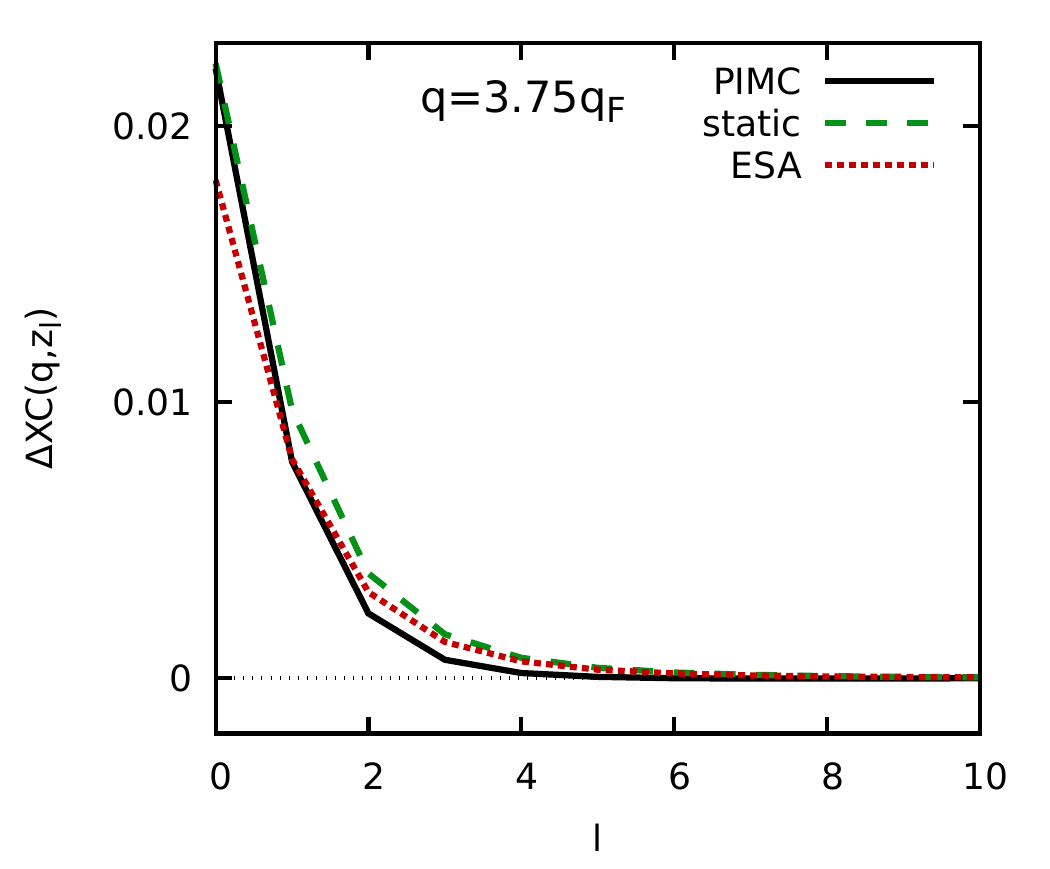}
\caption{\label{fig:XC_UEG_N34_rs20_theta1} Dynamic XC-effect metric [cf.~Eq.~(\ref{eq:Delta_XC})] as a function of the Matsubara order $l$ for two representative wave numbers $q$. Solid black: PIMC; dashed green: \emph{static approximation}; dotted red: ESA.
}
\end{figure} 

To trace this trend to the impact of dynamic XC-effects, we define the deviation measure
\begin{eqnarray}\label{eq:Delta_XC}
    \Delta\textnormal{XC}(\mathbf{q},z_l) = - \frac{1}{n\beta} \left[\widetilde{\chi}(\mathbf{q},z_l) - \widetilde{\chi}_\textnormal{RPA}(\mathbf{q},z_l) \right]\ ,
\end{eqnarray}
which corrects Eq.~(\ref{eq:Matsubara_Series}) with respect to the mean-field based RPA.
In Fig.~\ref{fig:XC_UEG_N34_rs20_theta1}, we show the dependence of Eq.~(\ref{eq:Delta_XC}) on $l$ for two representative wave numbers. For $q=1.25q_\textnormal{F}$ (top panel), both the \emph{static approximation $G(\mathbf{q},0)$} and the ESA reproduce the PIMC reference data equally well, and their corresponding result for $S(\mathbf{q})$ is highly accurate. For $q=3.75q_\textnormal{F}$, we find an error cancellation in the ESA; in contrast, the \emph{static approximation $G(\mathbf{q},0)$} is exact for $l=0$, but significantly overestimates the true XC-correction for $1\leq{l}\lesssim 7$. The accumulation of these terms then leads to the observed overestimation of the static structure factor for large $q$.

\subsection{Asymptotic high frequency behavior of the Matsubara local field correction}\label{sec:asymptotic_results}

\begin{figure}
\includegraphics[width=0.49\textwidth]{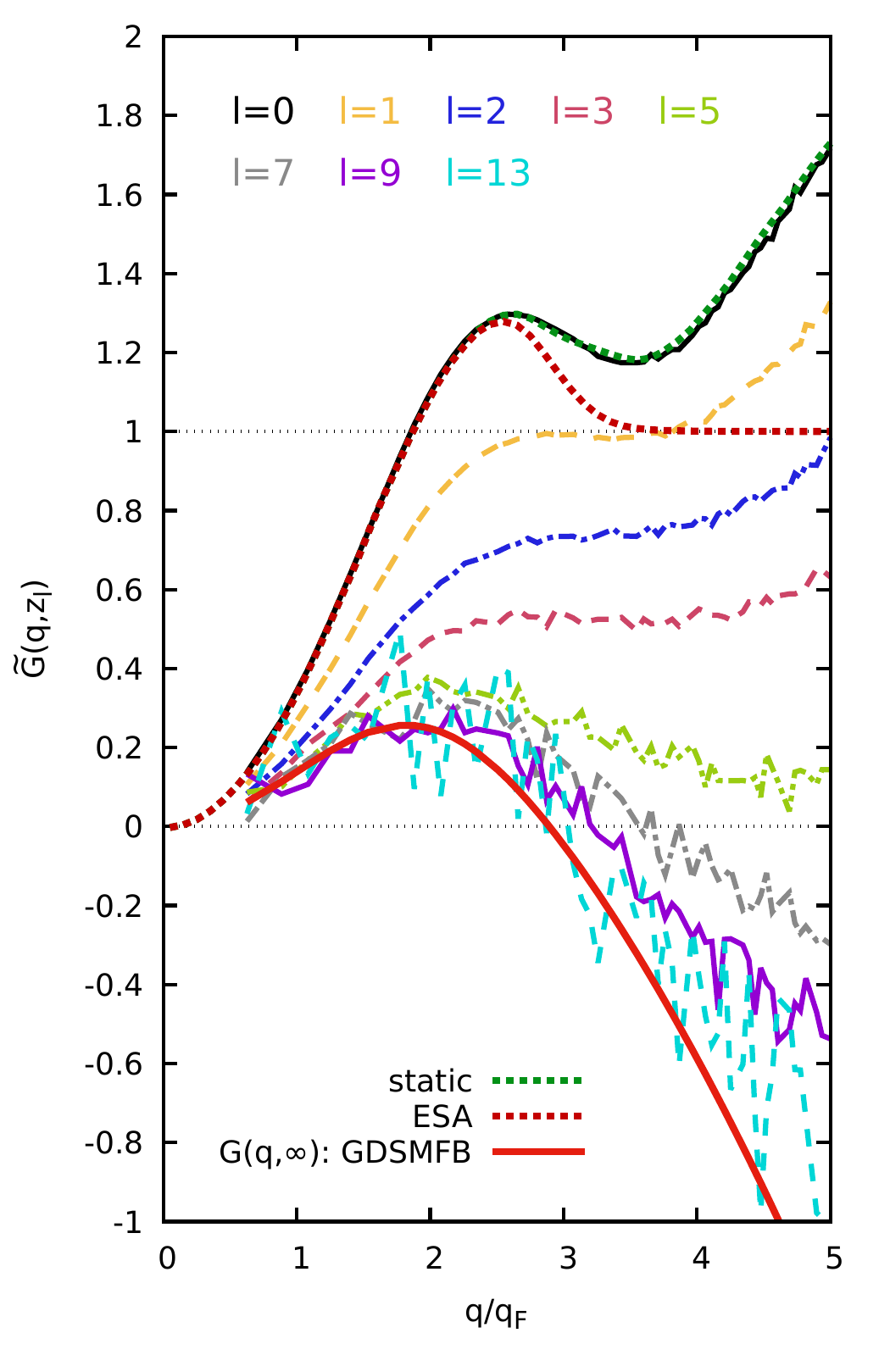}
\caption{\label{fig:other} Matsubara order ($l$) resolved wave number dependence of the dynamic local field correction $\widetilde{G}(\mathbf{q},z_l)$ of the UEG at $r_s=20$ and $\Theta=1$. Dotted green: neural network representation from Ref.~\cite{dornheim_ML}; dotted red: ESA parametrization from Ref.~\cite{Dornheim_PRB_ESA_2021}; solid red: high frequency limit, Eq.~(\ref{eq:Ginfty}), using $T_\textnormal{xc}$ computed from the XC-free energy parametrization by Groth \textit{et al.}~\cite{groth_prl} (GDSMFB).
}
\end{figure} 

Let us conclude by analyzing the high-frequency limit of the dynamic Matsubara local field correction $\widetilde{G}(\mathbf{q},z_l)$. In Fig.~\ref{fig:other}, we plot $\widetilde{G}(\mathbf{q},z_l)$ for various $l$ in the range of $0\leq l \leq 13$. It is evident that the data become increasingly noisy for large $l$, which is a consequence of the diminishing impact of the local field correction onto $\widetilde{\chi}(\mathbf{q},z_l)$. For every constant wave number, we observe a monotonic decrease of $\widetilde{G}(\mathbf{q},z_l)$ from the static limit towards the asymptotic high-frequency limit, Eq.~(\ref{eq:Ginfty}), that is shown as the bold solid red curve. In practice, the convergence of $\widetilde{G}(\mathbf{q},z_l)$ towards $\widetilde{G}(\mathbf{q},\infty)$ with $l$ becomes substantially slower with increasing wavenumber due to the more pronounced importance of quantum delocalization effects, as it has already been explained above. Overall, Fig.~\ref{fig:other} suggests that, at least for UEG state points for which the XC contribution to the kinetic energy is positive ($\tau_{\mathrm{xc}}>0$), the dynamic Matsubara local field correction $\widetilde{G}(\mathbf{q},z_l)$ is upper bounded by its static limit $\widetilde{G}(\mathbf{q},0)$ and lower bounded by its high frequency limit $\widetilde{G}(\mathbf{q},\infty)$. 

\begin{figure}
\includegraphics[width=0.49\textwidth]{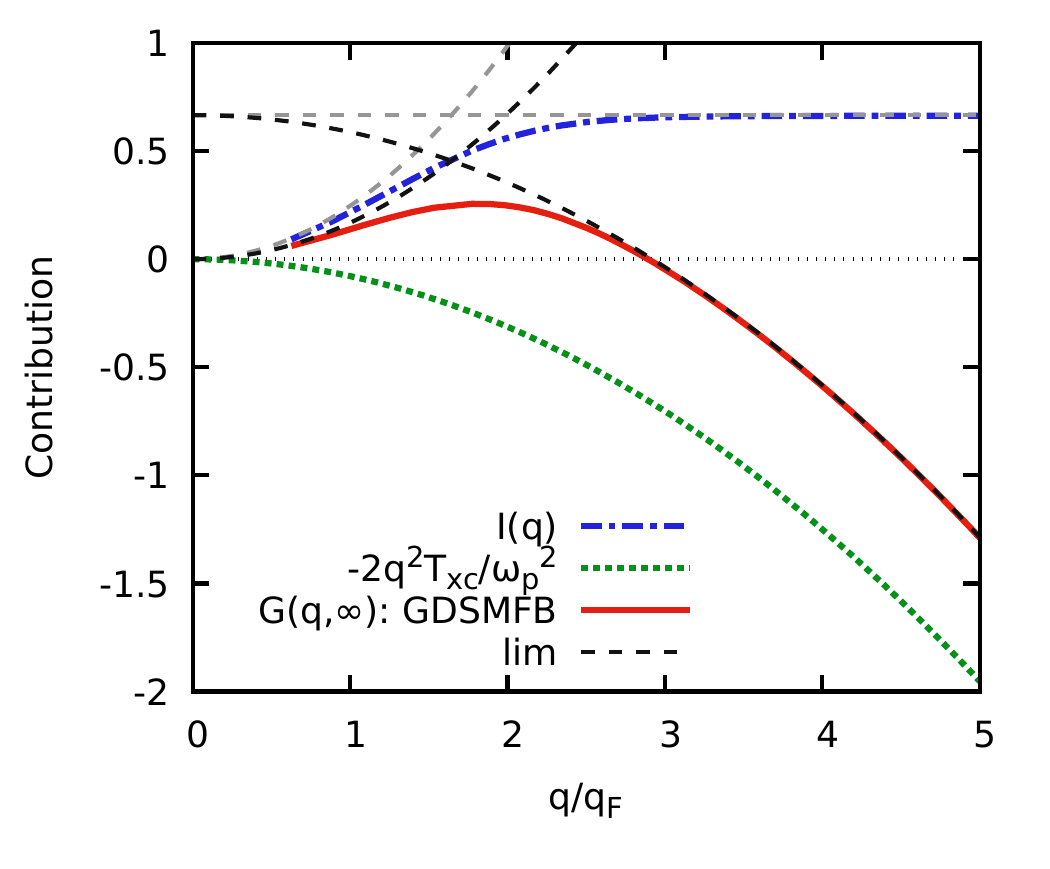}
\caption{\label{fig:contrib} Wave number dependence of the high frequency limit ($l\to\infty$) of the dynamic Matsubara local field correction $\widetilde{G}(\mathbf{q},z_l)$ of the UEG at $r_s=20$ and $\Theta=1$. Solid red: Full $G(\mathbf{q},\infty)$, Eq.~(\ref{eq:Ginfty}); dash-dotted blue: $I(\mathbf{q})$, Eq.~(\ref{eq:Iq}); dashed green: $T_\textnormal{xc}$ term evaluated using $T_\textnormal{xc}$ computed from the XC-free energy parametrization by Groth \textit{et al.}~\cite{groth_prl}. The dashed grey and black lines show the asymptotic short and long wavelength limits of $I(\mathbf{q})$ and $G(\mathbf{q},\infty)$, respectively, cf.~Sec.~\ref{sec:asymptoticfrequency}.
}
\end{figure} 

In Fig.~\ref{fig:contrib}, we decompose the individual contributions to $\widetilde{G}(\mathbf{q},\infty)$. The dotted green line corresponds to the kinetic energy contribution, which is parabolic for all $q$. The dash-dotted blue curve shows the Pathak-Vashishta functional [Eq.~(\ref{eq:Iq})] that depends on the static structure factor $S(\mathbf{q})$. It converges towards $2/3$ in the $q\to\infty$ limit as the on-top pair correlation function vanishes at these parameters [Eq.~(\ref{eq:Iqshort})], and attains a parabola in the limit of $q\to0$ [Eq.~(\ref{eq:Iqlong})]; both limits are shown as the dashed light grey curves in Fig.~\ref{fig:contrib} and are in excellent agreement with our numerical results for $I(\mathbf{q})$. Naturally, the same holds for the dashed black lines that depict the corresponding limits of $\widetilde{G}(\mathbf{q},\infty)$. 

The above two findings clearly hint at the intriguing possibility of constructing an analytic four-parameter representation of the full dynamic Matsubara local field correction $\widetilde{G}(\mathbf{q},z_l;r_s,\Theta)$ over a broad range of UEG parameters.

\section{Summary and Discussion\label{sec:summary}}

In this work, we have presented an analysis of dynamic XC-effects in the strongly coupled electron liquid. This has been achieved on the basis of highly accurate direct PIMC results for the Matsubara density response function $\widetilde{\chi}(\mathbf{q},z_l)$ that have been obtained from the ITCF $F(\mathbf{q},\tau)$ using the recently derived Fourier-Matsubara expansion by Tolias \emph{et al.}~\cite{tolias2024fouriermatsubara}. In particular, this approach allows us to obtain the full dynamic Matsubara local field correction $\widetilde{G}(\mathbf{q},z_l)$, which exhibits a nontrivial behaviour. Our results provide new insights into the complex interplay between XC-correlation and quantum delocalization effects, and explain the observed differences between the \emph{static approximation} and ESA at large wave numbers~\cite{Dornheim_PRL_2020_ESA,Dornheim_PRB_ESA_2021}.

We are convinced that the presented methodology opens up a number of possibilities for impactful future research. First, the dynamic LFC constitutes the key property in dielectric theories~\cite{arora,Tolias_JCP_2023,tolias2024_VS,dynamic_ii}; the availability of highly accurate PIMC results can thus give new insights into existing approximations, and guide the development of new approaches. Second, extensive future results for the dynamic LFC might be used as input to construct a parametrization $\widetilde{G}(\mathbf{q},z_l;r_s,\Theta)$ that covers (substantial parts of) the WDM regime. 

Such a parametrization would be key input for the construction of advanced, non-local XC-functionals for thermal DFT simulations based on the adiabatic--connection formula and the fluctuation--dissipation theorem~\cite{pribram}. Finally, we note that direct PIMC simulations that retain access to the full imaginary-time structure have recently been presented for warm dense hydrogen~\cite{dornheim2024ab,Dornheim_JCP_2024} and beryllium~\cite{Dornheim_JCP_2024,Dornheim_Science_2024}, which opens up the enticing possibility to study species-resolved dynamic XC-effects in real materials.

\section*{Acknowledgments}
This work was partially supported by the Center for Advanced Systems Understanding (CASUS), financed by Germany’s Federal Ministry of Education and Research (BMBF) and the Saxon state government out of the State budget approved by the Saxon State Parliament. Further support is acknowledged for the CASUS Open Project \emph{Guiding dielectric theories with ab initio quantum Monte Carlo simulations: from the strongly coupled electron liquid to warm dense matter}. This work has received funding from the European Research Council (ERC) under the European Union’s Horizon 2022 research and innovation programme
(Grant agreement No. 101076233, "PREXTREME"). Views and opinions expressed are however those of the authors only and do not necessarily reflect those of the European Union or the European Research Council Executive Agency. Neither the European Union nor the granting authority can be held responsible for them. Computations were performed on a Bull Cluster at the Center for Information Services and High-Performance Computing (ZIH) at Technische Universit\"at Dresden, at the Norddeutscher Verbund f\"ur Hoch- und H\"ochstleistungsrechnen (HLRN) under grant mvp00024, and on the HoreKa supercomputer funded by the Ministry of Science, Research and the Arts Baden-W\"urttemberg and
by the Federal Ministry of Education and Research.

\bibliography{bibliography}

\begin{thebibliography}{117}%
\makeatletter
\providecommand \@ifxundefined [1]{%
 \@ifx{#1\undefined}
}%
\providecommand \@ifnum [1]{%
 \ifnum #1\expandafter \@firstoftwo
 \else \expandafter \@secondoftwo
 \fi
}%
\providecommand \@ifx [1]{%
 \ifx #1\expandafter \@firstoftwo
 \else \expandafter \@secondoftwo
 \fi
}%
\providecommand \natexlab [1]{#1}%
\providecommand \enquote  [1]{``#1''}%
\providecommand \bibnamefont  [1]{#1}%
\providecommand \bibfnamefont [1]{#1}%
\providecommand \citenamefont [1]{#1}%
\providecommand \href@noop [0]{\@secondoftwo}%
\providecommand \href [0]{\begingroup \@sanitize@url \@href}%
\providecommand \@href[1]{\@@startlink{#1}\@@href}%
\providecommand \@@href[1]{\endgroup#1\@@endlink}%
\providecommand \@sanitize@url [0]{\catcode `\\12\catcode `\$12\catcode `\&12\catcode `\#12\catcode `\^12\catcode `\_12\catcode `\%12\relax}%
\providecommand \@@startlink[1]{}%
\providecommand \@@endlink[0]{}%
\providecommand \url  [0]{\begingroup\@sanitize@url \@url }%
\providecommand \@url [1]{\endgroup\@href {#1}{\urlprefix }}%
\providecommand \urlprefix  [0]{URL }%
\providecommand \Eprint [0]{\href }%
\providecommand \doibase [0]{http://dx.doi.org/}%
\providecommand \selectlanguage [0]{\@gobble}%
\providecommand \bibinfo  [0]{\@secondoftwo}%
\providecommand \bibfield  [0]{\@secondoftwo}%
\providecommand \translation [1]{[#1]}%
\providecommand \BibitemOpen [0]{}%
\providecommand \bibitemStop [0]{}%
\providecommand \bibitemNoStop [0]{.\EOS\space}%
\providecommand \EOS [0]{\spacefactor3000\relax}%
\providecommand \BibitemShut  [1]{\csname bibitem#1\endcsname}%
\let\auto@bib@innerbib\@empty
\bibitem [{\citenamefont {Giuliani}\ and\ \citenamefont {Vignale}(2008)}]{quantum_theory}%
  \BibitemOpen
  \bibfield  {author} {\bibinfo {author} {\bibfnamefont {G.}~\bibnamefont {Giuliani}}\ and\ \bibinfo {author} {\bibfnamefont {G.}~\bibnamefont {Vignale}},\ }\href@noop {} {\emph {\bibinfo {title} {Quantum Theory of the Electron Liquid}}}\ (\bibinfo  {publisher} {Cambridge University Press},\ \bibinfo {address} {Cambridge},\ \bibinfo {year} {2008})\BibitemShut {NoStop}%
\bibitem [{\citenamefont {Loos}\ and\ \citenamefont {Gill}(2016)}]{loos}%
  \BibitemOpen
  \bibfield  {author} {\bibinfo {author} {\bibfnamefont {P.-F.}\ \bibnamefont {Loos}}\ and\ \bibinfo {author} {\bibfnamefont {P.~M.~W.}\ \bibnamefont {Gill}},\ }\bibfield  {title} {\enquote {\bibinfo {title} {The uniform electron gas},}\ }\href {http://onlinelibrary.wiley.com/doi/10.1002/wcms.1257/abstract} {\bibfield  {journal} {\bibinfo  {journal} {Comput. Mol. Sci}\ }\textbf {\bibinfo {volume} {6}},\ \bibinfo {pages} {410--429} (\bibinfo {year} {2016})}\BibitemShut {NoStop}%
\bibitem [{\citenamefont {Dornheim}\ \emph {et~al.}(2018{\natexlab{a}})\citenamefont {Dornheim}, \citenamefont {Groth},\ and\ \citenamefont {Bonitz}}]{review}%
  \BibitemOpen
  \bibfield  {author} {\bibinfo {author} {\bibfnamefont {T.}~\bibnamefont {Dornheim}}, \bibinfo {author} {\bibfnamefont {S.}~\bibnamefont {Groth}}, \ and\ \bibinfo {author} {\bibfnamefont {M.}~\bibnamefont {Bonitz}},\ }\bibfield  {title} {\enquote {\bibinfo {title} {The uniform electron gas at warm dense matter conditions},}\ }\href {https://www.sciencedirect.com/science/article/abs/pii/S0370157318300516} {\bibfield  {journal} {\bibinfo  {journal} {Phys. Reports}\ }\textbf {\bibinfo {volume} {744}},\ \bibinfo {pages} {1--86} (\bibinfo {year} {2018}{\natexlab{a}})}\BibitemShut {NoStop}%
\bibitem [{\citenamefont {Mahan}(1990)}]{mahan1990many}%
  \BibitemOpen
  \bibfield  {author} {\bibinfo {author} {\bibfnamefont {G.D.}\ \bibnamefont {Mahan}},\ }\href {https://books.google.de/books?id=v8du6cp0vUAC} {\emph {\bibinfo {title} {Many-Particle Physics}}},\ Physics of Solids and Liquids\ (\bibinfo  {publisher} {Springer US},\ \bibinfo {year} {1990})\BibitemShut {NoStop}%
\bibitem [{\citenamefont {Takada}(2016)}]{Takada_PRB_2016}%
  \BibitemOpen
  \bibfield  {author} {\bibinfo {author} {\bibfnamefont {Yasutami}\ \bibnamefont {Takada}},\ }\bibfield  {title} {\enquote {\bibinfo {title} {Emergence of an excitonic collective mode in the dilute electron gas},}\ }\href {\doibase 10.1103/PhysRevB.94.245106} {\bibfield  {journal} {\bibinfo  {journal} {Phys. Rev. B}\ }\textbf {\bibinfo {volume} {94}},\ \bibinfo {pages} {245106} (\bibinfo {year} {2016})}\BibitemShut {NoStop}%
\bibitem [{\citenamefont {Dornheim}\ \emph {et~al.}(2018{\natexlab{b}})\citenamefont {Dornheim}, \citenamefont {Groth}, \citenamefont {Vorberger},\ and\ \citenamefont {Bonitz}}]{dornheim_dynamic}%
  \BibitemOpen
  \bibfield  {author} {\bibinfo {author} {\bibfnamefont {T.}~\bibnamefont {Dornheim}}, \bibinfo {author} {\bibfnamefont {S.}~\bibnamefont {Groth}}, \bibinfo {author} {\bibfnamefont {J.}~\bibnamefont {Vorberger}}, \ and\ \bibinfo {author} {\bibfnamefont {M.}~\bibnamefont {Bonitz}},\ }\bibfield  {title} {\enquote {\bibinfo {title} {Ab initio path integral {M}onte {C}arlo results for the dynamic structure factor of correlated electrons: From the electron liquid to warm dense matter},}\ }\href {https://journals.aps.org/prl/abstract/10.1103/PhysRevLett.121.255001} {\bibfield  {journal} {\bibinfo  {journal} {Phys. Rev. Lett.}\ }\textbf {\bibinfo {volume} {121}},\ \bibinfo {pages} {255001} (\bibinfo {year} {2018}{\natexlab{b}})}\BibitemShut {NoStop}%
\bibitem [{\citenamefont {Groth}\ \emph {et~al.}(2019)\citenamefont {Groth}, \citenamefont {Dornheim},\ and\ \citenamefont {Vorberger}}]{dynamic_folgepaper}%
  \BibitemOpen
  \bibfield  {author} {\bibinfo {author} {\bibfnamefont {S.}~\bibnamefont {Groth}}, \bibinfo {author} {\bibfnamefont {T.}~\bibnamefont {Dornheim}}, \ and\ \bibinfo {author} {\bibfnamefont {J.}~\bibnamefont {Vorberger}},\ }\bibfield  {title} {\enquote {\bibinfo {title} {Ab initio path integral {M}onte {C}arlo approach to the static and dynamic density response of the uniform electron gas},}\ }\href {https://link.aps.org/doi/10.1103/PhysRevB.99.235122} {\bibfield  {journal} {\bibinfo  {journal} {Phys. Rev. B}\ }\textbf {\bibinfo {volume} {99}},\ \bibinfo {pages} {235122} (\bibinfo {year} {2019})}\BibitemShut {NoStop}%
\bibitem [{\citenamefont {Dornheim}\ \emph {et~al.}(2022{\natexlab{a}})\citenamefont {Dornheim}, \citenamefont {Moldabekov}, \citenamefont {Vorberger}, \citenamefont {K{\"a}hlert},\ and\ \citenamefont {Bonitz}}]{Dornheim_Nature_2022}%
  \BibitemOpen
  \bibfield  {author} {\bibinfo {author} {\bibfnamefont {Tobias}\ \bibnamefont {Dornheim}}, \bibinfo {author} {\bibfnamefont {Zhandos}\ \bibnamefont {Moldabekov}}, \bibinfo {author} {\bibfnamefont {Jan}\ \bibnamefont {Vorberger}}, \bibinfo {author} {\bibfnamefont {Hanno}\ \bibnamefont {K{\"a}hlert}}, \ and\ \bibinfo {author} {\bibfnamefont {Michael}\ \bibnamefont {Bonitz}},\ }\bibfield  {title} {\enquote {\bibinfo {title} {Electronic pair alignment and roton feature in the warm dense electron gas},}\ }\href {\doibase 10.1038/s42005-022-01078-9} {\bibfield  {journal} {\bibinfo  {journal} {Communications Physics}\ }\textbf {\bibinfo {volume} {5}},\ \bibinfo {pages} {304} (\bibinfo {year} {2022}{\natexlab{a}})}\BibitemShut {NoStop}%
\bibitem [{\citenamefont {Dornheim}\ \emph {et~al.}(2022{\natexlab{b}})\citenamefont {Dornheim}, \citenamefont {Tolias}, \citenamefont {Moldabekov}, \citenamefont {Cangi},\ and\ \citenamefont {Vorberger}}]{Dornheim_JCP_Force_2022}%
  \BibitemOpen
  \bibfield  {author} {\bibinfo {author} {\bibfnamefont {Tobias}\ \bibnamefont {Dornheim}}, \bibinfo {author} {\bibfnamefont {Panagiotis}\ \bibnamefont {Tolias}}, \bibinfo {author} {\bibfnamefont {Zhandos~A.}\ \bibnamefont {Moldabekov}}, \bibinfo {author} {\bibfnamefont {Attila}\ \bibnamefont {Cangi}}, \ and\ \bibinfo {author} {\bibfnamefont {Jan}\ \bibnamefont {Vorberger}},\ }\bibfield  {title} {\enquote {\bibinfo {title} {{Effective electronic forces and potentials from ab initio path integral Monte Carlo simulations}},}\ }\href {\doibase 10.1063/5.0097768} {\bibfield  {journal} {\bibinfo  {journal} {The Journal of Chemical Physics}\ }\textbf {\bibinfo {volume} {156}},\ \bibinfo {pages} {244113} (\bibinfo {year} {2022}{\natexlab{b}})}\BibitemShut {NoStop}%
\bibitem [{\citenamefont {Koskelo}\ \emph {et~al.}(2023)\citenamefont {Koskelo}, \citenamefont {Reining},\ and\ \citenamefont {Gatti}}]{koskelo2023shortrange}%
  \BibitemOpen
  \bibfield  {author} {\bibinfo {author} {\bibfnamefont {Jaakko}\ \bibnamefont {Koskelo}}, \bibinfo {author} {\bibfnamefont {Lucia}\ \bibnamefont {Reining}}, \ and\ \bibinfo {author} {\bibfnamefont {Matteo}\ \bibnamefont {Gatti}},\ }\href@noop {} {\enquote {\bibinfo {title} {Short-range excitonic phenomena in low-density metals},}\ } (\bibinfo {year} {2023}),\ \Eprint {http://arxiv.org/abs/2301.00474} {arXiv:2301.00474 [cond-mat.str-el]} \BibitemShut {NoStop}%
\bibitem [{\citenamefont {Wigner}(1934)}]{Wigner_PhysRev_1934}%
  \BibitemOpen
  \bibfield  {author} {\bibinfo {author} {\bibfnamefont {E.}~\bibnamefont {Wigner}},\ }\bibfield  {title} {\enquote {\bibinfo {title} {On the interaction of electrons in metals},}\ }\href {\doibase 10.1103/PhysRev.46.1002} {\bibfield  {journal} {\bibinfo  {journal} {Phys. Rev.}\ }\textbf {\bibinfo {volume} {46}},\ \bibinfo {pages} {1002--1011} (\bibinfo {year} {1934})}\BibitemShut {NoStop}%
\bibitem [{\citenamefont {Drummond}\ \emph {et~al.}(2004)\citenamefont {Drummond}, \citenamefont {Radnai}, \citenamefont {Trail}, \citenamefont {Towler},\ and\ \citenamefont {Needs}}]{Drummond_PRB_2004}%
  \BibitemOpen
  \bibfield  {author} {\bibinfo {author} {\bibfnamefont {N.~D.}\ \bibnamefont {Drummond}}, \bibinfo {author} {\bibfnamefont {Z.}~\bibnamefont {Radnai}}, \bibinfo {author} {\bibfnamefont {J.~R.}\ \bibnamefont {Trail}}, \bibinfo {author} {\bibfnamefont {M.~D.}\ \bibnamefont {Towler}}, \ and\ \bibinfo {author} {\bibfnamefont {R.~J.}\ \bibnamefont {Needs}},\ }\bibfield  {title} {\enquote {\bibinfo {title} {{Diffusion quantum Monte Carlo study of three-dimensional Wigner crystals}},}\ }\href {\doibase 10.1103/PhysRevB.69.085116} {\bibfield  {journal} {\bibinfo  {journal} {Phys. Rev. B}\ }\textbf {\bibinfo {volume} {69}},\ \bibinfo {pages} {085116} (\bibinfo {year} {2004})}\BibitemShut {NoStop}%
\bibitem [{\citenamefont {Azadi}\ and\ \citenamefont {Drummond}(2022)}]{Azadi_Wigner_2022}%
  \BibitemOpen
  \bibfield  {author} {\bibinfo {author} {\bibfnamefont {Sam}\ \bibnamefont {Azadi}}\ and\ \bibinfo {author} {\bibfnamefont {N.~D.}\ \bibnamefont {Drummond}},\ }\bibfield  {title} {\enquote {\bibinfo {title} {Low-density phase diagram of the three-dimensional electron gas},}\ }\href {\doibase 10.1103/PhysRevB.105.245135} {\bibfield  {journal} {\bibinfo  {journal} {Phys. Rev. B}\ }\textbf {\bibinfo {volume} {105}},\ \bibinfo {pages} {245135} (\bibinfo {year} {2022})}\BibitemShut {NoStop}%
\bibitem [{\citenamefont {Vosko}\ \emph {et~al.}(1980)\citenamefont {Vosko}, \citenamefont {Wilk},\ and\ \citenamefont {Nusair}}]{vwn}%
  \BibitemOpen
  \bibfield  {author} {\bibinfo {author} {\bibfnamefont {S.~H.}\ \bibnamefont {Vosko}}, \bibinfo {author} {\bibfnamefont {L.}~\bibnamefont {Wilk}}, \ and\ \bibinfo {author} {\bibfnamefont {M.}~\bibnamefont {Nusair}},\ }\bibfield  {title} {\enquote {\bibinfo {title} {Accurate spin-dependent electron liquid correlation energies for local spin density calculations: a critical analysis},}\ }\href {\doibase 10.1139/p80-159} {\bibfield  {journal} {\bibinfo  {journal} {Canadian Journal of Physics}\ }\textbf {\bibinfo {volume} {58}},\ \bibinfo {pages} {1200--1211} (\bibinfo {year} {1980})}\BibitemShut {NoStop}%
\bibitem [{\citenamefont {Perdew}\ and\ \citenamefont {Wang}(1992)}]{Perdew_Wang_PRB_1992}%
  \BibitemOpen
  \bibfield  {author} {\bibinfo {author} {\bibfnamefont {John~P.}\ \bibnamefont {Perdew}}\ and\ \bibinfo {author} {\bibfnamefont {Yue}\ \bibnamefont {Wang}},\ }\bibfield  {title} {\enquote {\bibinfo {title} {Accurate and simple analytic representation of the electron-gas correlation energy},}\ }\href {\doibase 10.1103/PhysRevB.45.13244} {\bibfield  {journal} {\bibinfo  {journal} {Phys. Rev. B}\ }\textbf {\bibinfo {volume} {45}},\ \bibinfo {pages} {13244--13249} (\bibinfo {year} {1992})}\BibitemShut {NoStop}%
\bibitem [{\citenamefont {Perdew}\ and\ \citenamefont {Zunger}(1981)}]{Perdew_Zunger_PRB_1981}%
  \BibitemOpen
  \bibfield  {author} {\bibinfo {author} {\bibfnamefont {J.~P.}\ \bibnamefont {Perdew}}\ and\ \bibinfo {author} {\bibfnamefont {Alex}\ \bibnamefont {Zunger}},\ }\bibfield  {title} {\enquote {\bibinfo {title} {Self-interaction correction to density-functional approximations for many-electron systems},}\ }\href {\doibase 10.1103/PhysRevB.23.5048} {\bibfield  {journal} {\bibinfo  {journal} {Phys. Rev. B}\ }\textbf {\bibinfo {volume} {23}},\ \bibinfo {pages} {5048--5079} (\bibinfo {year} {1981})}\BibitemShut {NoStop}%
\bibitem [{\citenamefont {Gori-Giorgi}\ \emph {et~al.}(2000)\citenamefont {Gori-Giorgi}, \citenamefont {Sacchetti},\ and\ \citenamefont {Bachelet}}]{Gori-Giorgi_PRB_2000}%
  \BibitemOpen
  \bibfield  {author} {\bibinfo {author} {\bibfnamefont {Paola}\ \bibnamefont {Gori-Giorgi}}, \bibinfo {author} {\bibfnamefont {Francesco}\ \bibnamefont {Sacchetti}}, \ and\ \bibinfo {author} {\bibfnamefont {Giovanni~B.}\ \bibnamefont {Bachelet}},\ }\bibfield  {title} {\enquote {\bibinfo {title} {Analytic static structure factors and pair-correlation functions for the unpolarized homogeneous electron gas},}\ }\href {\doibase 10.1103/PhysRevB.61.7353} {\bibfield  {journal} {\bibinfo  {journal} {Phys. Rev. B}\ }\textbf {\bibinfo {volume} {61}},\ \bibinfo {pages} {7353--7363} (\bibinfo {year} {2000})}\BibitemShut {NoStop}%
\bibitem [{\citenamefont {Corradini}\ \emph {et~al.}(1998)\citenamefont {Corradini}, \citenamefont {Sole}, \citenamefont {Onida},\ and\ \citenamefont {Palummo}}]{cdop}%
  \BibitemOpen
  \bibfield  {author} {\bibinfo {author} {\bibfnamefont {M.}~\bibnamefont {Corradini}}, \bibinfo {author} {\bibfnamefont {R.~Del}\ \bibnamefont {Sole}}, \bibinfo {author} {\bibfnamefont {G.}~\bibnamefont {Onida}}, \ and\ \bibinfo {author} {\bibfnamefont {M.}~\bibnamefont {Palummo}},\ }\bibfield  {title} {\enquote {\bibinfo {title} {Analytical expressions for the local-field factor $g(q)$ and the exchange-correlation kernel ${K}_{\mathrm{xc}}(r)$ of the homogeneous electron gas},}\ }\href {http://link.aps.org/doi/10.1103/PhysRevB.57.14569} {\bibfield  {journal} {\bibinfo  {journal} {Phys. Rev. B}\ }\textbf {\bibinfo {volume} {57}},\ \bibinfo {pages} {14569} (\bibinfo {year} {1998})}\BibitemShut {NoStop}%
\bibitem [{\citenamefont {Ceperley}\ and\ \citenamefont {Alder}(1980)}]{Ceperley_Alder_PRL_1980}%
  \BibitemOpen
  \bibfield  {author} {\bibinfo {author} {\bibfnamefont {D.~M.}\ \bibnamefont {Ceperley}}\ and\ \bibinfo {author} {\bibfnamefont {B.~J.}\ \bibnamefont {Alder}},\ }\bibfield  {title} {\enquote {\bibinfo {title} {Ground state of the electron gas by a stochastic method},}\ }\href {\doibase 10.1103/PhysRevLett.45.566} {\bibfield  {journal} {\bibinfo  {journal} {Phys. Rev. Lett.}\ }\textbf {\bibinfo {volume} {45}},\ \bibinfo {pages} {566--569} (\bibinfo {year} {1980})}\BibitemShut {NoStop}%
\bibitem [{\citenamefont {Spink}\ \emph {et~al.}(2013)\citenamefont {Spink}, \citenamefont {Needs},\ and\ \citenamefont {Drummond}}]{Spink_PRB_2013}%
  \BibitemOpen
  \bibfield  {author} {\bibinfo {author} {\bibfnamefont {G.~G.}\ \bibnamefont {Spink}}, \bibinfo {author} {\bibfnamefont {R.~J.}\ \bibnamefont {Needs}}, \ and\ \bibinfo {author} {\bibfnamefont {N.~D.}\ \bibnamefont {Drummond}},\ }\bibfield  {title} {\enquote {\bibinfo {title} {{Quantum Monte Carlo study of the three-dimensional spin-polarized homogeneous electron gas}},}\ }\href {\doibase 10.1103/PhysRevB.88.085121} {\bibfield  {journal} {\bibinfo  {journal} {Phys. Rev. B}\ }\textbf {\bibinfo {volume} {88}},\ \bibinfo {pages} {085121} (\bibinfo {year} {2013})}\BibitemShut {NoStop}%
\bibitem [{\citenamefont {Moroni}\ \emph {et~al.}(1992)\citenamefont {Moroni}, \citenamefont {Ceperley},\ and\ \citenamefont {Senatore}}]{moroni}%
  \BibitemOpen
  \bibfield  {author} {\bibinfo {author} {\bibfnamefont {S.}~\bibnamefont {Moroni}}, \bibinfo {author} {\bibfnamefont {D.~M.}\ \bibnamefont {Ceperley}}, \ and\ \bibinfo {author} {\bibfnamefont {G.}~\bibnamefont {Senatore}},\ }\bibfield  {title} {\enquote {\bibinfo {title} {Static response from quantum {M}onte {C}arlo calculations},}\ }\href {https://journals.aps.org/prl/abstract/10.1103/PhysRevLett.69.1837} {\bibfield  {journal} {\bibinfo  {journal} {Phys. Rev. Lett}\ }\textbf {\bibinfo {volume} {69}},\ \bibinfo {pages} {1837} (\bibinfo {year} {1992})}\BibitemShut {NoStop}%
\bibitem [{\citenamefont {Moroni}\ \emph {et~al.}(1995)\citenamefont {Moroni}, \citenamefont {Ceperley},\ and\ \citenamefont {Senatore}}]{moroni2}%
  \BibitemOpen
  \bibfield  {author} {\bibinfo {author} {\bibfnamefont {S.}~\bibnamefont {Moroni}}, \bibinfo {author} {\bibfnamefont {D.~M.}\ \bibnamefont {Ceperley}}, \ and\ \bibinfo {author} {\bibfnamefont {G.}~\bibnamefont {Senatore}},\ }\bibfield  {title} {\enquote {\bibinfo {title} {Static response and local field factor of the electron gas},}\ }\href {http://link.aps.org/doi/10.1103/PhysRevLett.75.689} {\bibfield  {journal} {\bibinfo  {journal} {Phys. Rev. Lett}\ }\textbf {\bibinfo {volume} {75}},\ \bibinfo {pages} {689} (\bibinfo {year} {1995})}\BibitemShut {NoStop}%
\bibitem [{\citenamefont {Jones}(2015)}]{Jones_RMP_2015}%
  \BibitemOpen
  \bibfield  {author} {\bibinfo {author} {\bibfnamefont {R.~O.}\ \bibnamefont {Jones}},\ }\bibfield  {title} {\enquote {\bibinfo {title} {Density functional theory: Its origins, rise to prominence, and future},}\ }\href {\doibase 10.1103/RevModPhys.87.897} {\bibfield  {journal} {\bibinfo  {journal} {Rev. Mod. Phys.}\ }\textbf {\bibinfo {volume} {87}},\ \bibinfo {pages} {897--923} (\bibinfo {year} {2015})}\BibitemShut {NoStop}%
\bibitem [{\citenamefont {Betti}\ and\ \citenamefont {Hurricane}(2016)}]{Betti2016}%
  \BibitemOpen
  \bibfield  {author} {\bibinfo {author} {\bibfnamefont {R.}~\bibnamefont {Betti}}\ and\ \bibinfo {author} {\bibfnamefont {O.~A.}\ \bibnamefont {Hurricane}},\ }\bibfield  {title} {\enquote {\bibinfo {title} {Inertial-confinement fusion with lasers},}\ }\href {\doibase 10.1038/nphys3736} {\bibfield  {journal} {\bibinfo  {journal} {Nature Physics}\ }\textbf {\bibinfo {volume} {12}},\ \bibinfo {pages} {435--448} (\bibinfo {year} {2016})}\BibitemShut {NoStop}%
\bibitem [{\citenamefont {Hurricane}\ \emph {et~al.}(2023)\citenamefont {Hurricane}, \citenamefont {Patel}, \citenamefont {Betti}, \citenamefont {Froula}, \citenamefont {Regan}, \citenamefont {Slutz}, \citenamefont {Gomez},\ and\ \citenamefont {Sweeney}}]{Hurricane_RevModPhys_2023}%
  \BibitemOpen
  \bibfield  {author} {\bibinfo {author} {\bibfnamefont {O.~A.}\ \bibnamefont {Hurricane}}, \bibinfo {author} {\bibfnamefont {P.~K.}\ \bibnamefont {Patel}}, \bibinfo {author} {\bibfnamefont {R.}~\bibnamefont {Betti}}, \bibinfo {author} {\bibfnamefont {D.~H.}\ \bibnamefont {Froula}}, \bibinfo {author} {\bibfnamefont {S.~P.}\ \bibnamefont {Regan}}, \bibinfo {author} {\bibfnamefont {S.~A.}\ \bibnamefont {Slutz}}, \bibinfo {author} {\bibfnamefont {M.~R.}\ \bibnamefont {Gomez}}, \ and\ \bibinfo {author} {\bibfnamefont {M.~A.}\ \bibnamefont {Sweeney}},\ }\bibfield  {title} {\enquote {\bibinfo {title} {Physics principles of inertial confinement fusion and u.s. program overview},}\ }\href {\doibase 10.1103/RevModPhys.95.025005} {\bibfield  {journal} {\bibinfo  {journal} {Rev. Mod. Phys.}\ }\textbf {\bibinfo {volume} {95}},\ \bibinfo {pages} {025005} (\bibinfo {year} {2023})}\BibitemShut {NoStop}%
\bibitem [{\citenamefont {Abu-Shawareb}\ \emph {et~al.}(2024)\citenamefont {Abu-Shawareb} \emph {et~al.}}]{NIF_PRL_2024}%
  \BibitemOpen
  \bibfield  {author} {\bibinfo {author} {\bibnamefont {Abu-Shawareb}} \emph {et~al.} (\bibinfo {collaboration} {The Indirect Drive ICF Collaboration}),\ }\bibfield  {title} {\enquote {\bibinfo {title} {Achievement of target gain larger than unity in an inertial fusion experiment},}\ }\href {\doibase 10.1103/PhysRevLett.132.065102} {\bibfield  {journal} {\bibinfo  {journal} {Phys. Rev. Lett.}\ }\textbf {\bibinfo {volume} {132}},\ \bibinfo {pages} {065102} (\bibinfo {year} {2024})}\BibitemShut {NoStop}%
\bibitem [{\citenamefont {Benuzzi-Mounaix}\ \emph {et~al.}(2014)\citenamefont {Benuzzi-Mounaix}, \citenamefont {Mazevet}, \citenamefont {Ravasio}, \citenamefont {Vinci}, \citenamefont {Denoeud}, \citenamefont {Koenig}, \citenamefont {Amadou}, \citenamefont {Brambrink}, \citenamefont {Festa}, \citenamefont {Levy}, \citenamefont {Harmand}, \citenamefont {Brygoo}, \citenamefont {Huser}, \citenamefont {Recoules}, \citenamefont {Bouchet}, \citenamefont {Morard}, \citenamefont {Guyot}, \citenamefont {de~Resseguier}, \citenamefont {Myanishi}, \citenamefont {Ozaki}, \citenamefont {Dorchies}, \citenamefont {Gaudin}, \citenamefont {Leguay}, \citenamefont {Peyrusse}, \citenamefont {Henry}, \citenamefont {Raffestin}, \citenamefont {Pape}, \citenamefont {Smith},\ and\ \citenamefont {Musella}}]{Benuzzi_Mounaix_2014}%
  \BibitemOpen
  \bibfield  {author} {\bibinfo {author} {\bibfnamefont {Alessandra}\ \bibnamefont {Benuzzi-Mounaix}}, \bibinfo {author} {\bibfnamefont {St{\'{e}}phane}\ \bibnamefont {Mazevet}}, \bibinfo {author} {\bibfnamefont {Alessandra}\ \bibnamefont {Ravasio}}, \bibinfo {author} {\bibfnamefont {Tommaso}\ \bibnamefont {Vinci}}, \bibinfo {author} {\bibfnamefont {Adrien}\ \bibnamefont {Denoeud}}, \bibinfo {author} {\bibfnamefont {Michel}\ \bibnamefont {Koenig}}, \bibinfo {author} {\bibfnamefont {Nourou}\ \bibnamefont {Amadou}}, \bibinfo {author} {\bibfnamefont {Erik}\ \bibnamefont {Brambrink}}, \bibinfo {author} {\bibfnamefont {Floriane}\ \bibnamefont {Festa}}, \bibinfo {author} {\bibfnamefont {Anna}\ \bibnamefont {Levy}}, \bibinfo {author} {\bibfnamefont {Marion}\ \bibnamefont {Harmand}}, \bibinfo {author} {\bibfnamefont {St{\'{e}}phanie}\ \bibnamefont {Brygoo}}, \bibinfo {author} {\bibfnamefont {Gael}\ \bibnamefont {Huser}}, \bibinfo {author} {\bibfnamefont {Vanina}\ \bibnamefont {Recoules}}, \bibinfo {author}
  {\bibfnamefont {Johan}\ \bibnamefont {Bouchet}}, \bibinfo {author} {\bibfnamefont {Guillaume}\ \bibnamefont {Morard}}, \bibinfo {author} {\bibfnamefont {Fran{\c{c}}ois}\ \bibnamefont {Guyot}}, \bibinfo {author} {\bibfnamefont {Thibaut}\ \bibnamefont {de~Resseguier}}, \bibinfo {author} {\bibfnamefont {Kohei}\ \bibnamefont {Myanishi}}, \bibinfo {author} {\bibfnamefont {Norimasa}\ \bibnamefont {Ozaki}}, \bibinfo {author} {\bibfnamefont {Fabien}\ \bibnamefont {Dorchies}}, \bibinfo {author} {\bibfnamefont {Jer{\^{o}}me}\ \bibnamefont {Gaudin}}, \bibinfo {author} {\bibfnamefont {Pierre~Marie}\ \bibnamefont {Leguay}}, \bibinfo {author} {\bibfnamefont {Olivier}\ \bibnamefont {Peyrusse}}, \bibinfo {author} {\bibfnamefont {Olivier}\ \bibnamefont {Henry}}, \bibinfo {author} {\bibfnamefont {Didier}\ \bibnamefont {Raffestin}}, \bibinfo {author} {\bibfnamefont {Sebastien~Le}\ \bibnamefont {Pape}}, \bibinfo {author} {\bibfnamefont {Ray}\ \bibnamefont {Smith}}, \ and\ \bibinfo {author} {\bibfnamefont {Riccardo}\
  \bibnamefont {Musella}},\ }\bibfield  {title} {\enquote {\bibinfo {title} {Progress in warm dense matter study with applications to planetology},}\ }\href {\doibase 10.1088/0031-8949/2014/t161/014060} {\bibfield  {journal} {\bibinfo  {journal} {Phys. Scripta}\ }\textbf {\bibinfo {volume} {T161}},\ \bibinfo {pages} {014060} (\bibinfo {year} {2014})}\BibitemShut {NoStop}%
\bibitem [{\citenamefont {Becker}\ \emph {et~al.}(2014)\citenamefont {Becker}, \citenamefont {Lorenzen}, \citenamefont {Fortney}, \citenamefont {Nettelmann}, \citenamefont {Sch\"ottler},\ and\ \citenamefont {Redmer}}]{becker}%
  \BibitemOpen
  \bibfield  {author} {\bibinfo {author} {\bibfnamefont {A.}~\bibnamefont {Becker}}, \bibinfo {author} {\bibfnamefont {W.}~\bibnamefont {Lorenzen}}, \bibinfo {author} {\bibfnamefont {J.~J.}\ \bibnamefont {Fortney}}, \bibinfo {author} {\bibfnamefont {N.}~\bibnamefont {Nettelmann}}, \bibinfo {author} {\bibfnamefont {M.}~\bibnamefont {Sch\"ottler}}, \ and\ \bibinfo {author} {\bibfnamefont {R.}~\bibnamefont {Redmer}},\ }\bibfield  {title} {\enquote {\bibinfo {title} {Ab initio equations of state for hydrogen (h-reos.3) and helium (he-reos.3) and their implications for the interior of brown dwarfs},}\ }\href {https://iopscience.iop.org/article/10.1088/0067-0049/215/2/21/meta} {\bibfield  {journal} {\bibinfo  {journal} {Astrophys. J. Suppl. Ser}\ }\textbf {\bibinfo {volume} {215}},\ \bibinfo {pages} {21} (\bibinfo {year} {2014})}\BibitemShut {NoStop}%
\bibitem [{\citenamefont {Kritcher}\ \emph {et~al.}(2020)\citenamefont {Kritcher}, \citenamefont {Swift}, \citenamefont {D{\"o}ppner}, \citenamefont {Bachmann}, \citenamefont {Benedict}, \citenamefont {Collins}, \citenamefont {DuBois}, \citenamefont {Elsner}, \citenamefont {Fontaine}, \citenamefont {Gaffney}, \citenamefont {Hamel}, \citenamefont {Lazicki}, \citenamefont {Johnson}, \citenamefont {Kostinski}, \citenamefont {Kraus}, \citenamefont {MacDonald}, \citenamefont {Maddox}, \citenamefont {Martin}, \citenamefont {Neumayer}, \citenamefont {Nikroo}, \citenamefont {Nilsen}, \citenamefont {Remington}, \citenamefont {Saumon}, \citenamefont {Sterne}, \citenamefont {Sweet}, \citenamefont {Correa}, \citenamefont {Whitley}, \citenamefont {Falcone},\ and\ \citenamefont {Glenzer}}]{Kritcher_Nature_2020}%
  \BibitemOpen
  \bibfield  {author} {\bibinfo {author} {\bibfnamefont {Andrea~L.}\ \bibnamefont {Kritcher}}, \bibinfo {author} {\bibfnamefont {Damian~C.}\ \bibnamefont {Swift}}, \bibinfo {author} {\bibfnamefont {Tilo}\ \bibnamefont {D{\"o}ppner}}, \bibinfo {author} {\bibfnamefont {Benjamin}\ \bibnamefont {Bachmann}}, \bibinfo {author} {\bibfnamefont {Lorin~X.}\ \bibnamefont {Benedict}}, \bibinfo {author} {\bibfnamefont {Gilbert~W.}\ \bibnamefont {Collins}}, \bibinfo {author} {\bibfnamefont {Jonathan~L.}\ \bibnamefont {DuBois}}, \bibinfo {author} {\bibfnamefont {Fred}\ \bibnamefont {Elsner}}, \bibinfo {author} {\bibfnamefont {Gilles}\ \bibnamefont {Fontaine}}, \bibinfo {author} {\bibfnamefont {Jim~A.}\ \bibnamefont {Gaffney}}, \bibinfo {author} {\bibfnamefont {Sebastien}\ \bibnamefont {Hamel}}, \bibinfo {author} {\bibfnamefont {Amy}\ \bibnamefont {Lazicki}}, \bibinfo {author} {\bibfnamefont {Walter~R.}\ \bibnamefont {Johnson}}, \bibinfo {author} {\bibfnamefont {Natalie}\ \bibnamefont {Kostinski}}, \bibinfo {author}
  {\bibfnamefont {Dominik}\ \bibnamefont {Kraus}}, \bibinfo {author} {\bibfnamefont {Michael~J.}\ \bibnamefont {MacDonald}}, \bibinfo {author} {\bibfnamefont {Brian}\ \bibnamefont {Maddox}}, \bibinfo {author} {\bibfnamefont {Madison~E.}\ \bibnamefont {Martin}}, \bibinfo {author} {\bibfnamefont {Paul}\ \bibnamefont {Neumayer}}, \bibinfo {author} {\bibfnamefont {Abbas}\ \bibnamefont {Nikroo}}, \bibinfo {author} {\bibfnamefont {Joseph}\ \bibnamefont {Nilsen}}, \bibinfo {author} {\bibfnamefont {Bruce~A.}\ \bibnamefont {Remington}}, \bibinfo {author} {\bibfnamefont {Didier}\ \bibnamefont {Saumon}}, \bibinfo {author} {\bibfnamefont {Phillip~A.}\ \bibnamefont {Sterne}}, \bibinfo {author} {\bibfnamefont {Wendi}\ \bibnamefont {Sweet}}, \bibinfo {author} {\bibfnamefont {Alfredo~A.}\ \bibnamefont {Correa}}, \bibinfo {author} {\bibfnamefont {Heather~D.}\ \bibnamefont {Whitley}}, \bibinfo {author} {\bibfnamefont {Roger~W.}\ \bibnamefont {Falcone}}, \ and\ \bibinfo {author} {\bibfnamefont {Siegfried~H.}\ \bibnamefont
  {Glenzer}},\ }\bibfield  {title} {\enquote {\bibinfo {title} {A measurement of the equation of state of carbon envelopes of white dwarfs},}\ }\href {\doibase 10.1038/s41586-020-2535-y} {\bibfield  {journal} {\bibinfo  {journal} {Nature}\ }\textbf {\bibinfo {volume} {584}},\ \bibinfo {pages} {51--54} (\bibinfo {year} {2020})}\BibitemShut {NoStop}%
\bibitem [{\citenamefont {Malone}\ \emph {et~al.}(2016)\citenamefont {Malone}, \citenamefont {Blunt}, \citenamefont {Brown}, \citenamefont {Lee}, \citenamefont {Spencer}, \citenamefont {Foulkes},\ and\ \citenamefont {Shepherd}}]{Malone_PRL_2016}%
  \BibitemOpen
  \bibfield  {author} {\bibinfo {author} {\bibfnamefont {Fionn~D.}\ \bibnamefont {Malone}}, \bibinfo {author} {\bibfnamefont {N.~S.}\ \bibnamefont {Blunt}}, \bibinfo {author} {\bibfnamefont {Ethan~W.}\ \bibnamefont {Brown}}, \bibinfo {author} {\bibfnamefont {D.~K.~K.}\ \bibnamefont {Lee}}, \bibinfo {author} {\bibfnamefont {J.~S.}\ \bibnamefont {Spencer}}, \bibinfo {author} {\bibfnamefont {W.~M.~C.}\ \bibnamefont {Foulkes}}, \ and\ \bibinfo {author} {\bibfnamefont {James~J.}\ \bibnamefont {Shepherd}},\ }\bibfield  {title} {\enquote {\bibinfo {title} {Accurate exchange-correlation energies for the warm dense electron gas},}\ }\href {\doibase 10.1103/PhysRevLett.117.115701} {\bibfield  {journal} {\bibinfo  {journal} {Phys. Rev. Lett.}\ }\textbf {\bibinfo {volume} {117}},\ \bibinfo {pages} {115701} (\bibinfo {year} {2016})}\BibitemShut {NoStop}%
\bibitem [{\citenamefont {Dornheim}\ \emph {et~al.}(2016)\citenamefont {Dornheim}, \citenamefont {Groth}, \citenamefont {Sjostrom}, \citenamefont {Malone}, \citenamefont {Foulkes},\ and\ \citenamefont {Bonitz}}]{dornheim_prl}%
  \BibitemOpen
  \bibfield  {author} {\bibinfo {author} {\bibfnamefont {T.}~\bibnamefont {Dornheim}}, \bibinfo {author} {\bibfnamefont {S.}~\bibnamefont {Groth}}, \bibinfo {author} {\bibfnamefont {T.}~\bibnamefont {Sjostrom}}, \bibinfo {author} {\bibfnamefont {F.~D.}\ \bibnamefont {Malone}}, \bibinfo {author} {\bibfnamefont {W.~M.~C.}\ \bibnamefont {Foulkes}}, \ and\ \bibinfo {author} {\bibfnamefont {M.}~\bibnamefont {Bonitz}},\ }\bibfield  {title} {\enquote {\bibinfo {title} {Ab initio quantum {M}onte {C}arlo simulation of the warm dense electron gas in the thermodynamic limit},}\ }\href {http://link.aps.org/doi/10.1103/PhysRevLett.117.156403} {\bibfield  {journal} {\bibinfo  {journal} {Phys. Rev. Lett.}\ }\textbf {\bibinfo {volume} {117}},\ \bibinfo {pages} {156403} (\bibinfo {year} {2016})}\BibitemShut {NoStop}%
\bibitem [{\citenamefont {Groth}\ \emph {et~al.}(2017{\natexlab{a}})\citenamefont {Groth}, \citenamefont {Dornheim}, \citenamefont {Sjostrom}, \citenamefont {Malone}, \citenamefont {Foulkes},\ and\ \citenamefont {Bonitz}}]{groth_prl}%
  \BibitemOpen
  \bibfield  {author} {\bibinfo {author} {\bibfnamefont {S.}~\bibnamefont {Groth}}, \bibinfo {author} {\bibfnamefont {T.}~\bibnamefont {Dornheim}}, \bibinfo {author} {\bibfnamefont {T.}~\bibnamefont {Sjostrom}}, \bibinfo {author} {\bibfnamefont {F.~D.}\ \bibnamefont {Malone}}, \bibinfo {author} {\bibfnamefont {W.~M.~C.}\ \bibnamefont {Foulkes}}, \ and\ \bibinfo {author} {\bibfnamefont {M.}~\bibnamefont {Bonitz}},\ }\bibfield  {title} {\enquote {\bibinfo {title} {Ab initio exchange--correlation free energy of the uniform electron gas at warm dense matter conditions},}\ }\href {https://journals.aps.org/prl/abstract/10.1103/PhysRevLett.119.135001} {\bibfield  {journal} {\bibinfo  {journal} {Phys. Rev. Lett.}\ }\textbf {\bibinfo {volume} {119}},\ \bibinfo {pages} {135001} (\bibinfo {year} {2017}{\natexlab{a}})}\BibitemShut {NoStop}%
\bibitem [{\citenamefont {Karasiev}\ \emph {et~al.}(2014)\citenamefont {Karasiev}, \citenamefont {Sjostrom}, \citenamefont {Dufty},\ and\ \citenamefont {Trickey}}]{ksdt}%
  \BibitemOpen
  \bibfield  {author} {\bibinfo {author} {\bibfnamefont {Valentin~V.}\ \bibnamefont {Karasiev}}, \bibinfo {author} {\bibfnamefont {Travis}\ \bibnamefont {Sjostrom}}, \bibinfo {author} {\bibfnamefont {James}\ \bibnamefont {Dufty}}, \ and\ \bibinfo {author} {\bibfnamefont {S.~B.}\ \bibnamefont {Trickey}},\ }\bibfield  {title} {\enquote {\bibinfo {title} {Accurate homogeneous electron gas exchange-correlation free energy for local spin-density calculations},}\ }\href {\doibase 10.1103/PhysRevLett.112.076403} {\bibfield  {journal} {\bibinfo  {journal} {Phys. Rev. Lett.}\ }\textbf {\bibinfo {volume} {112}},\ \bibinfo {pages} {076403} (\bibinfo {year} {2014})}\BibitemShut {NoStop}%
\bibitem [{\citenamefont {Karasiev}\ \emph {et~al.}(2019)\citenamefont {Karasiev}, \citenamefont {Trickey},\ and\ \citenamefont {Dufty}}]{Karasiev_status_2019}%
  \BibitemOpen
  \bibfield  {author} {\bibinfo {author} {\bibfnamefont {Valentin~V.}\ \bibnamefont {Karasiev}}, \bibinfo {author} {\bibfnamefont {S.~B.}\ \bibnamefont {Trickey}}, \ and\ \bibinfo {author} {\bibfnamefont {James~W.}\ \bibnamefont {Dufty}},\ }\bibfield  {title} {\enquote {\bibinfo {title} {Status of free-energy representations for the homogeneous electron gas},}\ }\href {\doibase 10.1103/PhysRevB.99.195134} {\bibfield  {journal} {\bibinfo  {journal} {Phys. Rev. B}\ }\textbf {\bibinfo {volume} {99}},\ \bibinfo {pages} {195134} (\bibinfo {year} {2019})}\BibitemShut {NoStop}%
\bibitem [{\citenamefont {Dornheim}\ \emph {et~al.}(2017)\citenamefont {Dornheim}, \citenamefont {Groth}, \citenamefont {Vorberger},\ and\ \citenamefont {Bonitz}}]{dornheim_pre}%
  \BibitemOpen
  \bibfield  {author} {\bibinfo {author} {\bibfnamefont {T.}~\bibnamefont {Dornheim}}, \bibinfo {author} {\bibfnamefont {S.}~\bibnamefont {Groth}}, \bibinfo {author} {\bibfnamefont {J.}~\bibnamefont {Vorberger}}, \ and\ \bibinfo {author} {\bibfnamefont {M.}~\bibnamefont {Bonitz}},\ }\bibfield  {title} {\enquote {\bibinfo {title} {Permutation blocking path integral {M}onte {C}arlo approach to the static density response of the warm dense electron gas},}\ }\href {https://journals.aps.org/pre/abstract/10.1103/PhysRevE.96.023203} {\bibfield  {journal} {\bibinfo  {journal} {Phys. Rev. E}\ }\textbf {\bibinfo {volume} {96}},\ \bibinfo {pages} {023203} (\bibinfo {year} {2017})}\BibitemShut {NoStop}%
\bibitem [{\citenamefont {Groth}\ \emph {et~al.}(2017{\natexlab{b}})\citenamefont {Groth}, \citenamefont {Dornheim},\ and\ \citenamefont {Bonitz}}]{groth_jcp}%
  \BibitemOpen
  \bibfield  {author} {\bibinfo {author} {\bibfnamefont {S.}~\bibnamefont {Groth}}, \bibinfo {author} {\bibfnamefont {T.}~\bibnamefont {Dornheim}}, \ and\ \bibinfo {author} {\bibfnamefont {M.}~\bibnamefont {Bonitz}},\ }\bibfield  {title} {\enquote {\bibinfo {title} {Configuration path integral {M}onte {C}arlo approach to the static density response of the warm dense electron gas},}\ }\href {https://aip.scitation.org/doi/abs/10.1063/1.4999907} {\bibfield  {journal} {\bibinfo  {journal} {J. Chem. Phys}\ }\textbf {\bibinfo {volume} {147}},\ \bibinfo {pages} {164108} (\bibinfo {year} {2017}{\natexlab{b}})}\BibitemShut {NoStop}%
\bibitem [{\citenamefont {Bonitz}\ \emph {et~al.}(2020)\citenamefont {Bonitz}, \citenamefont {Dornheim}, \citenamefont {Moldabekov}, \citenamefont {Zhang}, \citenamefont {Hamann}, \citenamefont {Kählert}, \citenamefont {Filinov}, \citenamefont {Ramakrishna},\ and\ \citenamefont {Vorberger}}]{new_POP}%
  \BibitemOpen
  \bibfield  {author} {\bibinfo {author} {\bibfnamefont {M.}~\bibnamefont {Bonitz}}, \bibinfo {author} {\bibfnamefont {T.}~\bibnamefont {Dornheim}}, \bibinfo {author} {\bibfnamefont {Zh.~A.}\ \bibnamefont {Moldabekov}}, \bibinfo {author} {\bibfnamefont {S.}~\bibnamefont {Zhang}}, \bibinfo {author} {\bibfnamefont {P.}~\bibnamefont {Hamann}}, \bibinfo {author} {\bibfnamefont {H.}~\bibnamefont {Kählert}}, \bibinfo {author} {\bibfnamefont {A.}~\bibnamefont {Filinov}}, \bibinfo {author} {\bibfnamefont {K.}~\bibnamefont {Ramakrishna}}, \ and\ \bibinfo {author} {\bibfnamefont {J.}~\bibnamefont {Vorberger}},\ }\bibfield  {title} {\enquote {\bibinfo {title} {Ab initio simulation of warm dense matter},}\ }\href {\doibase 10.1063/1.5143225} {\bibfield  {journal} {\bibinfo  {journal} {Physics of Plasmas}\ }\textbf {\bibinfo {volume} {27}},\ \bibinfo {pages} {042710} (\bibinfo {year} {2020})}\BibitemShut {NoStop}%
\bibitem [{\citenamefont {Graziani}\ \emph {et~al.}(2014)\citenamefont {Graziani}, \citenamefont {Desjarlais}, \citenamefont {Redmer},\ and\ \citenamefont {Trickey}}]{wdm_book}%
  \BibitemOpen
  \bibinfo {editor} {\bibfnamefont {F.}~\bibnamefont {Graziani}}, \bibinfo {editor} {\bibfnamefont {M.~P.}\ \bibnamefont {Desjarlais}}, \bibinfo {editor} {\bibfnamefont {R.}~\bibnamefont {Redmer}}, \ and\ \bibinfo {editor} {\bibfnamefont {S.~B.}\ \bibnamefont {Trickey}},\ eds.,\ \href@noop {} {\emph {\bibinfo {title} {Frontiers and Challenges in Warm Dense Matter}}}\ (\bibinfo  {publisher} {Springer},\ \bibinfo {address} {International Publishing},\ \bibinfo {year} {2014})\BibitemShut {NoStop}%
\bibitem [{\citenamefont {Ceperley}(1995)}]{cep}%
  \BibitemOpen
  \bibfield  {author} {\bibinfo {author} {\bibfnamefont {D.~M.}\ \bibnamefont {Ceperley}},\ }\bibfield  {title} {\enquote {\bibinfo {title} {Path integrals in the theory of condensed helium},}\ }\href {https://journals.aps.org/rmp/abstract/10.1103/RevModPhys.67.279} {\bibfield  {journal} {\bibinfo  {journal} {Rev. Mod. Phys}\ }\textbf {\bibinfo {volume} {67}},\ \bibinfo {pages} {279} (\bibinfo {year} {1995})}\BibitemShut {NoStop}%
\bibitem [{\citenamefont {Herman}\ \emph {et~al.}(1982)\citenamefont {Herman}, \citenamefont {Bruskin},\ and\ \citenamefont {Berne}}]{Berne_JCP_1982}%
  \BibitemOpen
  \bibfield  {author} {\bibinfo {author} {\bibfnamefont {M.~F.}\ \bibnamefont {Herman}}, \bibinfo {author} {\bibfnamefont {E.~J.}\ \bibnamefont {Bruskin}}, \ and\ \bibinfo {author} {\bibfnamefont {B.~J.}\ \bibnamefont {Berne}},\ }\bibfield  {title} {\enquote {\bibinfo {title} {{On path integral Monte Carlo simulations}},}\ }\href {\doibase 10.1063/1.442815} {\bibfield  {journal} {\bibinfo  {journal} {The Journal of Chemical Physics}\ }\textbf {\bibinfo {volume} {76}},\ \bibinfo {pages} {5150--5155} (\bibinfo {year} {1982})}\BibitemShut {NoStop}%
\bibitem [{\citenamefont {Ott}\ \emph {et~al.}(2018)\citenamefont {Ott}, \citenamefont {Thomsen}, \citenamefont {Abraham}, \citenamefont {Dornheim},\ and\ \citenamefont {Bonitz}}]{Ott2018}%
  \BibitemOpen
  \bibfield  {author} {\bibinfo {author} {\bibfnamefont {Torben}\ \bibnamefont {Ott}}, \bibinfo {author} {\bibfnamefont {Hauke}\ \bibnamefont {Thomsen}}, \bibinfo {author} {\bibfnamefont {Jan~Willem}\ \bibnamefont {Abraham}}, \bibinfo {author} {\bibfnamefont {Tobias}\ \bibnamefont {Dornheim}}, \ and\ \bibinfo {author} {\bibfnamefont {Michael}\ \bibnamefont {Bonitz}},\ }\bibfield  {title} {\enquote {\bibinfo {title} {Recent progress in the theory and simulation of strongly correlated plasmas: phase transitions, transport, quantum, and magnetic field effects},}\ }\href {\doibase 10.1140/epjd/e2018-80385-7} {\bibfield  {journal} {\bibinfo  {journal} {The European Physical Journal D}\ }\textbf {\bibinfo {volume} {72}},\ \bibinfo {pages} {84} (\bibinfo {year} {2018})}\BibitemShut {NoStop}%
\bibitem [{\citenamefont {Dornheim}\ \emph {et~al.}(2023{\natexlab{a}})\citenamefont {Dornheim}, \citenamefont {Moldabekov}, \citenamefont {Ramakrishna}, \citenamefont {Tolias}, \citenamefont {Baczewski}, \citenamefont {Kraus}, \citenamefont {Preston}, \citenamefont {Chapman}, \citenamefont {Böhme}, \citenamefont {Döppner}, \citenamefont {Graziani}, \citenamefont {Bonitz}, \citenamefont {Cangi},\ and\ \citenamefont {Vorberger}}]{Dornheim_review}%
  \BibitemOpen
  \bibfield  {author} {\bibinfo {author} {\bibfnamefont {Tobias}\ \bibnamefont {Dornheim}}, \bibinfo {author} {\bibfnamefont {Zhandos~A.}\ \bibnamefont {Moldabekov}}, \bibinfo {author} {\bibfnamefont {Kushal}\ \bibnamefont {Ramakrishna}}, \bibinfo {author} {\bibfnamefont {Panagiotis}\ \bibnamefont {Tolias}}, \bibinfo {author} {\bibfnamefont {Andrew~D.}\ \bibnamefont {Baczewski}}, \bibinfo {author} {\bibfnamefont {Dominik}\ \bibnamefont {Kraus}}, \bibinfo {author} {\bibfnamefont {Thomas~R.}\ \bibnamefont {Preston}}, \bibinfo {author} {\bibfnamefont {David~A.}\ \bibnamefont {Chapman}}, \bibinfo {author} {\bibfnamefont {Maximilian~P.}\ \bibnamefont {Böhme}}, \bibinfo {author} {\bibfnamefont {Tilo}\ \bibnamefont {Döppner}}, \bibinfo {author} {\bibfnamefont {Frank}\ \bibnamefont {Graziani}}, \bibinfo {author} {\bibfnamefont {Michael}\ \bibnamefont {Bonitz}}, \bibinfo {author} {\bibfnamefont {Attila}\ \bibnamefont {Cangi}}, \ and\ \bibinfo {author} {\bibfnamefont {Jan}\ \bibnamefont {Vorberger}},\ }\bibfield
  {title} {\enquote {\bibinfo {title} {{Electronic density response of warm dense matter}},}\ }\href {\doibase 10.1063/5.0138955} {\bibfield  {journal} {\bibinfo  {journal} {Physics of Plasmas}\ }\textbf {\bibinfo {volume} {30}},\ \bibinfo {pages} {032705} (\bibinfo {year} {2023}{\natexlab{a}})}\BibitemShut {NoStop}%
\bibitem [{\citenamefont {Kugler}(1975)}]{kugler1}%
  \BibitemOpen
  \bibfield  {author} {\bibinfo {author} {\bibfnamefont {A.~A.}\ \bibnamefont {Kugler}},\ }\bibfield  {title} {\enquote {\bibinfo {title} {Theory of the local field correction in an electron gas},}\ }\href {http://link.springer.com/article/10.1007/BF01024183} {\bibfield  {journal} {\bibinfo  {journal} {J. Stat. Phys}\ }\textbf {\bibinfo {volume} {12}},\ \bibinfo {pages} {35} (\bibinfo {year} {1975})}\BibitemShut {NoStop}%
\bibitem [{\citenamefont {Zan}\ \emph {et~al.}(2021)\citenamefont {Zan}, \citenamefont {Lin}, \citenamefont {Hou},\ and\ \citenamefont {Yuan}}]{Zan_PRE_2021}%
  \BibitemOpen
  \bibfield  {author} {\bibinfo {author} {\bibfnamefont {Xiaolei}\ \bibnamefont {Zan}}, \bibinfo {author} {\bibfnamefont {Chengliang}\ \bibnamefont {Lin}}, \bibinfo {author} {\bibfnamefont {Yong}\ \bibnamefont {Hou}}, \ and\ \bibinfo {author} {\bibfnamefont {Jianmin}\ \bibnamefont {Yuan}},\ }\bibfield  {title} {\enquote {\bibinfo {title} {Local field correction to ionization potential depression of ions in warm or hot dense matter},}\ }\href {\doibase 10.1103/PhysRevE.104.025203} {\bibfield  {journal} {\bibinfo  {journal} {Phys. Rev. E}\ }\textbf {\bibinfo {volume} {104}},\ \bibinfo {pages} {025203} (\bibinfo {year} {2021})}\BibitemShut {NoStop}%
\bibitem [{\citenamefont {Moldabekov}\ \emph {et~al.}(2020)\citenamefont {Moldabekov}, \citenamefont {Dornheim}, \citenamefont {Bonitz},\ and\ \citenamefont {Ramazanov}}]{Moldabekov_PRE_2020}%
  \BibitemOpen
  \bibfield  {author} {\bibinfo {author} {\bibfnamefont {Zh.~A.}\ \bibnamefont {Moldabekov}}, \bibinfo {author} {\bibfnamefont {T.}~\bibnamefont {Dornheim}}, \bibinfo {author} {\bibfnamefont {M.}~\bibnamefont {Bonitz}}, \ and\ \bibinfo {author} {\bibfnamefont {T.~S.}\ \bibnamefont {Ramazanov}},\ }\bibfield  {title} {\enquote {\bibinfo {title} {Ion energy-loss characteristics and friction in a free-electron gas at warm dense matter and nonideal dense plasma conditions},}\ }\href {\doibase 10.1103/PhysRevE.101.053203} {\bibfield  {journal} {\bibinfo  {journal} {Phys. Rev. E}\ }\textbf {\bibinfo {volume} {101}},\ \bibinfo {pages} {053203} (\bibinfo {year} {2020})}\BibitemShut {NoStop}%
\bibitem [{\citenamefont {Fortmann}\ \emph {et~al.}(2010)\citenamefont {Fortmann}, \citenamefont {Wierling},\ and\ \citenamefont {R\"opke}}]{Fortmann_PRE_2010}%
  \BibitemOpen
  \bibfield  {author} {\bibinfo {author} {\bibfnamefont {Carsten}\ \bibnamefont {Fortmann}}, \bibinfo {author} {\bibfnamefont {August}\ \bibnamefont {Wierling}}, \ and\ \bibinfo {author} {\bibfnamefont {Gerd}\ \bibnamefont {R\"opke}},\ }\bibfield  {title} {\enquote {\bibinfo {title} {Influence of local-field corrections on thomson scattering in collision-dominated two-component plasmas},}\ }\href {\doibase 10.1103/PhysRevE.81.026405} {\bibfield  {journal} {\bibinfo  {journal} {Phys. Rev. E}\ }\textbf {\bibinfo {volume} {81}},\ \bibinfo {pages} {026405} (\bibinfo {year} {2010})}\BibitemShut {NoStop}%
\bibitem [{\citenamefont {Ullrich}(2011)}]{book_Ullrich}%
  \BibitemOpen
  \bibfield  {author} {\bibinfo {author} {\bibfnamefont {Carsten~A.}\ \bibnamefont {Ullrich}},\ }\href {\doibase 10.1093/acprof:oso/9780199563029.001.0001} {\emph {\bibinfo {title} {{Time-Dependent Density-Functional Theory: Concepts and Applications}}}}\ (\bibinfo  {publisher} {Oxford University Press},\ \bibinfo {year} {2011})\BibitemShut {NoStop}%
\bibitem [{\citenamefont {Moldabekov}\ \emph {et~al.}(2023{\natexlab{a}})\citenamefont {Moldabekov}, \citenamefont {Pavanello}, \citenamefont {B\"ohme}, \citenamefont {Vorberger},\ and\ \citenamefont {Dornheim}}]{Moldabekov_PRR_2023}%
  \BibitemOpen
  \bibfield  {author} {\bibinfo {author} {\bibfnamefont {Zhandos~A.}\ \bibnamefont {Moldabekov}}, \bibinfo {author} {\bibfnamefont {Michele}\ \bibnamefont {Pavanello}}, \bibinfo {author} {\bibfnamefont {Maximilian~P.}\ \bibnamefont {B\"ohme}}, \bibinfo {author} {\bibfnamefont {Jan}\ \bibnamefont {Vorberger}}, \ and\ \bibinfo {author} {\bibfnamefont {Tobias}\ \bibnamefont {Dornheim}},\ }\bibfield  {title} {\enquote {\bibinfo {title} {Linear-response time-dependent density functional theory approach to warm dense matter with adiabatic exchange-correlation kernels},}\ }\href {\doibase 10.1103/PhysRevResearch.5.023089} {\bibfield  {journal} {\bibinfo  {journal} {Phys. Rev. Res.}\ }\textbf {\bibinfo {volume} {5}},\ \bibinfo {pages} {023089} (\bibinfo {year} {2023}{\natexlab{a}})}\BibitemShut {NoStop}%
\bibitem [{\citenamefont {Moldabekov}\ \emph {et~al.}(2023{\natexlab{b}})\citenamefont {Moldabekov}, \citenamefont {B{\"o}hme}, \citenamefont {Vorberger}, \citenamefont {Blaschke},\ and\ \citenamefont {Dornheim}}]{Moldabekov_JCTC_2023}%
  \BibitemOpen
  \bibfield  {author} {\bibinfo {author} {\bibfnamefont {Zhandos}\ \bibnamefont {Moldabekov}}, \bibinfo {author} {\bibfnamefont {Maximilian}\ \bibnamefont {B{\"o}hme}}, \bibinfo {author} {\bibfnamefont {Jan}\ \bibnamefont {Vorberger}}, \bibinfo {author} {\bibfnamefont {David}\ \bibnamefont {Blaschke}}, \ and\ \bibinfo {author} {\bibfnamefont {Tobias}\ \bibnamefont {Dornheim}},\ }\bibfield  {title} {\enquote {\bibinfo {title} {Ab initio static exchange--correlation kernel across jacob's ladder without functional derivatives},}\ }\href {\doibase 10.1021/acs.jctc.2c01180} {\bibfield  {journal} {\bibinfo  {journal} {Journal of Chemical Theory and Computation}\ }\textbf {\bibinfo {volume} {19}},\ \bibinfo {pages} {1286--1299} (\bibinfo {year} {2023}{\natexlab{b}})}\BibitemShut {NoStop}%
\bibitem [{\citenamefont {Glenzer}\ and\ \citenamefont {Redmer}(2009)}]{siegfried_review}%
  \BibitemOpen
  \bibfield  {author} {\bibinfo {author} {\bibfnamefont {S.~H.}\ \bibnamefont {Glenzer}}\ and\ \bibinfo {author} {\bibfnamefont {R.}~\bibnamefont {Redmer}},\ }\bibfield  {title} {\enquote {\bibinfo {title} {X-ray thomson scattering in high energy density plasmas},}\ }\href {https://journals.aps.org/rmp/abstract/10.1103/RevModPhys.81.1625} {\bibfield  {journal} {\bibinfo  {journal} {Rev. Mod. Phys}\ }\textbf {\bibinfo {volume} {81}},\ \bibinfo {pages} {1625} (\bibinfo {year} {2009})}\BibitemShut {NoStop}%
\bibitem [{\citenamefont {D{\"o}ppner}\ \emph {et~al.}(2023)\citenamefont {D{\"o}ppner}, \citenamefont {Bethkenhagen}, \citenamefont {Kraus}, \citenamefont {Neumayer}, \citenamefont {Chapman}, \citenamefont {Bachmann}, \citenamefont {Baggott}, \citenamefont {B{\"o}hme}, \citenamefont {Divol}, \citenamefont {Falcone}, \citenamefont {Fletcher}, \citenamefont {Landen}, \citenamefont {MacDonald}, \citenamefont {Saunders}, \citenamefont {Sch{\"o}rner}, \citenamefont {Sterne}, \citenamefont {Vorberger}, \citenamefont {Witte}, \citenamefont {Yi}, \citenamefont {Redmer}, \citenamefont {Glenzer},\ and\ \citenamefont {Gericke}}]{Tilo_Nature_2023}%
  \BibitemOpen
  \bibfield  {author} {\bibinfo {author} {\bibfnamefont {T.}~\bibnamefont {D{\"o}ppner}}, \bibinfo {author} {\bibfnamefont {M.}~\bibnamefont {Bethkenhagen}}, \bibinfo {author} {\bibfnamefont {D.}~\bibnamefont {Kraus}}, \bibinfo {author} {\bibfnamefont {P.}~\bibnamefont {Neumayer}}, \bibinfo {author} {\bibfnamefont {D.~A.}\ \bibnamefont {Chapman}}, \bibinfo {author} {\bibfnamefont {B.}~\bibnamefont {Bachmann}}, \bibinfo {author} {\bibfnamefont {R.~A.}\ \bibnamefont {Baggott}}, \bibinfo {author} {\bibfnamefont {M.~P.}\ \bibnamefont {B{\"o}hme}}, \bibinfo {author} {\bibfnamefont {L.}~\bibnamefont {Divol}}, \bibinfo {author} {\bibfnamefont {R.~W.}\ \bibnamefont {Falcone}}, \bibinfo {author} {\bibfnamefont {L.~B.}\ \bibnamefont {Fletcher}}, \bibinfo {author} {\bibfnamefont {O.~L.}\ \bibnamefont {Landen}}, \bibinfo {author} {\bibfnamefont {M.~J.}\ \bibnamefont {MacDonald}}, \bibinfo {author} {\bibfnamefont {A.~M.}\ \bibnamefont {Saunders}}, \bibinfo {author} {\bibfnamefont {M.}~\bibnamefont {Sch{\"o}rner}},
  \bibinfo {author} {\bibfnamefont {P.~A.}\ \bibnamefont {Sterne}}, \bibinfo {author} {\bibfnamefont {J.}~\bibnamefont {Vorberger}}, \bibinfo {author} {\bibfnamefont {B.~B.~L.}\ \bibnamefont {Witte}}, \bibinfo {author} {\bibfnamefont {A.}~\bibnamefont {Yi}}, \bibinfo {author} {\bibfnamefont {R.}~\bibnamefont {Redmer}}, \bibinfo {author} {\bibfnamefont {S.~H.}\ \bibnamefont {Glenzer}}, \ and\ \bibinfo {author} {\bibfnamefont {D.~O.}\ \bibnamefont {Gericke}},\ }\bibfield  {title} {\enquote {\bibinfo {title} {Observing the onset of pressure-driven k-shell delocalization},}\ }\href {\doibase 10.1038/s41586-023-05996-8} {\bibfield  {journal} {\bibinfo  {journal} {Nature}\ }\textbf {\bibinfo {volume} {618}},\ \bibinfo {pages} {270–275} (\bibinfo {year} {2023})}\BibitemShut {NoStop}%
\bibitem [{\citenamefont {Dornheim}\ \emph {et~al.}(2022{\natexlab{c}})\citenamefont {Dornheim}, \citenamefont {B{\"o}hme}, \citenamefont {Kraus}, \citenamefont {D{\"o}ppner}, \citenamefont {Preston}, \citenamefont {Moldabekov},\ and\ \citenamefont {Vorberger}}]{Dornheim_T_2022}%
  \BibitemOpen
  \bibfield  {author} {\bibinfo {author} {\bibfnamefont {Tobias}\ \bibnamefont {Dornheim}}, \bibinfo {author} {\bibfnamefont {Maximilian}\ \bibnamefont {B{\"o}hme}}, \bibinfo {author} {\bibfnamefont {Dominik}\ \bibnamefont {Kraus}}, \bibinfo {author} {\bibfnamefont {Tilo}\ \bibnamefont {D{\"o}ppner}}, \bibinfo {author} {\bibfnamefont {Thomas~R.}\ \bibnamefont {Preston}}, \bibinfo {author} {\bibfnamefont {Zhandos~A.}\ \bibnamefont {Moldabekov}}, \ and\ \bibinfo {author} {\bibfnamefont {Jan}\ \bibnamefont {Vorberger}},\ }\bibfield  {title} {\enquote {\bibinfo {title} {Accurate temperature diagnostics for matter under extreme conditions},}\ }\href {\doibase 10.1038/s41467-022-35578-7} {\bibfield  {journal} {\bibinfo  {journal} {Nature Communications}\ }\textbf {\bibinfo {volume} {13}},\ \bibinfo {pages} {7911} (\bibinfo {year} {2022}{\natexlab{c}})}\BibitemShut {NoStop}%
\bibitem [{\citenamefont {Dornheim}\ \emph {et~al.}(2023{\natexlab{b}})\citenamefont {Dornheim}, \citenamefont {Böhme}, \citenamefont {Chapman}, \citenamefont {Kraus}, \citenamefont {Preston}, \citenamefont {Moldabekov}, \citenamefont {Schlünzen}, \citenamefont {Cangi}, \citenamefont {Döppner},\ and\ \citenamefont {Vorberger}}]{Dornheim_T2_2022}%
  \BibitemOpen
  \bibfield  {author} {\bibinfo {author} {\bibfnamefont {Tobias}\ \bibnamefont {Dornheim}}, \bibinfo {author} {\bibfnamefont {Maximilian~P.}\ \bibnamefont {Böhme}}, \bibinfo {author} {\bibfnamefont {David~A.}\ \bibnamefont {Chapman}}, \bibinfo {author} {\bibfnamefont {Dominik}\ \bibnamefont {Kraus}}, \bibinfo {author} {\bibfnamefont {Thomas~R.}\ \bibnamefont {Preston}}, \bibinfo {author} {\bibfnamefont {Zhandos~A.}\ \bibnamefont {Moldabekov}}, \bibinfo {author} {\bibfnamefont {Niclas}\ \bibnamefont {Schlünzen}}, \bibinfo {author} {\bibfnamefont {Attila}\ \bibnamefont {Cangi}}, \bibinfo {author} {\bibfnamefont {Tilo}\ \bibnamefont {Döppner}}, \ and\ \bibinfo {author} {\bibfnamefont {Jan}\ \bibnamefont {Vorberger}},\ }\bibfield  {title} {\enquote {\bibinfo {title} {{Imaginary-time correlation function thermometry: A new, high-accuracy and model-free temperature analysis technique for x-ray Thomson scattering data}},}\ }\href {\doibase 10.1063/5.0139560} {\bibfield  {journal} {\bibinfo  {journal} {Physics of
  Plasmas}\ }\textbf {\bibinfo {volume} {30}},\ \bibinfo {pages} {042707} (\bibinfo {year} {2023}{\natexlab{b}})}\BibitemShut {NoStop}%
\bibitem [{\citenamefont {Frydrych}\ \emph {et~al.}(2020)\citenamefont {Frydrych}, \citenamefont {Vorberger}, \citenamefont {Hartley}, \citenamefont {Schuster}, \citenamefont {Ramakrishna}, \citenamefont {Saunders}, \citenamefont {van Driel}, \citenamefont {Falcone}, \citenamefont {Fletcher}, \citenamefont {Galtier}, \citenamefont {Gamboa}, \citenamefont {Glenzer}, \citenamefont {Granados}, \citenamefont {MacDonald}, \citenamefont {MacKinnon}, \citenamefont {McBride}, \citenamefont {Nam}, \citenamefont {Neumayer}, \citenamefont {Pak}, \citenamefont {Voigt}, \citenamefont {Roth}, \citenamefont {Sun}, \citenamefont {Gericke}, \citenamefont {D{\"o}ppner},\ and\ \citenamefont {Kraus}}]{Frydrych2020}%
  \BibitemOpen
  \bibfield  {author} {\bibinfo {author} {\bibfnamefont {S.}~\bibnamefont {Frydrych}}, \bibinfo {author} {\bibfnamefont {J.}~\bibnamefont {Vorberger}}, \bibinfo {author} {\bibfnamefont {N.~J.}\ \bibnamefont {Hartley}}, \bibinfo {author} {\bibfnamefont {A.~K.}\ \bibnamefont {Schuster}}, \bibinfo {author} {\bibfnamefont {K.}~\bibnamefont {Ramakrishna}}, \bibinfo {author} {\bibfnamefont {A.~M.}\ \bibnamefont {Saunders}}, \bibinfo {author} {\bibfnamefont {T.}~\bibnamefont {van Driel}}, \bibinfo {author} {\bibfnamefont {R.~W.}\ \bibnamefont {Falcone}}, \bibinfo {author} {\bibfnamefont {L.~B.}\ \bibnamefont {Fletcher}}, \bibinfo {author} {\bibfnamefont {E.}~\bibnamefont {Galtier}}, \bibinfo {author} {\bibfnamefont {E.~J.}\ \bibnamefont {Gamboa}}, \bibinfo {author} {\bibfnamefont {S.~H.}\ \bibnamefont {Glenzer}}, \bibinfo {author} {\bibfnamefont {E.}~\bibnamefont {Granados}}, \bibinfo {author} {\bibfnamefont {M.~J.}\ \bibnamefont {MacDonald}}, \bibinfo {author} {\bibfnamefont {A.~J.}\ \bibnamefont {MacKinnon}},
  \bibinfo {author} {\bibfnamefont {E.~E.}\ \bibnamefont {McBride}}, \bibinfo {author} {\bibfnamefont {I.}~\bibnamefont {Nam}}, \bibinfo {author} {\bibfnamefont {P.}~\bibnamefont {Neumayer}}, \bibinfo {author} {\bibfnamefont {A.}~\bibnamefont {Pak}}, \bibinfo {author} {\bibfnamefont {K.}~\bibnamefont {Voigt}}, \bibinfo {author} {\bibfnamefont {M.}~\bibnamefont {Roth}}, \bibinfo {author} {\bibfnamefont {P.}~\bibnamefont {Sun}}, \bibinfo {author} {\bibfnamefont {D.~O.}\ \bibnamefont {Gericke}}, \bibinfo {author} {\bibfnamefont {T.}~\bibnamefont {D{\"o}ppner}}, \ and\ \bibinfo {author} {\bibfnamefont {D.}~\bibnamefont {Kraus}},\ }\bibfield  {title} {\enquote {\bibinfo {title} {Demonstration of x-ray thomson scattering as diagnostics for miscibility in warm dense matter},}\ }\href {\doibase 10.1038/s41467-020-16426-y} {\bibfield  {journal} {\bibinfo  {journal} {Nature Communications}\ }\textbf {\bibinfo {volume} {11}},\ \bibinfo {pages} {2620} (\bibinfo {year} {2020})}\BibitemShut {NoStop}%
\bibitem [{\citenamefont {Kraus}\ \emph {et~al.}(2019)\citenamefont {Kraus}, \citenamefont {Bachmann}, \citenamefont {Barbrel}, \citenamefont {Falcone}, \citenamefont {Fletcher}, \citenamefont {Frydrych}, \citenamefont {Gamboa}, \citenamefont {Gauthier}, \citenamefont {Gericke}, \citenamefont {Glenzer}, \citenamefont {G\"ode}, \citenamefont {Granados}, \citenamefont {Hartley}, \citenamefont {Helfrich}, \citenamefont {Lee}, \citenamefont {Nagler}, \citenamefont {Ravasio}, \citenamefont {Schumaker}, \citenamefont {Vorberger},\ and\ \citenamefont {D\"oppner}}]{kraus_xrts}%
  \BibitemOpen
  \bibfield  {author} {\bibinfo {author} {\bibfnamefont {D.}~\bibnamefont {Kraus}}, \bibinfo {author} {\bibfnamefont {B.}~\bibnamefont {Bachmann}}, \bibinfo {author} {\bibfnamefont {B.}~\bibnamefont {Barbrel}}, \bibinfo {author} {\bibfnamefont {R.~W.}\ \bibnamefont {Falcone}}, \bibinfo {author} {\bibfnamefont {L.~B.}\ \bibnamefont {Fletcher}}, \bibinfo {author} {\bibfnamefont {S.}~\bibnamefont {Frydrych}}, \bibinfo {author} {\bibfnamefont {E.~J.}\ \bibnamefont {Gamboa}}, \bibinfo {author} {\bibfnamefont {M.}~\bibnamefont {Gauthier}}, \bibinfo {author} {\bibfnamefont {D.~O.}\ \bibnamefont {Gericke}}, \bibinfo {author} {\bibfnamefont {S.~H.}\ \bibnamefont {Glenzer}}, \bibinfo {author} {\bibfnamefont {S.}~\bibnamefont {G\"ode}}, \bibinfo {author} {\bibfnamefont {E.}~\bibnamefont {Granados}}, \bibinfo {author} {\bibfnamefont {N.~J.}\ \bibnamefont {Hartley}}, \bibinfo {author} {\bibfnamefont {J.}~\bibnamefont {Helfrich}}, \bibinfo {author} {\bibfnamefont {H.~J.}\ \bibnamefont {Lee}}, \bibinfo {author}
  {\bibfnamefont {B.}~\bibnamefont {Nagler}}, \bibinfo {author} {\bibfnamefont {A.}~\bibnamefont {Ravasio}}, \bibinfo {author} {\bibfnamefont {W.}~\bibnamefont {Schumaker}}, \bibinfo {author} {\bibfnamefont {J.}~\bibnamefont {Vorberger}}, \ and\ \bibinfo {author} {\bibfnamefont {T.}~\bibnamefont {D\"oppner}},\ }\bibfield  {title} {\enquote {\bibinfo {title} {Characterizing the ionization potential depression in dense carbon plasmas with high-precision spectrally resolved x-ray scattering},}\ }\href {https://iopscience.iop.org/article/10.1088/1361-6587/aadd6c/meta} {\bibfield  {journal} {\bibinfo  {journal} {Plasma Phys. Control Fusion}\ }\textbf {\bibinfo {volume} {61}},\ \bibinfo {pages} {014015} (\bibinfo {year} {2019})}\BibitemShut {NoStop}%
\bibitem [{\citenamefont {Böhme}\ \emph {et~al.}(2023)\citenamefont {Böhme}, \citenamefont {Fletcher}, \citenamefont {Döppner}, \citenamefont {Kraus}, \citenamefont {Baczewski}, \citenamefont {Preston}, \citenamefont {MacDonald}, \citenamefont {Graziani}, \citenamefont {Moldabekov}, \citenamefont {Vorberger},\ and\ \citenamefont {Dornheim}}]{boehme2023evidence}%
  \BibitemOpen
  \bibfield  {author} {\bibinfo {author} {\bibfnamefont {Maximilian~P.}\ \bibnamefont {Böhme}}, \bibinfo {author} {\bibfnamefont {Luke~B.}\ \bibnamefont {Fletcher}}, \bibinfo {author} {\bibfnamefont {Tilo}\ \bibnamefont {Döppner}}, \bibinfo {author} {\bibfnamefont {Dominik}\ \bibnamefont {Kraus}}, \bibinfo {author} {\bibfnamefont {Andrew~D.}\ \bibnamefont {Baczewski}}, \bibinfo {author} {\bibfnamefont {Thomas~R.}\ \bibnamefont {Preston}}, \bibinfo {author} {\bibfnamefont {Michael~J.}\ \bibnamefont {MacDonald}}, \bibinfo {author} {\bibfnamefont {Frank~R.}\ \bibnamefont {Graziani}}, \bibinfo {author} {\bibfnamefont {Zhandos~A.}\ \bibnamefont {Moldabekov}}, \bibinfo {author} {\bibfnamefont {Jan}\ \bibnamefont {Vorberger}}, \ and\ \bibinfo {author} {\bibfnamefont {Tobias}\ \bibnamefont {Dornheim}},\ }\bibfield  {title} {\enquote {\bibinfo {title} {Evidence of free-bound transitions in warm dense matter and their impact on equation-of-state measurements},}\ }\href@noop {} {\  (\bibinfo {year} {2023})},\ \Eprint
  {http://arxiv.org/abs/2306.17653} {arXiv:2306.17653 [physics.plasm-ph]} \BibitemShut {NoStop}%
\bibitem [{\citenamefont {Gregori}\ \emph {et~al.}(2003)\citenamefont {Gregori}, \citenamefont {Glenzer}, \citenamefont {Rozmus}, \citenamefont {Lee},\ and\ \citenamefont {Landen}}]{Gregori_PRE_2003}%
  \BibitemOpen
  \bibfield  {author} {\bibinfo {author} {\bibfnamefont {G.}~\bibnamefont {Gregori}}, \bibinfo {author} {\bibfnamefont {S.~H.}\ \bibnamefont {Glenzer}}, \bibinfo {author} {\bibfnamefont {W.}~\bibnamefont {Rozmus}}, \bibinfo {author} {\bibfnamefont {R.~W.}\ \bibnamefont {Lee}}, \ and\ \bibinfo {author} {\bibfnamefont {O.~L.}\ \bibnamefont {Landen}},\ }\bibfield  {title} {\enquote {\bibinfo {title} {Theoretical model of x-ray scattering as a dense matter probe},}\ }\href {\doibase 10.1103/PhysRevE.67.026412} {\bibfield  {journal} {\bibinfo  {journal} {Phys. Rev. E}\ }\textbf {\bibinfo {volume} {67}},\ \bibinfo {pages} {026412} (\bibinfo {year} {2003})}\BibitemShut {NoStop}%
\bibitem [{\citenamefont {Pribram-Jones}\ \emph {et~al.}(2016)\citenamefont {Pribram-Jones}, \citenamefont {Grabowski},\ and\ \citenamefont {Burke}}]{pribram}%
  \BibitemOpen
  \bibfield  {author} {\bibinfo {author} {\bibfnamefont {A.}~\bibnamefont {Pribram-Jones}}, \bibinfo {author} {\bibfnamefont {P.~E.}\ \bibnamefont {Grabowski}}, \ and\ \bibinfo {author} {\bibfnamefont {K.}~\bibnamefont {Burke}},\ }\bibfield  {title} {\enquote {\bibinfo {title} {Thermal density functional theory: Time-dependent linear response and approximate functionals from the fluctuation-dissipation theorem},}\ }\href {https://journals.aps.org/prl/abstract/10.1103/PhysRevLett.116.233001} {\bibfield  {journal} {\bibinfo  {journal} {Phys. Rev. Lett}\ }\textbf {\bibinfo {volume} {116}},\ \bibinfo {pages} {233001} (\bibinfo {year} {2016})}\BibitemShut {NoStop}%
\bibitem [{\citenamefont {Gross}\ and\ \citenamefont {Kohn}(1985)}]{dynamic1}%
  \BibitemOpen
  \bibfield  {author} {\bibinfo {author} {\bibfnamefont {E.~K.~U.}\ \bibnamefont {Gross}}\ and\ \bibinfo {author} {\bibfnamefont {W.}~\bibnamefont {Kohn}},\ }\bibfield  {title} {\enquote {\bibinfo {title} {Local density-functional theory of frequency-dependent linear response},}\ }\href {https://journals.aps.org/prl/abstract/10.1103/PhysRevLett.55.2850} {\bibfield  {journal} {\bibinfo  {journal} {Phys. Rev. Lett}\ }\textbf {\bibinfo {volume} {55}},\ \bibinfo {pages} {2850} (\bibinfo {year} {1985})}\BibitemShut {NoStop}%
\bibitem [{\citenamefont {Constantin}\ and\ \citenamefont {Pitarke}(2007)}]{Constantin_PRB_2007}%
  \BibitemOpen
  \bibfield  {author} {\bibinfo {author} {\bibfnamefont {Lucian~A.}\ \bibnamefont {Constantin}}\ and\ \bibinfo {author} {\bibfnamefont {J.~M.}\ \bibnamefont {Pitarke}},\ }\bibfield  {title} {\enquote {\bibinfo {title} {Simple dynamic exchange-correlation kernel of a uniform electron gas},}\ }\href {\doibase 10.1103/PhysRevB.75.245127} {\bibfield  {journal} {\bibinfo  {journal} {Phys. Rev. B}\ }\textbf {\bibinfo {volume} {75}},\ \bibinfo {pages} {245127} (\bibinfo {year} {2007})}\BibitemShut {NoStop}%
\bibitem [{\citenamefont {Qian}\ and\ \citenamefont {Vignale}(2002)}]{PhysRevB.65.235121}%
  \BibitemOpen
  \bibfield  {author} {\bibinfo {author} {\bibfnamefont {Zhixin}\ \bibnamefont {Qian}}\ and\ \bibinfo {author} {\bibfnamefont {Giovanni}\ \bibnamefont {Vignale}},\ }\bibfield  {title} {\enquote {\bibinfo {title} {Dynamical exchange-correlation potentials for an electron liquid},}\ }\href {\doibase 10.1103/PhysRevB.65.235121} {\bibfield  {journal} {\bibinfo  {journal} {Phys. Rev. B}\ }\textbf {\bibinfo {volume} {65}},\ \bibinfo {pages} {235121} (\bibinfo {year} {2002})}\BibitemShut {NoStop}%
\bibitem [{\citenamefont {Dabrowski}(1986)}]{Dabrowski_PRB_1986}%
  \BibitemOpen
  \bibfield  {author} {\bibinfo {author} {\bibfnamefont {Bogdan}\ \bibnamefont {Dabrowski}},\ }\bibfield  {title} {\enquote {\bibinfo {title} {Dynamical local-field factor in the response function of an electron gas},}\ }\href {\doibase 10.1103/PhysRevB.34.4989} {\bibfield  {journal} {\bibinfo  {journal} {Phys. Rev. B}\ }\textbf {\bibinfo {volume} {34}},\ \bibinfo {pages} {4989--4995} (\bibinfo {year} {1986})}\BibitemShut {NoStop}%
\bibitem [{\citenamefont {Ruzsinszky}\ \emph {et~al.}(2020)\citenamefont {Ruzsinszky}, \citenamefont {Nepal}, \citenamefont {Pitarke},\ and\ \citenamefont {Perdew}}]{Adrienn_PRB_2020}%
  \BibitemOpen
  \bibfield  {author} {\bibinfo {author} {\bibfnamefont {Adrienn}\ \bibnamefont {Ruzsinszky}}, \bibinfo {author} {\bibfnamefont {Niraj~K.}\ \bibnamefont {Nepal}}, \bibinfo {author} {\bibfnamefont {J.~M.}\ \bibnamefont {Pitarke}}, \ and\ \bibinfo {author} {\bibfnamefont {John~P.}\ \bibnamefont {Perdew}},\ }\bibfield  {title} {\enquote {\bibinfo {title} {Constraint-based wave vector and frequency dependent exchange-correlation kernel of the uniform electron gas},}\ }\href {\doibase 10.1103/PhysRevB.101.245135} {\bibfield  {journal} {\bibinfo  {journal} {Phys. Rev. B}\ }\textbf {\bibinfo {volume} {101}},\ \bibinfo {pages} {245135} (\bibinfo {year} {2020})}\BibitemShut {NoStop}%
\bibitem [{\citenamefont {Holas}\ and\ \citenamefont {Rahman}(1987)}]{dynamic_ii}%
  \BibitemOpen
  \bibfield  {author} {\bibinfo {author} {\bibfnamefont {A.}~\bibnamefont {Holas}}\ and\ \bibinfo {author} {\bibfnamefont {S.}~\bibnamefont {Rahman}},\ }\bibfield  {title} {\enquote {\bibinfo {title} {Dynamic local-field factor of an electron liquid in the quantum versions of the {S}ingwi-{T}osi-{L}and-{S}j\"olander and {V}ashishta-{S}ingwi theories},}\ }\href {https://journals.aps.org/prb/abstract/10.1103/PhysRevB.35.2720} {\bibfield  {journal} {\bibinfo  {journal} {Phys. Rev. B}\ }\textbf {\bibinfo {volume} {35}},\ \bibinfo {pages} {2720} (\bibinfo {year} {1987})}\BibitemShut {NoStop}%
\bibitem [{\citenamefont {Dornheim}\ and\ \citenamefont {Vorberger}(2020)}]{Dornheim_PRE_2020}%
  \BibitemOpen
  \bibfield  {author} {\bibinfo {author} {\bibfnamefont {Tobias}\ \bibnamefont {Dornheim}}\ and\ \bibinfo {author} {\bibfnamefont {Jan}\ \bibnamefont {Vorberger}},\ }\bibfield  {title} {\enquote {\bibinfo {title} {{Finite-size effects in the reconstruction of dynamic properties from ab initio path integral Monte Carlo simulations}},}\ }\href {\doibase 10.1103/PhysRevE.102.063301} {\bibfield  {journal} {\bibinfo  {journal} {Phys. Rev. E}\ }\textbf {\bibinfo {volume} {102}},\ \bibinfo {pages} {063301} (\bibinfo {year} {2020})}\BibitemShut {NoStop}%
\bibitem [{\citenamefont {Hamann}\ \emph {et~al.}(2020{\natexlab{a}})\citenamefont {Hamann}, \citenamefont {Dornheim}, \citenamefont {Vorberger}, \citenamefont {Moldabekov},\ and\ \citenamefont {Bonitz}}]{Hamann_PRB_2020}%
  \BibitemOpen
  \bibfield  {author} {\bibinfo {author} {\bibfnamefont {Paul}\ \bibnamefont {Hamann}}, \bibinfo {author} {\bibfnamefont {Tobias}\ \bibnamefont {Dornheim}}, \bibinfo {author} {\bibfnamefont {Jan}\ \bibnamefont {Vorberger}}, \bibinfo {author} {\bibfnamefont {Zhandos~A.}\ \bibnamefont {Moldabekov}}, \ and\ \bibinfo {author} {\bibfnamefont {Michael}\ \bibnamefont {Bonitz}},\ }\bibfield  {title} {\enquote {\bibinfo {title} {Dynamic properties of the warm dense electron gas based on $ab initio$ path integral monte carlo simulations},}\ }\href {\doibase 10.1103/PhysRevB.102.125150} {\bibfield  {journal} {\bibinfo  {journal} {Phys. Rev. B}\ }\textbf {\bibinfo {volume} {102}},\ \bibinfo {pages} {125150} (\bibinfo {year} {2020}{\natexlab{a}})}\BibitemShut {NoStop}%
\bibitem [{\citenamefont {Jarrell}\ and\ \citenamefont {Gubernatis}(1996)}]{JARRELL1996133}%
  \BibitemOpen
  \bibfield  {author} {\bibinfo {author} {\bibfnamefont {Mark}\ \bibnamefont {Jarrell}}\ and\ \bibinfo {author} {\bibfnamefont {J.E.}\ \bibnamefont {Gubernatis}},\ }\bibfield  {title} {\enquote {\bibinfo {title} {{Bayesian inference and the analytic continuation of imaginary-time quantum Monte Carlo data}},}\ }\href {\doibase https://doi.org/10.1016/0370-1573(95)00074-7} {\bibfield  {journal} {\bibinfo  {journal} {Physics Reports}\ }\textbf {\bibinfo {volume} {269}},\ \bibinfo {pages} {133--195} (\bibinfo {year} {1996})}\BibitemShut {NoStop}%
\bibitem [{\citenamefont {LeBlanc}\ \emph {et~al.}(2022)\citenamefont {LeBlanc}, \citenamefont {Chen}, \citenamefont {Haule}, \citenamefont {Prokof'ev},\ and\ \citenamefont {Tupitsyn}}]{LeBlanc_PRL_2022}%
  \BibitemOpen
  \bibfield  {author} {\bibinfo {author} {\bibfnamefont {James P.~F.}\ \bibnamefont {LeBlanc}}, \bibinfo {author} {\bibfnamefont {Kun}\ \bibnamefont {Chen}}, \bibinfo {author} {\bibfnamefont {Kristjan}\ \bibnamefont {Haule}}, \bibinfo {author} {\bibfnamefont {Nikolay~V.}\ \bibnamefont {Prokof'ev}}, \ and\ \bibinfo {author} {\bibfnamefont {Igor~S.}\ \bibnamefont {Tupitsyn}},\ }\bibfield  {title} {\enquote {\bibinfo {title} {Dynamic response of an electron gas: Towards the exact exchange-correlation kernel},}\ }\href {\doibase 10.1103/PhysRevLett.129.246401} {\bibfield  {journal} {\bibinfo  {journal} {Phys. Rev. Lett.}\ }\textbf {\bibinfo {volume} {129}},\ \bibinfo {pages} {246401} (\bibinfo {year} {2022})}\BibitemShut {NoStop}%
\bibitem [{\citenamefont {Hamann}\ \emph {et~al.}(2020{\natexlab{b}})\citenamefont {Hamann}, \citenamefont {Vorberger}, \citenamefont {Dornheim}, \citenamefont {Moldabekov},\ and\ \citenamefont {Bonitz}}]{Hamann_CPP_2020}%
  \BibitemOpen
  \bibfield  {author} {\bibinfo {author} {\bibfnamefont {Paul}\ \bibnamefont {Hamann}}, \bibinfo {author} {\bibfnamefont {Jan}\ \bibnamefont {Vorberger}}, \bibinfo {author} {\bibfnamefont {Tobias}\ \bibnamefont {Dornheim}}, \bibinfo {author} {\bibfnamefont {Zhandos~A.}\ \bibnamefont {Moldabekov}}, \ and\ \bibinfo {author} {\bibfnamefont {Michael}\ \bibnamefont {Bonitz}},\ }\bibfield  {title} {\enquote {\bibinfo {title} {Ab initio results for the plasmon dispersion and damping of the warm dense electron gas},}\ }\href {\doibase https://doi.org/10.1002/ctpp.202000147} {\bibfield  {journal} {\bibinfo  {journal} {Contributions to Plasma Physics}\ }\textbf {\bibinfo {volume} {60}},\ \bibinfo {pages} {e202000147} (\bibinfo {year} {2020}{\natexlab{b}})}\BibitemShut {NoStop}%
\bibitem [{\citenamefont {Dornheim}\ \emph {et~al.}(2020{\natexlab{a}})\citenamefont {Dornheim}, \citenamefont {Cangi}, \citenamefont {Ramakrishna}, \citenamefont {B\"ohme}, \citenamefont {Tanaka},\ and\ \citenamefont {Vorberger}}]{Dornheim_PRL_2020_ESA}%
  \BibitemOpen
  \bibfield  {author} {\bibinfo {author} {\bibfnamefont {Tobias}\ \bibnamefont {Dornheim}}, \bibinfo {author} {\bibfnamefont {Attila}\ \bibnamefont {Cangi}}, \bibinfo {author} {\bibfnamefont {Kushal}\ \bibnamefont {Ramakrishna}}, \bibinfo {author} {\bibfnamefont {Maximilian}\ \bibnamefont {B\"ohme}}, \bibinfo {author} {\bibfnamefont {Shigenori}\ \bibnamefont {Tanaka}}, \ and\ \bibinfo {author} {\bibfnamefont {Jan}\ \bibnamefont {Vorberger}},\ }\bibfield  {title} {\enquote {\bibinfo {title} {Effective static approximation: A fast and reliable tool for warm-dense matter theory},}\ }\href {\doibase 10.1103/PhysRevLett.125.235001} {\bibfield  {journal} {\bibinfo  {journal} {Phys. Rev. Lett.}\ }\textbf {\bibinfo {volume} {125}},\ \bibinfo {pages} {235001} (\bibinfo {year} {2020}{\natexlab{a}})}\BibitemShut {NoStop}%
\bibitem [{\citenamefont {Dornheim}\ \emph {et~al.}(2019)\citenamefont {Dornheim}, \citenamefont {Vorberger}, \citenamefont {Groth}, \citenamefont {Hoffmann}, \citenamefont {Moldabekov},\ and\ \citenamefont {Bonitz}}]{dornheim_ML}%
  \BibitemOpen
  \bibfield  {author} {\bibinfo {author} {\bibfnamefont {T.}~\bibnamefont {Dornheim}}, \bibinfo {author} {\bibfnamefont {J.}~\bibnamefont {Vorberger}}, \bibinfo {author} {\bibfnamefont {S.}~\bibnamefont {Groth}}, \bibinfo {author} {\bibfnamefont {N.}~\bibnamefont {Hoffmann}}, \bibinfo {author} {\bibfnamefont {Zh.A.}\ \bibnamefont {Moldabekov}}, \ and\ \bibinfo {author} {\bibfnamefont {M.}~\bibnamefont {Bonitz}},\ }\bibfield  {title} {\enquote {\bibinfo {title} {The static local field correction of the warm dense electron gas: An ab initio path integral {M}onte {C}arlo study and machine learning representation},}\ }\href {https://aip.scitation.org/doi/full/10.1063/1.5123013} {\bibfield  {journal} {\bibinfo  {journal} {J. Chem. Phys}\ }\textbf {\bibinfo {volume} {151}},\ \bibinfo {pages} {194104} (\bibinfo {year} {2019})}\BibitemShut {NoStop}%
\bibitem [{\citenamefont {Brown}\ \emph {et~al.}(2013)\citenamefont {Brown}, \citenamefont {Clark}, \citenamefont {DuBois},\ and\ \citenamefont {Ceperley}}]{Brown_PRL_2013}%
  \BibitemOpen
  \bibfield  {author} {\bibinfo {author} {\bibfnamefont {Ethan~W.}\ \bibnamefont {Brown}}, \bibinfo {author} {\bibfnamefont {Bryan~K.}\ \bibnamefont {Clark}}, \bibinfo {author} {\bibfnamefont {Jonathan~L.}\ \bibnamefont {DuBois}}, \ and\ \bibinfo {author} {\bibfnamefont {David~M.}\ \bibnamefont {Ceperley}},\ }\bibfield  {title} {\enquote {\bibinfo {title} {{Path-Integral Monte Carlo Simulation of the Warm Dense Homogeneous Electron Gas}},}\ }\href {\doibase 10.1103/PhysRevLett.110.146405} {\bibfield  {journal} {\bibinfo  {journal} {Phys. Rev. Lett.}\ }\textbf {\bibinfo {volume} {110}},\ \bibinfo {pages} {146405} (\bibinfo {year} {2013})}\BibitemShut {NoStop}%
\bibitem [{\citenamefont {Ichimaru}\ \emph {et~al.}(1987)\citenamefont {Ichimaru}, \citenamefont {Iyetomi},\ and\ \citenamefont {Tanaka}}]{IIT}%
  \BibitemOpen
  \bibfield  {author} {\bibinfo {author} {\bibfnamefont {Setsuo}\ \bibnamefont {Ichimaru}}, \bibinfo {author} {\bibfnamefont {Hiroshi}\ \bibnamefont {Iyetomi}}, \ and\ \bibinfo {author} {\bibfnamefont {Shigenori}\ \bibnamefont {Tanaka}},\ }\bibfield  {title} {\enquote {\bibinfo {title} {Statistical physics of dense plasmas: Thermodynamics, transport coefficients and dynamic correlations},}\ }\href {\doibase https://doi.org/10.1016/0370-1573(87)90125-6} {\bibfield  {journal} {\bibinfo  {journal} {Physics Reports}\ }\textbf {\bibinfo {volume} {149}},\ \bibinfo {pages} {91--205} (\bibinfo {year} {1987})}\BibitemShut {NoStop}%
\bibitem [{\citenamefont {Tanaka}\ and\ \citenamefont {Ichimaru}(1986)}]{stls}%
  \BibitemOpen
  \bibfield  {author} {\bibinfo {author} {\bibfnamefont {S.}~\bibnamefont {Tanaka}}\ and\ \bibinfo {author} {\bibfnamefont {S.}~\bibnamefont {Ichimaru}},\ }\bibfield  {title} {\enquote {\bibinfo {title} {Thermodynamics and correlational properties of finite-temperature electron liquids in the {S}ingwi-{T}osi-{Land}-{S}j\"olander approximation},}\ }\href {http://journals.jps.jp/doi/abs/10.1143/JPSJ.55.2278} {\bibfield  {journal} {\bibinfo  {journal} {J. Phys. Soc. Jpn}\ }\textbf {\bibinfo {volume} {55}},\ \bibinfo {pages} {2278--2289} (\bibinfo {year} {1986})}\BibitemShut {NoStop}%
\bibitem [{\citenamefont {Singwi}\ \emph {et~al.}(1968)\citenamefont {Singwi}, \citenamefont {Tosi}, \citenamefont {Land},\ and\ \citenamefont {Sj\"olander}}]{stls_original}%
  \BibitemOpen
  \bibfield  {author} {\bibinfo {author} {\bibfnamefont {K.~S.}\ \bibnamefont {Singwi}}, \bibinfo {author} {\bibfnamefont {M.~P.}\ \bibnamefont {Tosi}}, \bibinfo {author} {\bibfnamefont {R.~H.}\ \bibnamefont {Land}}, \ and\ \bibinfo {author} {\bibfnamefont {A.}~\bibnamefont {Sj\"olander}},\ }\bibfield  {title} {\enquote {\bibinfo {title} {Electron correlations at metallic densities},}\ }\href {http://link.aps.org/doi/10.1103/PhysRev.176.589} {\bibfield  {journal} {\bibinfo  {journal} {Phys. Rev}\ }\textbf {\bibinfo {volume} {176}},\ \bibinfo {pages} {589} (\bibinfo {year} {1968})}\BibitemShut {NoStop}%
\bibitem [{\citenamefont {Sjostrom}\ and\ \citenamefont {Dufty}(2013)}]{stls2}%
  \BibitemOpen
  \bibfield  {author} {\bibinfo {author} {\bibfnamefont {T.}~\bibnamefont {Sjostrom}}\ and\ \bibinfo {author} {\bibfnamefont {J.}~\bibnamefont {Dufty}},\ }\bibfield  {title} {\enquote {\bibinfo {title} {Uniform electron gas at finite temperatures},}\ }\href {http://link.aps.org/doi/10.1103/PhysRevB.88.115123} {\bibfield  {journal} {\bibinfo  {journal} {Phys. Rev. B}\ }\textbf {\bibinfo {volume} {88}},\ \bibinfo {pages} {115123} (\bibinfo {year} {2013})}\BibitemShut {NoStop}%
\bibitem [{\citenamefont {Dornheim}\ \emph {et~al.}(2020{\natexlab{b}})\citenamefont {Dornheim}, \citenamefont {Sjostrom}, \citenamefont {Tanaka},\ and\ \citenamefont {Vorberger}}]{dornheim_electron_liquid}%
  \BibitemOpen
  \bibfield  {author} {\bibinfo {author} {\bibfnamefont {Tobias}\ \bibnamefont {Dornheim}}, \bibinfo {author} {\bibfnamefont {Travis}\ \bibnamefont {Sjostrom}}, \bibinfo {author} {\bibfnamefont {Shigenori}\ \bibnamefont {Tanaka}}, \ and\ \bibinfo {author} {\bibfnamefont {Jan}\ \bibnamefont {Vorberger}},\ }\bibfield  {title} {\enquote {\bibinfo {title} {{Strongly coupled electron liquid: Ab initio path integral Monte Carlo simulations and dielectric theories}},}\ }\href {\doibase 10.1103/PhysRevB.101.045129} {\bibfield  {journal} {\bibinfo  {journal} {Phys. Rev. B}\ }\textbf {\bibinfo {volume} {101}},\ \bibinfo {pages} {045129} (\bibinfo {year} {2020}{\natexlab{b}})}\BibitemShut {NoStop}%
\bibitem [{\citenamefont {Tolias}\ \emph {et~al.}(2024{\natexlab{a}})\citenamefont {Tolias}, \citenamefont {Castello}, \citenamefont {Kalkavouras},\ and\ \citenamefont {Dornheim}}]{tolias2024_VS}%
  \BibitemOpen
  \bibfield  {author} {\bibinfo {author} {\bibfnamefont {Panagiotis}\ \bibnamefont {Tolias}}, \bibinfo {author} {\bibfnamefont {Federico~Lucco}\ \bibnamefont {Castello}}, \bibinfo {author} {\bibfnamefont {Fotios}\ \bibnamefont {Kalkavouras}}, \ and\ \bibinfo {author} {\bibfnamefont {Tobias}\ \bibnamefont {Dornheim}},\ }\bibfield  {title} {\enquote {\bibinfo {title} {Revisiting the {Vashishta-Singwi} dielectric scheme for the warm dense uniform electron fluid},}\ }\href {\doibase 10.1103/PhysRevB.109.125134} {\bibfield  {journal} {\bibinfo  {journal} {Phys. Rev. B}\ }\textbf {\bibinfo {volume} {109}},\ \bibinfo {pages} {125134} (\bibinfo {year} {2024}{\natexlab{a}})}\BibitemShut {NoStop}%
\bibitem [{\citenamefont {Tolias}\ \emph {et~al.}(2021)\citenamefont {Tolias}, \citenamefont {Lucco~Castello},\ and\ \citenamefont {Dornheim}}]{Tolias_JCP_2021}%
  \BibitemOpen
  \bibfield  {author} {\bibinfo {author} {\bibfnamefont {P.}~\bibnamefont {Tolias}}, \bibinfo {author} {\bibfnamefont {F.}~\bibnamefont {Lucco~Castello}}, \ and\ \bibinfo {author} {\bibfnamefont {T.}~\bibnamefont {Dornheim}},\ }\bibfield  {title} {\enquote {\bibinfo {title} {Integral equation theory based dielectric scheme for strongly coupled electron liquids},}\ }\href {\doibase 10.1063/5.0065988} {\bibfield  {journal} {\bibinfo  {journal} {The Journal of Chemical Physics}\ }\textbf {\bibinfo {volume} {155}},\ \bibinfo {pages} {134115} (\bibinfo {year} {2021})}\BibitemShut {NoStop}%
\bibitem [{\citenamefont {Tolias}\ \emph {et~al.}(2023)\citenamefont {Tolias}, \citenamefont {Lucco~Castello},\ and\ \citenamefont {Dornheim}}]{Tolias_JCP_2023}%
  \BibitemOpen
  \bibfield  {author} {\bibinfo {author} {\bibfnamefont {Panagiotis}\ \bibnamefont {Tolias}}, \bibinfo {author} {\bibfnamefont {Federico}\ \bibnamefont {Lucco~Castello}}, \ and\ \bibinfo {author} {\bibfnamefont {Tobias}\ \bibnamefont {Dornheim}},\ }\bibfield  {title} {\enquote {\bibinfo {title} {{Quantum version of the integral equation theory-based dielectric scheme for strongly coupled electron liquids}},}\ }\href {\doibase 10.1063/5.0145687} {\bibfield  {journal} {\bibinfo  {journal} {The Journal of Chemical Physics}\ }\textbf {\bibinfo {volume} {158}},\ \bibinfo {pages} {141102} (\bibinfo {year} {2023})}\BibitemShut {NoStop}%
\bibitem [{\citenamefont {Tanaka}(2016)}]{tanaka_hnc}%
  \BibitemOpen
  \bibfield  {author} {\bibinfo {author} {\bibfnamefont {S.}~\bibnamefont {Tanaka}},\ }\bibfield  {title} {\enquote {\bibinfo {title} {Correlational and thermodynamic properties of finite-temperature electron liquids in the hypernetted-chain approximation},}\ }\href {https://aip.scitation.org/doi/abs/10.1063/1.4969071} {\bibfield  {journal} {\bibinfo  {journal} {J. Chem. Phys}\ }\textbf {\bibinfo {volume} {145}},\ \bibinfo {pages} {214104} (\bibinfo {year} {2016})}\BibitemShut {NoStop}%
\bibitem [{\citenamefont {Tanaka}(2017)}]{Tanaka_CPP_2017}%
  \BibitemOpen
  \bibfield  {author} {\bibinfo {author} {\bibfnamefont {Shigenori}\ \bibnamefont {Tanaka}},\ }\bibfield  {title} {\enquote {\bibinfo {title} {Improved equation of state for finite-temperature spin-polarized electron liquids on the basis of singwi–tosi–land–sjölander approximation},}\ }\href {\doibase https://doi.org/10.1002/ctpp.201600096} {\bibfield  {journal} {\bibinfo  {journal} {Contributions to Plasma Physics}\ }\textbf {\bibinfo {volume} {57}},\ \bibinfo {pages} {126--136} (\bibinfo {year} {2017})}\BibitemShut {NoStop}%
\bibitem [{\citenamefont {Arora}\ \emph {et~al.}(2017)\citenamefont {Arora}, \citenamefont {Kumar},\ and\ \citenamefont {Moudgil}}]{arora}%
  \BibitemOpen
  \bibfield  {author} {\bibinfo {author} {\bibfnamefont {P.}~\bibnamefont {Arora}}, \bibinfo {author} {\bibfnamefont {K.}~\bibnamefont {Kumar}}, \ and\ \bibinfo {author} {\bibfnamefont {R.~K.}\ \bibnamefont {Moudgil}},\ }\bibfield  {title} {\enquote {\bibinfo {title} {Spin-resolved correlations in the warm-dense homogeneous electron gas},}\ }\href {https://link.springer.com/article/10.1140/epjb/e2017-70532-y} {\bibfield  {journal} {\bibinfo  {journal} {Eur. Phys. J. B}\ }\textbf {\bibinfo {volume} {90}},\ \bibinfo {pages} {76} (\bibinfo {year} {2017})}\BibitemShut {NoStop}%
\bibitem [{\citenamefont {Tolias}\ \emph {et~al.}(2024{\natexlab{b}})\citenamefont {Tolias}, \citenamefont {Kalkavouras},\ and\ \citenamefont {Dornheim}}]{tolias2024fouriermatsubara}%
  \BibitemOpen
  \bibfield  {author} {\bibinfo {author} {\bibfnamefont {Panagiotis}\ \bibnamefont {Tolias}}, \bibinfo {author} {\bibfnamefont {Fotios}\ \bibnamefont {Kalkavouras}}, \ and\ \bibinfo {author} {\bibfnamefont {Tobias}\ \bibnamefont {Dornheim}},\ }\bibfield  {title} {\enquote {\bibinfo {title} {{Fourier-Matsubara} series expansion for imaginary-time correlation functions},}\ }\href {\doibase 10.1063/5.0211814} {\bibfield  {journal} {\bibinfo  {journal} {J. Chem. Phys.}\ }\textbf {\bibinfo {volume} {160}},\ \bibinfo {pages} {181102} (\bibinfo {year} {2024}{\natexlab{b}})}\BibitemShut {NoStop}%
\bibitem [{\citenamefont {Kleinert}(2009)}]{kleinert2009path}%
  \BibitemOpen
  \bibfield  {author} {\bibinfo {author} {\bibfnamefont {H.}~\bibnamefont {Kleinert}},\ }\href {https://books.google.de/books?id=VJ1qNz5xYzkC} {\emph {\bibinfo {title} {Path Integrals in Quantum Mechanics, Statistics, Polymer Physics, and Financial Markets}}},\ EBL-Schweitzer\ (\bibinfo  {publisher} {World Scientific},\ \bibinfo {year} {2009})\BibitemShut {NoStop}%
\bibitem [{\citenamefont {Boninsegni}\ \emph {et~al.}(2006{\natexlab{a}})\citenamefont {Boninsegni}, \citenamefont {Prokofev},\ and\ \citenamefont {Svistunov}}]{boninsegni1}%
  \BibitemOpen
  \bibfield  {author} {\bibinfo {author} {\bibfnamefont {M.}~\bibnamefont {Boninsegni}}, \bibinfo {author} {\bibfnamefont {N.~V.}\ \bibnamefont {Prokofev}}, \ and\ \bibinfo {author} {\bibfnamefont {B.~V.}\ \bibnamefont {Svistunov}},\ }\bibfield  {title} {\enquote {\bibinfo {title} {Worm algorithm and diagrammatic {M}onte {C}arlo: A new approach to continuous-space path integral {M}onte {C}arlo simulations},}\ }\href {https://journals.aps.org/pre/abstract/10.1103/PhysRevE.74.036701} {\bibfield  {journal} {\bibinfo  {journal} {Phys. Rev. E}\ }\textbf {\bibinfo {volume} {74}},\ \bibinfo {pages} {036701} (\bibinfo {year} {2006}{\natexlab{a}})}\BibitemShut {NoStop}%
\bibitem [{\citenamefont {Boninsegni}\ \emph {et~al.}(2006{\natexlab{b}})\citenamefont {Boninsegni}, \citenamefont {Prokofev},\ and\ \citenamefont {Svistunov}}]{boninsegni2}%
  \BibitemOpen
  \bibfield  {author} {\bibinfo {author} {\bibfnamefont {M.}~\bibnamefont {Boninsegni}}, \bibinfo {author} {\bibfnamefont {N.~V.}\ \bibnamefont {Prokofev}}, \ and\ \bibinfo {author} {\bibfnamefont {B.~V.}\ \bibnamefont {Svistunov}},\ }\bibfield  {title} {\enquote {\bibinfo {title} {Worm algorithm for continuous-space path integral {M}onte {C}arlo simulations},}\ }\href {https://journals.aps.org/prl/abstract/10.1103/PhysRevLett.96.070601} {\bibfield  {journal} {\bibinfo  {journal} {Phys. Rev. Lett}\ }\textbf {\bibinfo {volume} {96}},\ \bibinfo {pages} {070601} (\bibinfo {year} {2006}{\natexlab{b}})}\BibitemShut {NoStop}%
\bibitem [{\citenamefont {Dornheim}\ \emph {et~al.}(2021{\natexlab{a}})\citenamefont {Dornheim}, \citenamefont {B\"ohme}, \citenamefont {Militzer},\ and\ \citenamefont {Vorberger}}]{Dornheim_PRB_nk_2021}%
  \BibitemOpen
  \bibfield  {author} {\bibinfo {author} {\bibfnamefont {Tobias}\ \bibnamefont {Dornheim}}, \bibinfo {author} {\bibfnamefont {Maximilian}\ \bibnamefont {B\"ohme}}, \bibinfo {author} {\bibfnamefont {Burkhard}\ \bibnamefont {Militzer}}, \ and\ \bibinfo {author} {\bibfnamefont {Jan}\ \bibnamefont {Vorberger}},\ }\bibfield  {title} {\enquote {\bibinfo {title} {Ab initio path integral monte carlo approach to the momentum distribution of the uniform electron gas at finite temperature without fixed nodes},}\ }\href {\doibase 10.1103/PhysRevB.103.205142} {\bibfield  {journal} {\bibinfo  {journal} {Phys. Rev. B}\ }\textbf {\bibinfo {volume} {103}},\ \bibinfo {pages} {205142} (\bibinfo {year} {2021}{\natexlab{a}})}\BibitemShut {NoStop}%
\bibitem [{\citenamefont {Dornheim}\ \emph {et~al.}(2021{\natexlab{b}})\citenamefont {Dornheim}, \citenamefont {Moldabekov},\ and\ \citenamefont {Vorberger}}]{Dornheim_JCP_ITCF_2021}%
  \BibitemOpen
  \bibfield  {author} {\bibinfo {author} {\bibfnamefont {Tobias}\ \bibnamefont {Dornheim}}, \bibinfo {author} {\bibfnamefont {Zhandos~A.}\ \bibnamefont {Moldabekov}}, \ and\ \bibinfo {author} {\bibfnamefont {Jan}\ \bibnamefont {Vorberger}},\ }\bibfield  {title} {\enquote {\bibinfo {title} {{Nonlinear density response from imaginary-time correlation functions: Ab initio path integral Monte Carlo simulations of the warm dense electron gas}},}\ }\href {\doibase 10.1063/5.0058988} {\bibfield  {journal} {\bibinfo  {journal} {The Journal of Chemical Physics}\ }\textbf {\bibinfo {volume} {155}},\ \bibinfo {pages} {054110} (\bibinfo {year} {2021}{\natexlab{b}})}\BibitemShut {NoStop}%
\bibitem [{\citenamefont {Militzer}\ \emph {et~al.}(2019)\citenamefont {Militzer}, \citenamefont {Pollock},\ and\ \citenamefont {Ceperley}}]{MILITZER201913}%
  \BibitemOpen
  \bibfield  {author} {\bibinfo {author} {\bibfnamefont {B.}~\bibnamefont {Militzer}}, \bibinfo {author} {\bibfnamefont {E.L.}\ \bibnamefont {Pollock}}, \ and\ \bibinfo {author} {\bibfnamefont {D.M.}\ \bibnamefont {Ceperley}},\ }\bibfield  {title} {\enquote {\bibinfo {title} {Path integral monte carlo calculation of the momentum distribution of the homogeneous electron gas at finite temperature},}\ }\href {\doibase https://doi.org/10.1016/j.hedp.2018.12.004} {\bibfield  {journal} {\bibinfo  {journal} {High Energy Density Physics}\ }\textbf {\bibinfo {volume} {30}},\ \bibinfo {pages} {13--20} (\bibinfo {year} {2019})}\BibitemShut {NoStop}%
\bibitem [{\citenamefont {Dornheim}\ \emph {et~al.}(2021{\natexlab{c}})\citenamefont {Dornheim}, \citenamefont {Vorberger}, \citenamefont {Militzer},\ and\ \citenamefont {Moldabekov}}]{Dornheim_PRE_2021}%
  \BibitemOpen
  \bibfield  {author} {\bibinfo {author} {\bibfnamefont {Tobias}\ \bibnamefont {Dornheim}}, \bibinfo {author} {\bibfnamefont {Jan}\ \bibnamefont {Vorberger}}, \bibinfo {author} {\bibfnamefont {Burkhard}\ \bibnamefont {Militzer}}, \ and\ \bibinfo {author} {\bibfnamefont {Zhandos~A.}\ \bibnamefont {Moldabekov}},\ }\bibfield  {title} {\enquote {\bibinfo {title} {Momentum distribution of the uniform electron gas at finite temperature: Effects of spin polarization},}\ }\href {\doibase 10.1103/PhysRevE.104.055206} {\bibfield  {journal} {\bibinfo  {journal} {Phys. Rev. E}\ }\textbf {\bibinfo {volume} {104}},\ \bibinfo {pages} {055206} (\bibinfo {year} {2021}{\natexlab{c}})}\BibitemShut {NoStop}%
\bibitem [{\citenamefont {Dornheim}\ \emph {et~al.}(2023{\natexlab{c}})\citenamefont {Dornheim}, \citenamefont {Moldabekov}, \citenamefont {Tolias}, \citenamefont {Böhme},\ and\ \citenamefont {Vorberger}}]{Dornheim_insight_2022}%
  \BibitemOpen
  \bibfield  {author} {\bibinfo {author} {\bibfnamefont {Tobias}\ \bibnamefont {Dornheim}}, \bibinfo {author} {\bibfnamefont {Zhandos}\ \bibnamefont {Moldabekov}}, \bibinfo {author} {\bibfnamefont {Panagiotis}\ \bibnamefont {Tolias}}, \bibinfo {author} {\bibfnamefont {Maximilian}\ \bibnamefont {Böhme}}, \ and\ \bibinfo {author} {\bibfnamefont {Jan}\ \bibnamefont {Vorberger}},\ }\bibfield  {title} {\enquote {\bibinfo {title} {Physical insights from imaginary-time density--density correlation functions},}\ }\href {\doibase 10.1063/5.0149638} {\bibfield  {journal} {\bibinfo  {journal} {Matter and Radiation at Extremes}\ }\textbf {\bibinfo {volume} {8}},\ \bibinfo {pages} {056601} (\bibinfo {year} {2023}{\natexlab{c}})}\BibitemShut {NoStop}%
\bibitem [{\citenamefont {Dornheim}\ \emph {et~al.}(2020{\natexlab{c}})\citenamefont {Dornheim}, \citenamefont {Moldabekov}, \citenamefont {Vorberger},\ and\ \citenamefont {Groth}}]{dornheim_HEDP}%
  \BibitemOpen
  \bibfield  {author} {\bibinfo {author} {\bibfnamefont {Tobias}\ \bibnamefont {Dornheim}}, \bibinfo {author} {\bibfnamefont {Zhandos~A}\ \bibnamefont {Moldabekov}}, \bibinfo {author} {\bibfnamefont {Jan}\ \bibnamefont {Vorberger}}, \ and\ \bibinfo {author} {\bibfnamefont {Simon}\ \bibnamefont {Groth}},\ }\bibfield  {title} {\enquote {\bibinfo {title} {{Ab initio path integral monte carlo simulation of the uniform electron gas in the high energy density regime}},}\ }\href {\doibase 10.1088/1361-6587/ab8bb4} {\bibfield  {journal} {\bibinfo  {journal} {Plasma Physics and Controlled Fusion}\ }\textbf {\bibinfo {volume} {62}},\ \bibinfo {pages} {075003} (\bibinfo {year} {2020}{\natexlab{c}})}\BibitemShut {NoStop}%
\bibitem [{\citenamefont {Dornheim}\ \emph {et~al.}(2022{\natexlab{d}})\citenamefont {Dornheim}, \citenamefont {Vorberger}, \citenamefont {Moldabekov}, \citenamefont {Röpke},\ and\ \citenamefont {Kraeft}}]{Dornheim_HEDP_2022}%
  \BibitemOpen
  \bibfield  {author} {\bibinfo {author} {\bibfnamefont {Tobias}\ \bibnamefont {Dornheim}}, \bibinfo {author} {\bibfnamefont {Jan}\ \bibnamefont {Vorberger}}, \bibinfo {author} {\bibfnamefont {Zhandos}\ \bibnamefont {Moldabekov}}, \bibinfo {author} {\bibfnamefont {Gerd}\ \bibnamefont {Röpke}}, \ and\ \bibinfo {author} {\bibfnamefont {Wolf-Dietrich}\ \bibnamefont {Kraeft}},\ }\bibfield  {title} {\enquote {\bibinfo {title} {{The uniform electron gas at high temperatures: ab initio path integral Monte Carlo simulations and analytical theory}},}\ }\href {\doibase https://doi.org/10.1016/j.hedp.2022.101015} {\bibfield  {journal} {\bibinfo  {journal} {High Energy Density Physics}\ }\textbf {\bibinfo {volume} {45}},\ \bibinfo {pages} {101015} (\bibinfo {year} {2022}{\natexlab{d}})}\BibitemShut {NoStop}%
\bibitem [{\citenamefont {Dornheim}\ \emph {et~al.}(2024{\natexlab{a}})\citenamefont {Dornheim}, \citenamefont {Schwalbe}, \citenamefont {Tolias}, \citenamefont {Böhme}, \citenamefont {Moldabekov},\ and\ \citenamefont {Vorberger}}]{dornheim2024ab}%
  \BibitemOpen
  \bibfield  {author} {\bibinfo {author} {\bibfnamefont {Tobias}\ \bibnamefont {Dornheim}}, \bibinfo {author} {\bibfnamefont {Sebastian}\ \bibnamefont {Schwalbe}}, \bibinfo {author} {\bibfnamefont {Panagiotis}\ \bibnamefont {Tolias}}, \bibinfo {author} {\bibfnamefont {Maximilan}\ \bibnamefont {Böhme}}, \bibinfo {author} {\bibfnamefont {Zhandos}\ \bibnamefont {Moldabekov}}, \ and\ \bibinfo {author} {\bibfnamefont {Jan}\ \bibnamefont {Vorberger}},\ }\href@noop {} {\enquote {\bibinfo {title} {Ab initio density response and local field factor of warm dense hydrogen},}\ } (\bibinfo {year} {2024}{\natexlab{a}}),\ \Eprint {http://arxiv.org/abs/2403.08570} {arXiv:2403.08570 [physics.plasm-ph]} \BibitemShut {NoStop}%
\bibitem [{\citenamefont {Dornheim}\ \emph {et~al.}(2022{\natexlab{e}})\citenamefont {Dornheim}, \citenamefont {Vorberger}, \citenamefont {Moldabekov},\ and\ \citenamefont {Bonitz}}]{Dornheim_CPP_2022}%
  \BibitemOpen
  \bibfield  {author} {\bibinfo {author} {\bibfnamefont {Tobias}\ \bibnamefont {Dornheim}}, \bibinfo {author} {\bibfnamefont {Jan}\ \bibnamefont {Vorberger}}, \bibinfo {author} {\bibfnamefont {Zhandos~A.}\ \bibnamefont {Moldabekov}}, \ and\ \bibinfo {author} {\bibfnamefont {Michael}\ \bibnamefont {Bonitz}},\ }\bibfield  {title} {\enquote {\bibinfo {title} {Nonlinear interaction of external perturbations in warm dense matter},}\ }\href {\doibase https://doi.org/10.1002/ctpp.202100247} {\bibfield  {journal} {\bibinfo  {journal} {Contributions to Plasma Physics}\ }\textbf {\bibinfo {volume} {n/a}},\ \bibinfo {pages} {e202100247} (\bibinfo {year} {2022}{\natexlab{e}})}\BibitemShut {NoStop}%
\bibitem [{\citenamefont {Dornheim}\ \emph {et~al.}(2023{\natexlab{d}})\citenamefont {Dornheim}, \citenamefont {Döppner}, \citenamefont {Baczewski}, \citenamefont {Tolias}, \citenamefont {Böhme}, \citenamefont {Moldabekov}, \citenamefont {Ranjan}, \citenamefont {Chapman}, \citenamefont {MacDonald}, \citenamefont {Preston}, \citenamefont {Kraus},\ and\ \citenamefont {Vorberger}}]{dornheim2023xray}%
  \BibitemOpen
  \bibfield  {author} {\bibinfo {author} {\bibfnamefont {Tobias}\ \bibnamefont {Dornheim}}, \bibinfo {author} {\bibfnamefont {Tilo}\ \bibnamefont {Döppner}}, \bibinfo {author} {\bibfnamefont {Andrew~D.}\ \bibnamefont {Baczewski}}, \bibinfo {author} {\bibfnamefont {Panagiotis}\ \bibnamefont {Tolias}}, \bibinfo {author} {\bibfnamefont {Maximilian~P.}\ \bibnamefont {Böhme}}, \bibinfo {author} {\bibfnamefont {Zhandos~A.}\ \bibnamefont {Moldabekov}}, \bibinfo {author} {\bibfnamefont {Divyanshu}\ \bibnamefont {Ranjan}}, \bibinfo {author} {\bibfnamefont {David~A.}\ \bibnamefont {Chapman}}, \bibinfo {author} {\bibfnamefont {Michael~J.}\ \bibnamefont {MacDonald}}, \bibinfo {author} {\bibfnamefont {Thomas~R.}\ \bibnamefont {Preston}}, \bibinfo {author} {\bibfnamefont {Dominik}\ \bibnamefont {Kraus}}, \ and\ \bibinfo {author} {\bibfnamefont {Jan}\ \bibnamefont {Vorberger}},\ }\bibfield  {title} {\enquote {\bibinfo {title} {{X-ray Thomson scattering absolute intensity from the f-sum rule in the imaginary-time
  domain}},}\ }\href@noop {} {\bibfield  {journal} {\bibinfo  {journal} {arXiv}\ } (\bibinfo {year} {2023}{\natexlab{d}})},\ \Eprint {http://arxiv.org/abs/2305.15305} {2305.15305 [physics.plasm-ph]} \BibitemShut {NoStop}%
\bibitem [{\citenamefont {Sch\"orner}\ \emph {et~al.}(2023)\citenamefont {Sch\"orner}, \citenamefont {Bethkenhagen}, \citenamefont {D\"oppner}, \citenamefont {Kraus}, \citenamefont {Fletcher}, \citenamefont {Glenzer},\ and\ \citenamefont {Redmer}}]{Schoerner_PRE_2023}%
  \BibitemOpen
  \bibfield  {author} {\bibinfo {author} {\bibfnamefont {Maximilian}\ \bibnamefont {Sch\"orner}}, \bibinfo {author} {\bibfnamefont {Mandy}\ \bibnamefont {Bethkenhagen}}, \bibinfo {author} {\bibfnamefont {Tilo}\ \bibnamefont {D\"oppner}}, \bibinfo {author} {\bibfnamefont {Dominik}\ \bibnamefont {Kraus}}, \bibinfo {author} {\bibfnamefont {Luke~B.}\ \bibnamefont {Fletcher}}, \bibinfo {author} {\bibfnamefont {Siegfried~H.}\ \bibnamefont {Glenzer}}, \ and\ \bibinfo {author} {\bibfnamefont {Ronald}\ \bibnamefont {Redmer}},\ }\bibfield  {title} {\enquote {\bibinfo {title} {X-ray thomson scattering spectra from density functional theory molecular dynamics simulations based on a modified chihara formula},}\ }\href {\doibase 10.1103/PhysRevE.107.065207} {\bibfield  {journal} {\bibinfo  {journal} {Phys. Rev. E}\ }\textbf {\bibinfo {volume} {107}},\ \bibinfo {pages} {065207} (\bibinfo {year} {2023})}\BibitemShut {NoStop}%
\bibitem [{\citenamefont {Dornheim}\ \emph {et~al.}(2024{\natexlab{b}})\citenamefont {Dornheim}, \citenamefont {Döppner}, \citenamefont {Tolias}, \citenamefont {Böhme}, \citenamefont {Fletcher}, \citenamefont {Gawne}, \citenamefont {Graziani}, \citenamefont {Kraus}, \citenamefont {MacDonald}, \citenamefont {Moldabekov}, \citenamefont {Schwalbe}, \citenamefont {Gericke},\ and\ \citenamefont {Vorberger}}]{Dornheim_Science_2024}%
  \BibitemOpen
  \bibfield  {author} {\bibinfo {author} {\bibfnamefont {Tobias}\ \bibnamefont {Dornheim}}, \bibinfo {author} {\bibfnamefont {Tilo}\ \bibnamefont {Döppner}}, \bibinfo {author} {\bibfnamefont {Panagiotis}\ \bibnamefont {Tolias}}, \bibinfo {author} {\bibfnamefont {Maximilian}\ \bibnamefont {Böhme}}, \bibinfo {author} {\bibfnamefont {Luke}\ \bibnamefont {Fletcher}}, \bibinfo {author} {\bibfnamefont {Thomas}\ \bibnamefont {Gawne}}, \bibinfo {author} {\bibfnamefont {Frank}\ \bibnamefont {Graziani}}, \bibinfo {author} {\bibfnamefont {Dominik}\ \bibnamefont {Kraus}}, \bibinfo {author} {\bibfnamefont {Michael}\ \bibnamefont {MacDonald}}, \bibinfo {author} {\bibfnamefont {Zhandos}\ \bibnamefont {Moldabekov}}, \bibinfo {author} {\bibfnamefont {Sebastian}\ \bibnamefont {Schwalbe}}, \bibinfo {author} {\bibfnamefont {Dirk}\ \bibnamefont {Gericke}}, \ and\ \bibinfo {author} {\bibfnamefont {Jan}\ \bibnamefont {Vorberger}},\ }\href@noop {} {\enquote {\bibinfo {title} {Unraveling electronic correlations in warm dense
  quantum plasmas},}\ } (\bibinfo {year} {2024}{\natexlab{b}}),\ \Eprint {http://arxiv.org/abs/2402.19113} {arXiv:2402.19113 [physics.plasm-ph]} \BibitemShut {NoStop}%
\bibitem [{\citenamefont {Vorberger}\ \emph {et~al.}(2024)\citenamefont {Vorberger}, \citenamefont {Preston}, \citenamefont {Medvedev}, \citenamefont {Böhme}, \citenamefont {Moldabekov}, \citenamefont {Kraus},\ and\ \citenamefont {Dornheim}}]{Vorberger_PLA_2024}%
  \BibitemOpen
  \bibfield  {author} {\bibinfo {author} {\bibfnamefont {Jan}\ \bibnamefont {Vorberger}}, \bibinfo {author} {\bibfnamefont {Thomas~R.}\ \bibnamefont {Preston}}, \bibinfo {author} {\bibfnamefont {Nikita}\ \bibnamefont {Medvedev}}, \bibinfo {author} {\bibfnamefont {Maximilian~P.}\ \bibnamefont {Böhme}}, \bibinfo {author} {\bibfnamefont {Zhandos~A.}\ \bibnamefont {Moldabekov}}, \bibinfo {author} {\bibfnamefont {Dominik}\ \bibnamefont {Kraus}}, \ and\ \bibinfo {author} {\bibfnamefont {Tobias}\ \bibnamefont {Dornheim}},\ }\bibfield  {title} {\enquote {\bibinfo {title} {Revealing non-equilibrium and relaxation in laser heated matter},}\ }\href {\doibase https://doi.org/10.1016/j.physleta.2024.129362} {\bibfield  {journal} {\bibinfo  {journal} {Physics Letters A}\ }\textbf {\bibinfo {volume} {499}},\ \bibinfo {pages} {129362} (\bibinfo {year} {2024})}\BibitemShut {NoStop}%
\bibitem [{\citenamefont {Dornheim}\ \emph {et~al.}(2023{\natexlab{e}})\citenamefont {Dornheim}, \citenamefont {Wicaksono}, \citenamefont {Suarez-Cardona}, \citenamefont {Tolias}, \citenamefont {B\"ohme}, \citenamefont {Moldabekov}, \citenamefont {Hecht},\ and\ \citenamefont {Vorberger}}]{Dornheim_PRB_2023}%
  \BibitemOpen
  \bibfield  {author} {\bibinfo {author} {\bibfnamefont {Tobias}\ \bibnamefont {Dornheim}}, \bibinfo {author} {\bibfnamefont {Damar~C.}\ \bibnamefont {Wicaksono}}, \bibinfo {author} {\bibfnamefont {Juan~E.}\ \bibnamefont {Suarez-Cardona}}, \bibinfo {author} {\bibfnamefont {Panagiotis}\ \bibnamefont {Tolias}}, \bibinfo {author} {\bibfnamefont {Maximilian~P.}\ \bibnamefont {B\"ohme}}, \bibinfo {author} {\bibfnamefont {Zhandos~A.}\ \bibnamefont {Moldabekov}}, \bibinfo {author} {\bibfnamefont {Michael}\ \bibnamefont {Hecht}}, \ and\ \bibinfo {author} {\bibfnamefont {Jan}\ \bibnamefont {Vorberger}},\ }\bibfield  {title} {\enquote {\bibinfo {title} {Extraction of the frequency moments of spectral densities from imaginary-time correlation function data},}\ }\href {\doibase 10.1103/PhysRevB.107.155148} {\bibfield  {journal} {\bibinfo  {journal} {Phys. Rev. B}\ }\textbf {\bibinfo {volume} {107}},\ \bibinfo {pages} {155148} (\bibinfo {year} {2023}{\natexlab{e}})}\BibitemShut {NoStop}%
\bibitem [{\citenamefont {Ceperley}(1991)}]{Ceperley1991}%
  \BibitemOpen
  \bibfield  {author} {\bibinfo {author} {\bibfnamefont {D.~M.}\ \bibnamefont {Ceperley}},\ }\bibfield  {title} {\enquote {\bibinfo {title} {Fermion nodes},}\ }\href {\doibase 10.1007/BF01030009} {\bibfield  {journal} {\bibinfo  {journal} {Journal of Statistical Physics}\ }\textbf {\bibinfo {volume} {63}},\ \bibinfo {pages} {1237--1267} (\bibinfo {year} {1991})}\BibitemShut {NoStop}%
\bibitem [{\citenamefont {Dornheim}(2019)}]{dornheim_sign_problem}%
  \BibitemOpen
  \bibfield  {author} {\bibinfo {author} {\bibfnamefont {T.}~\bibnamefont {Dornheim}},\ }\bibfield  {title} {\enquote {\bibinfo {title} {Fermion sign problem in path integral {M}onte {C}arlo simulations: Quantum dots, ultracold atoms, and warm dense matter},}\ }\href {https://journals.aps.org/pre/abstract/10.1103/PhysRevE.100.023307} {\bibfield  {journal} {\bibinfo  {journal} {Phys. Rev. E}\ }\textbf {\bibinfo {volume} {100}},\ \bibinfo {pages} {023307} (\bibinfo {year} {2019})}\BibitemShut {NoStop}%
\bibitem [{\citenamefont {Holas}(1987)}]{holas_limit}%
  \BibitemOpen
  \bibfield  {author} {\bibinfo {author} {\bibfnamefont {A.}~\bibnamefont {Holas}},\ }\bibfield  {title} {\enquote {\bibinfo {title} {Exact asymptotic expression for the static dielectric function of a uniform electron liquid at large wave vector},}\ }in\ \href@noop {} {\emph {\bibinfo {booktitle} {Strongly Coupled Plasma Physics}}},\ \bibinfo {editor} {edited by\ \bibinfo {editor} {\bibfnamefont {F.J.}\ \bibnamefont {Rogers}}\ and\ \bibinfo {editor} {\bibfnamefont {H.E.}\ \bibnamefont {DeWitt}}}\ (\bibinfo  {publisher} {Plenum},\ \bibinfo {address} {New York},\ \bibinfo {year} {1987})\BibitemShut {NoStop}%
\bibitem [{\citenamefont {Pathak}\ and\ \citenamefont {Vashishta}(1973)}]{PathakVashishtaScheme}%
  \BibitemOpen
  \bibfield  {author} {\bibinfo {author} {\bibfnamefont {K.~N.}\ \bibnamefont {Pathak}}\ and\ \bibinfo {author} {\bibfnamefont {P.}~\bibnamefont {Vashishta}},\ }\bibfield  {title} {\enquote {\bibinfo {title} {Electron correlations and moment sum rules},}\ }\href {\doibase 10.1103/PhysRevB.7.3649} {\bibfield  {journal} {\bibinfo  {journal} {Phys. Rev. B}\ }\textbf {\bibinfo {volume} {7}},\ \bibinfo {pages} {3649--3656} (\bibinfo {year} {1973})}\BibitemShut {NoStop}%
\bibitem [{\citenamefont {Niklasson}(1974)}]{NiklassonLimit}%
  \BibitemOpen
  \bibfield  {author} {\bibinfo {author} {\bibfnamefont {G\"oran}\ \bibnamefont {Niklasson}},\ }\bibfield  {title} {\enquote {\bibinfo {title} {Dielectric function of the uniform electron gas for large frequencies or wave vectors},}\ }\href {\doibase 10.1103/PhysRevB.10.3052} {\bibfield  {journal} {\bibinfo  {journal} {Phys. Rev. B}\ }\textbf {\bibinfo {volume} {10}},\ \bibinfo {pages} {3052--3061} (\bibinfo {year} {1974})}\BibitemShut {NoStop}%
\bibitem [{\citenamefont {Singwi}\ and\ \citenamefont {Tosi}(1981)}]{SingwiTosi_Review}%
  \BibitemOpen
  \bibfield  {author} {\bibinfo {author} {\bibfnamefont {K.~S.}\ \bibnamefont {Singwi}}\ and\ \bibinfo {author} {\bibfnamefont {M.~P.}\ \bibnamefont {Tosi}},\ }\bibfield  {title} {\enquote {\bibinfo {title} {Correlations in electron liquids},}\ }\href {\doibase 10.1016/S0081-1947(08)60116-2} {\bibfield  {journal} {\bibinfo  {journal} {Solid State Physics}\ }\textbf {\bibinfo {volume} {36}},\ \bibinfo {pages} {177--266} (\bibinfo {year} {1981})}\BibitemShut {NoStop}%
\bibitem [{\citenamefont {Hou}\ \emph {et~al.}(2022)\citenamefont {Hou}, \citenamefont {Wang}, \citenamefont {Haule}, \citenamefont {Deng},\ and\ \citenamefont {Chen}}]{Hou_PRB_2022}%
  \BibitemOpen
  \bibfield  {author} {\bibinfo {author} {\bibfnamefont {Peng-Cheng}\ \bibnamefont {Hou}}, \bibinfo {author} {\bibfnamefont {Bao-Zong}\ \bibnamefont {Wang}}, \bibinfo {author} {\bibfnamefont {Kristjan}\ \bibnamefont {Haule}}, \bibinfo {author} {\bibfnamefont {Youjin}\ \bibnamefont {Deng}}, \ and\ \bibinfo {author} {\bibfnamefont {Kun}\ \bibnamefont {Chen}},\ }\bibfield  {title} {\enquote {\bibinfo {title} {Exchange-correlation effect in the charge response of a warm dense electron gas},}\ }\href {\doibase 10.1103/PhysRevB.106.L081126} {\bibfield  {journal} {\bibinfo  {journal} {Phys. Rev. B}\ }\textbf {\bibinfo {volume} {106}},\ \bibinfo {pages} {L081126} (\bibinfo {year} {2022})}\BibitemShut {NoStop}%
\bibitem [{\citenamefont {Dornheim}\ \emph {et~al.}(2021{\natexlab{d}})\citenamefont {Dornheim}, \citenamefont {Moldabekov},\ and\ \citenamefont {Tolias}}]{Dornheim_PRB_ESA_2021}%
  \BibitemOpen
  \bibfield  {author} {\bibinfo {author} {\bibfnamefont {Tobias}\ \bibnamefont {Dornheim}}, \bibinfo {author} {\bibfnamefont {Zhandos~A.}\ \bibnamefont {Moldabekov}}, \ and\ \bibinfo {author} {\bibfnamefont {Panagiotis}\ \bibnamefont {Tolias}},\ }\bibfield  {title} {\enquote {\bibinfo {title} {Analytical representation of the local field correction of the uniform electron gas within the effective static approximation},}\ }\href {\doibase 10.1103/PhysRevB.103.165102} {\bibfield  {journal} {\bibinfo  {journal} {Phys. Rev. B}\ }\textbf {\bibinfo {volume} {103}},\ \bibinfo {pages} {165102} (\bibinfo {year} {2021}{\natexlab{d}})}\BibitemShut {NoStop}%
\bibitem [{\citenamefont {Dornheim}\ \emph {et~al.}(2024{\natexlab{c}})\citenamefont {Dornheim}, \citenamefont {Böhme},\ and\ \citenamefont {Schwalbe}}]{ISHTAR}%
  \BibitemOpen
  \bibfield  {author} {\bibinfo {author} {\bibfnamefont {Tobias}\ \bibnamefont {Dornheim}}, \bibinfo {author} {\bibfnamefont {Maximilian}\ \bibnamefont {Böhme}}, \ and\ \bibinfo {author} {\bibfnamefont {Sebastian}\ \bibnamefont {Schwalbe}},\ }\href {\doibase 10.5281/zenodo.10497098} {\enquote {\bibinfo {title} {{ISHTAR - Imaginary-time Stochastic High- performance Tool for Ab initio Research}},}\ } (\bibinfo {year} {2024}{\natexlab{c}})\BibitemShut {NoStop}%
\bibitem [{rep()}]{repo}%
  \BibitemOpen
  \href@noop {} {}\bibinfo {note} {A link to a repository containing all PIMC raw data will be made available upon publication.}\BibitemShut {Stop}%
\bibitem [{\citenamefont {Dornheim}\ \emph {et~al.}(2023{\natexlab{f}})\citenamefont {Dornheim}, \citenamefont {Vorberger}, \citenamefont {Moldabekov},\ and\ \citenamefont {Böhme}}]{Dornheim_PTR_2022}%
  \BibitemOpen
  \bibfield  {author} {\bibinfo {author} {\bibfnamefont {Tobias}\ \bibnamefont {Dornheim}}, \bibinfo {author} {\bibfnamefont {Jan}\ \bibnamefont {Vorberger}}, \bibinfo {author} {\bibfnamefont {Zhandos~A.}\ \bibnamefont {Moldabekov}}, \ and\ \bibinfo {author} {\bibfnamefont {Maximilian}\ \bibnamefont {Böhme}},\ }\bibfield  {title} {\enquote {\bibinfo {title} {Analysing the dynamic structure of warm dense matter in the imaginary-time domain: theoretical models and simulations},}\ }\href {\doibase 10.1098/rsta.2022.0217} {\bibfield  {journal} {\bibinfo  {journal} {Philosophical Transactions of the Royal Society A: Mathematical, Physical and Engineering Sciences}\ }\textbf {\bibinfo {volume} {381}},\ \bibinfo {pages} {20220217} (\bibinfo {year} {2023}{\natexlab{f}})}\BibitemShut {NoStop}%
\bibitem [{\citenamefont {Kugler}(1970)}]{kugler_bounds}%
  \BibitemOpen
  \bibfield  {author} {\bibinfo {author} {\bibfnamefont {A.~A.}\ \bibnamefont {Kugler}},\ }\bibfield  {title} {\enquote {\bibinfo {title} {Bounds for some equilibrium properties of an electron gas},}\ }\href {https://journals.aps.org/pra/abstract/10.1103/PhysRevA.1.1688} {\bibfield  {journal} {\bibinfo  {journal} {Phys. Rev. A}\ }\textbf {\bibinfo {volume} {1}},\ \bibinfo {pages} {1688} (\bibinfo {year} {1970})}\BibitemShut {NoStop}%
\bibitem [{\citenamefont {Kugler}(1973)}]{kugler_classical}%
  \BibitemOpen
  \bibfield  {author} {\bibinfo {author} {\bibfnamefont {A.~A.}\ \bibnamefont {Kugler}},\ }\bibfield  {title} {\enquote {\bibinfo {title} {Collective modes, damping, and the scattering function in classical liquids},}\ }\href {https://doi.org/10.1007/BF01008535} {\bibfield  {journal} {\bibinfo  {journal} {J. Stat. Phys.}\ }\textbf {\bibinfo {volume} {8}},\ \bibinfo {pages} {107--153} (\bibinfo {year} {1973})}\BibitemShut {NoStop}%
\bibitem [{\citenamefont {Tolias}\ and\ \citenamefont {Castello}(2021)}]{tolias2021modes}%
  \BibitemOpen
  \bibfield  {author} {\bibinfo {author} {\bibfnamefont {P.}~\bibnamefont {Tolias}}\ and\ \bibinfo {author} {\bibfnamefont {F.~Lucco}\ \bibnamefont {Castello}},\ }\bibfield  {title} {\enquote {\bibinfo {title} {Description of longitudinal modes in moderately coupled {Yukawa} systems with the static local field correction},}\ }\href {\doibase 10.1063/5.0044871} {\bibfield  {journal} {\bibinfo  {journal} {Phys. Plasmas}\ }\textbf {\bibinfo {volume} {28}},\ \bibinfo {pages} {034502} (\bibinfo {year} {2021})}\BibitemShut {NoStop}%
\bibitem [{\citenamefont {Tanaka}\ \emph {et~al.}(1985)\citenamefont {Tanaka}, \citenamefont {Mitake},\ and\ \citenamefont {Ichimaru}}]{stlspra}%
  \BibitemOpen
  \bibfield  {author} {\bibinfo {author} {\bibfnamefont {Shigenori}\ \bibnamefont {Tanaka}}, \bibinfo {author} {\bibfnamefont {Shinichi}\ \bibnamefont {Mitake}}, \ and\ \bibinfo {author} {\bibfnamefont {Setsuo}\ \bibnamefont {Ichimaru}},\ }\bibfield  {title} {\enquote {\bibinfo {title} {Parametrized equation of state for electron liquids in the {S}ingwi-{T}osi-{Land}-{S}j\"olander approximation},}\ }\href {\doibase 10.1103/PhysRevA.32.1896} {\bibfield  {journal} {\bibinfo  {journal} {Phys. Rev. A}\ }\textbf {\bibinfo {volume} {32}},\ \bibinfo {pages} {1896} (\bibinfo {year} {1985})}\BibitemShut {NoStop}%
\bibitem [{\citenamefont {Dornheim}\ \emph {et~al.}(2024{\natexlab{d}})\citenamefont {Dornheim}, \citenamefont {Schwalbe}, \citenamefont {Böhme}, \citenamefont {Moldabekov}, \citenamefont {Vorberger},\ and\ \citenamefont {Tolias}}]{Dornheim_JCP_2024}%
  \BibitemOpen
  \bibfield  {author} {\bibinfo {author} {\bibfnamefont {Tobias}\ \bibnamefont {Dornheim}}, \bibinfo {author} {\bibfnamefont {Sebastian}\ \bibnamefont {Schwalbe}}, \bibinfo {author} {\bibfnamefont {Maximilian}\ \bibnamefont {Böhme}}, \bibinfo {author} {\bibfnamefont {Zhandos}\ \bibnamefont {Moldabekov}}, \bibinfo {author} {\bibfnamefont {Jan}\ \bibnamefont {Vorberger}}, \ and\ \bibinfo {author} {\bibfnamefont {Panagiotis}\ \bibnamefont {Tolias}},\ }\bibfield  {title} {\enquote {\bibinfo {title} {Ab initio path integral monte carlo simulations of warm dense two-component systems without fixed nodes: structural properties},}\ }\href {\doibase 10.1063/5.0206787} {\bibfield  {journal} {\bibinfo  {journal} {J. Chem. Phys.}\ }\textbf {\bibinfo {volume} {160}},\ \bibinfo {pages} {164111} (\bibinfo {year} {2024}{\natexlab{d}})}\BibitemShut {NoStop}%
\end{thebibliography}%
\end{document}